\documentclass[
  aps,prd,twocolumn,
  nofootinbib,
  superscriptaddress,
  amsmath,amssymb,
  floatfix
]{revtex4-2}

\usepackage{graphicx}
\usepackage{booktabs}
\usepackage{diagbox}
\usepackage{dsfont}
\usepackage{slashed}
\usepackage{amsthm}      % optional; REVTeX has its own theorem styles too
\usepackage{mathtools}   % nice to have; extends amsmath
\usepackage{wrapfig}
\usepackage{enumitem}
\usepackage[caption=false]{subfig} % subfigures compatible with REVTeX
\usepackage[colorlinks=true,linkcolor=blue,citecolor=blue,urlcolor=blue]{hyperref}
\usepackage[nameinlink,capitalise]{cleveref} % load AFTER hyperref

\makeatletter
\renewcommand{\p@section}{}
\renewcommand{\p@subsection}{\thesection.}
\makeatother

\usepackage[normalem]{ulem}
\usepackage{comment}

\newcommand{\Kahler}{\text{K\"ahler} }
\newcommand{\Poincare}{\text{Poincar\'e} }

\usepackage{orcidlink}

\usepackage{fontawesome5}
\usepackage{hyperref}

\makeatletter
\newcommand{\github}[1]{%
   \href{#1}{{\normalsize \color{black}\faGithub}}%
}
\makeatother

\allowdisplaybreaks[2]
\numberwithin{equation}{section}
\begin{document}

\title{Bayesian inference on Calabi--Yau moduli spaces and the axiverse: \\ experimental data meets string theory}

\author{Mudit Jain
\orcidlink{0000-0002-5223-3071}}
% \email{mudit.jain@kcl.ac.uk}
\email{mudit.jain@rice.edu}
\affiliation{Physics Department, King's College London, Strand, London WC2R 2LS, UK}
\affiliation{Department of Physics and Astronomy, Rice University, Houston, Texas 77005, U.S.A.}

\author{Elijah Sheridan \orcidlink{0000-0003-0961-0844}}
\email{es888@cornell.edu}
\affiliation{Department of Physics, Cornell University, Ithaca, NY 14853, USA}

\author{David J.~E.~Marsh \orcidlink{0000-0002-4690-3016}}
\email{david.j.marsh@kcl.ac.uk}
\affiliation{Physics Department, King's College London, Strand, London WC2R 2LS, UK}

\author{Elli Heyes \orcidlink{0000-0002-7608-3485}}
\email{e.heyes@imperial.ac.uk}
\affiliation{Abdus Salam Centre for Theoretical Physics, Imperial College London, London SW7 2AZ, UK}

\author{Keir K. Rogers \orcidlink{0000-0003-1601-8144}}
\email{keir.rogers@kcl.ac.uk}%k.rogers24@imperial.ac.uk
\affiliation{Physics Department, King's College London, Strand, London WC2R 2LS, UK}
\affiliation{Department of Physics, Imperial College London, Blackett Laboratory, Prince Consort Road, London SW7 2AZ, UK}

\author{Andreas Schachner \orcidlink{0000-0002-7287-1476}}
\email{as3475@cornell.edu}
\affiliation{Department of Physics, Cornell University, Ithaca, NY 14853, USA}
\affiliation{ASC for Theoretical Physics, LMU Munich, 80333 Munich, Germany}

\begin{abstract}

We develop tools of Bayesian inference on the moduli space of Calabi--Yau (CY) manifolds. We sample from the invariant Weil--Petersson (WP) measure using Markov Chain Monte Carlo and normalising flows on \Kahler moduli space with dimension up to $h^{1,1}=30$, and present results on the spectrum of the CY volume and properties of divisors when the measure is restricted in physically meaningful ways. We furthermore present a theory-informed prior on axion masses and decay constants $(m_a,f_a)$ marginalised over the WP measure for \emph{all} inequivalent CYs constructable from the Kreuzer--Skarke database with $h^{1,1}\leq 5$. We then impose likelihoods based on axion physics. We demonstrate how detection of a relatively heavy QCD axion at small $h^{1,1}$, e.g. by ADMX, provides detailed information about CY geometry and topology. Finally, we compute a full forward model incorporating likelihoods from the cosmic microwave background and Lyman-alpha forest and find the maximum posterior probability region on the moduli space of a given CY favoured by a resolution of the tension in these data by an ultralight axion composing $\mathcal{O}(1\%)$ of the dark matter. This demonstration serves as a blueprint for future statistical analyses within string phenomenology. \github{https://github.com/AndreasSchachner/kahler_cone_sampler}

\end{abstract}

%\keywords{Calabi--Yau manifolds, axions, Bayesian inference, moduli spaces}

\maketitle
\bigskip
\newpage

%\tableofcontents

\section{Introduction}\label{sec:intro}

String theory compactified on Calabi--Yau (CY) orientifolds is one of the best candidate theories to describe quantum gravity and particle physics in a unified manner~\cite{Candelas:1985en}. Recent years have seen great advances in our ability to study CY manifolds, and the resulting physics, numerically~\cite{Demirtas:2022hqf}. Axions~\cite{Peccei:1977hh,Weinberg:1977ma,Wilczek:1977pj} are a ubiquitous feature of string theory compactifications~\cite{Witten:1984dg,Svrcek:2006yi,Conlon:2006tq}, known as the \textit{axiverse}~\cite{Arvanitaki:2009fg,Acharya:2010zx,Cicoli:2012sz}, and the hunt for axions in the laboratory and in cosmology is currently undergoing intense progress~\cite{Chadha-Day:2021szb,Adams:2022pbo,OHare:2024nmr}. Advances in the explicit constructions of CYs have led to similar progress in our detailed understanding of the axiverse~\cite{Demirtas:2018akl,Sheridan:2024vtt,Gendler:2023kjt,Gendler:2024adn,Mehta:2021pwf,Demirtas:2021gsq,Benabou:2025kgx,Yin:2025amn,Cheng:2025ggf,MacFadden:2024him,Fallon:2025lvn}. The basic $C_4$ axion effective field theory (EFT) in type IIB string theory is well-understood: its kinetic matrix is polynomial in subvariety volumes and its potential is dominated by exponentials of divisor volumes. Divisor volumes are not, however, independent free parameters: in explicit CY geometries they are often strongly correlated, meaning that not all things possible from an EFT perspective are realised in string theory at the ensemble level. Outside specific CY models and beyond type IIB, other important recent progress includes Refs.~\cite{Petrossian-Byrne:2025mto,Reig:2025dqb,Loladze:2025uvf,Leedom:2025mlr}.

The progress on axions in explicit CY ensembles has identified important general principles. Axion masses are approximately log-uniformly distributed and axion decay constants are approximately log-normally distributed. The scale of the largest mass and mean decay constant are set by the Hodge number $h^{1,1}$ of the CY (which also sets the number of axions), with both decreasing as $h^{1,1}$ increases~\cite{Mehta:2021pwf}. To date, most of this research has been performed at a canonical but specific region in the moduli space: namely, along the ray generated by the ``tip of the stretched \Kahler cone'' (see \cref{sec:preliminaries}) and relatively close to the origin (e.g., \cite{Demirtas:2018akl,Demirtas:2021gsq,Mehta:2021pwf,Gendler:2023kjt,Gendler:2024adn,Benabou:2025kgx}). In this work we lay the groundwork for a systematic and statistically rigorous approach to studying $C_4$ axion models in the full stretched \Kahler cone.

%%%%%%%%%%%%%%
\begin{figure*}
    \centering
    \includegraphics[width=0.99\linewidth]{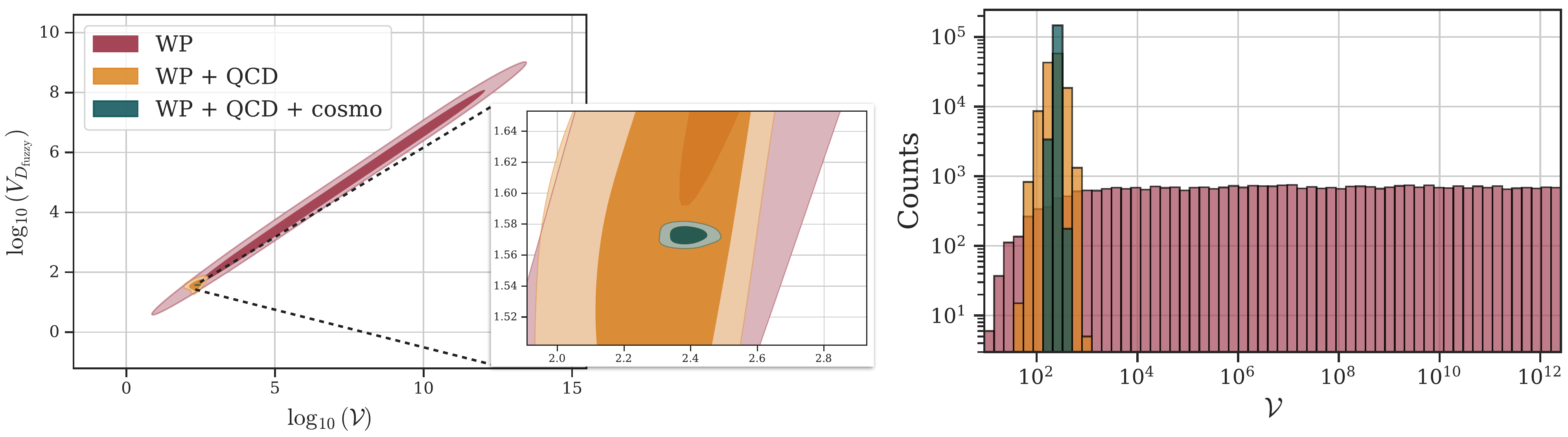}
    \caption{Joint posterior distribution of $\log_{10}\mathcal{V}$ and $\log_{10}(\tau_{\rm fuzzy})$ obtained by sampling the \Kahler moduli of a Calabi--Yau with $h^{1,1}=7$, where $\tau_{\rm fuzzy}$ is the volume of the prime toric divisor hosting the instanton associated with the ultralight (fuzzy) axion. We show three cases: (i) the Weil--Petersson prior alone (with IR/UV cutoffs from EFT control and the KK scale), (ii) the Weil--Petersson prior supplemented by the requirement that at least one prime toric divisor attains a volume suitable for a QCD-like gauge coupling, and (iii) the same prior further weighted by CMB and Lyman--$\alpha$ likelihoods. The inset zooms into the small region where the cosmology-selected posterior concentrates. The right panel shows the corresponding marginal posterior of $\log_{10}\mathcal{V}$ for all three cases, illustrating how cosmological data lead to a sharply peaked posterior and a strongly localised region in moduli space.}
    \label{fig:money_plot}
\end{figure*}
%%%%%%%%%%%%%%%%

A central challenge arising from such a program of study is the problem of \textit{moduli}. The \Kahler and complex structure moduli are scalar fields representing geometric deformations of the CY that preserve the defining property of Ricci flatness~\cite{Candelas:1990pi}. They are massless in the classical theory, but are believed to be stabilised, i.e., given masses and vacuum expectation values (vevs), by quantum effects (see Ref.~\cite{McAllister:2023vgy} for a review). There has also been significant progress in understanding moduli stabilisation in explicit examples~\cite{Louis:2012nb,Cicoli:2013cha,Crino:2020qwk,AbdusSalam:2020ywo,Demirtas:2021nlu,McAllister:2024lnt,AbdusSalam:2025twp}, yet even the most advanced models are still far from answering key questions about supersymmetry breaking and the cosmological constant in phenomenologically acceptable ways.

The values of the stabilised moduli fields (moduli vevs) can be seen as constant parameters of the four-dimensional EFT. The physics giving rise to moduli stabilisation, such as fluxes \cite{Giddings:2001yu} and D-brane instantons \cite{Witten:1996bn} is, however, fundamentally discrete, meaning that the true spectrum of allowed vacua must also be discrete. Nevertheless, it has long been appreciated that the available choices are so numerous that this discrete spectrum can effectively be treated as continuous, or more precisely, as forming a dense ``discretuum''~\cite{Bousso:2000xa,Denef:2004ze}. Following this reasoning, we approximate the vacuum expectation values of the \Kahler moduli, $\langle T_i \rangle$, as continuous variables. There exist physical effects that may contribute to moduli stabilisation but cannot currently be computed with reliability, i.e., known unknowns, such as loop corrections to the \Kahler potential; see, however, \cite{Berg:2004ek,Berg:2005ja}. In addition, there may be mechanisms for moduli stabilisation that remain entirely undiscovered, i.e., unknown unknowns.

Given the formidable challenge of determining where acceptable vacua populate moduli space, it may seem fruitless to address further questions of physics. In this work, however, we take the ``modular approach'' to phenomenology~\cite{McAllister:2023vgy}. Although we are presently unable to solve the problem of vacuum distributions across moduli space for large ensembles of CYs, we can nevertheless compute certain features of low energy EFTs throughout moduli space. Our aim is therefore to \emph{statistically sample} the physical properties of string compactifications. Provided that such sampling is computationally efficient, as well as \emph{fair and unbiased} (in the Bayesian sense), moduli stabilisation may then be imposed \emph{ex post facto}, allowing the resulting phenomenological consequences to be inferred. 

We thus set aside the problem of explicit moduli stabilisation and instead endeavor to map out \Kahler moduli space and the resulting physics statistically, for which we introduce tools of Bayesian inference and machine learning.\footnote{We note that the complex structure moduli, at our level of approximation, do not affect the physics likelihoods studied here, see however Refs.~\cite{Dubey:2023dvu,Ebelt:2023clh,Krippendorf:2023idy,Chauhan:2025rdj} for progress in this direction.} The \textit{Weil--Petersson} (WP) measure, induced by the WP metric for \Kahler moduli, is the natural measure on \Kahler moduli space (see \cref{app:kahler_metric} for some details). This metric exhibits several useful features for our purposes. For example, in addition to being diffeomorphism invariant, the WP measure is also scale invariant, as expected for an uninformative prior in the absence of a preferred scale, as parametrised by e.g. the CY volume $\mathcal{V}$ (which could later be set by, for example, a particular moduli stabilisation scheme). The WP measure is thus similar to the Jeffreys prior, and is therefore attractive from a purely Bayesian perspective. Consequently, in this work we develop efficient methods for sampling the WP measure. Using the sampling techniques of Markov Chain Monte Carlo (MCMC) and normalising flows (NF), our goal is to compute axion physics observables, compare them with experimental or observational likelihoods, and perform Bayesian inference.

Since we are concerned with phenomenology, we impose extra restrictions on top of the WP prior. In particular, we impose \textit{stretching}: i.e., the constraint that all holomorphic submanifolds have volume larger than unity in string units, thereby controlling the EFT. We also impose that the Kaluza-Klein (KK) scale must be above some physical energy scale to ensure consistency with observations, which implies a maximum CY volume. These two EFT restrictions break WP scale invariance, but have the welcome property of rendering the WP measure a proper (normalisable) prior. The shape of the region in moduli space satisfying our upper and lower bounds on the overall volume as well as our stretching conditions is sensitive to the topology of the CY manifold (such as its fibrations, see \cref{sec:geom_preq}). In particular, this region features limits near the boundary of the \Kahler cone where divisors with particular topological properties have distinctive volume spectra (see \cref{sec:geometry_results}), which we will connect to axion physics in \cref{sec:top_from_halo}. This also extends prior work relating divisor topology to statistics of volumes~\cite{Cheng:2025ggf}.

We assume that the Standard Model gauge group, and in particular quantum chromodynamics (QCD), is realised by a stack of D7-branes on some prime toric divisor --- distinguished divisors on toric hypersurface Calabi--Yau threefolds --- and we demand that the volume distribution of this divisor is relatively sharply peaked at $40$. This ensures that the running of the SM gauge couplings to the KK scale is not drastically altered compared to that implied by the observed particle content~\cite{ParticleDataGroup:2024cfk}. Allowing any such choice of prime toric divisor implicitly assumes the existence of an orientifold projection and a tadpole-cancelling D7/O7 configuration; given the variety of gauge realisations in F-theory, this is a reasonable idealisation. In this work we \textit{assume} $h^{1,1}_-=0$ and therefore restrict to the orientifold-even sector. This assumption should not be taken to imply that superpotential-relevant O(1) instantons are generically available: their existence is subject to consistency requirements arising from fermionic zero modes and the Freed--Witten condition~\cite{Grimm:2007xm,Grimm:2011dj,Cicoli:2021tzt}. For the purposes of this work we remain agnostic about these details and focus on developing and testing tools for efficiently sampling \Kahler moduli space for Bayesian inference. A systematic treatment of fluxed instantons and the general case with $h^{1,1}_-\neq 0$ will be presented in forthcoming work~\cite{2-form}.

We develop tools to efficiently sample the WP measure in geometries with $h^{1,1}$ as large as 30, and show that modest computational resources could extend this sampling to much larger values. We furthermore construct a theory-informed prior on axion parameters $(m_a,f_a)$ from an \textit{exhaustive} treatment of CYs from the Kreuzer--Skarke (KS) database with $h^{1,1}\leq 5$ under the WP measure.\footnote{The Kreuzer--Skarke database of CY threefolds arising from fine, regular, star triangulations (FRSTs) of four-dimensional reflexive polytopes is quite degenerate, in that many FRSTs induce the same CY manifold. Using the methods of \cite{MacFadden:2023cyf}, we can efficiently remove a large fraction of this degeneracy, but not all of it. Removing all degeneracy is a difficult problem \cite{Gendler:2023ujl,Chandra:2023afu} that we will not undertake in this work.} We impose model likelihoods for future QCD axion detection in the laboratory, and demonstrate how this can restrict CY topology and the distribution on moduli space. 

As a representative of our results, Fig.~\ref{fig:money_plot} summarizes a key achievement of this work. The figure shows the posterior probability on CY moduli space derived from three separate combinations of priors and likelihoods for a specific model with $h^{1,1}=7$ (see Sec. 4.2.3 in Ref.~\cite{Sheridan:2024vtt} for more details of the CY). Sampling from only the WP prior gives a log-flat spectrum for the Calabi-Yau volume $\mathcal{V}$ -- scale invariant -- between a minimum value set by control over the EFT, and a maximum set by the KK scale. The second set of samples adds to this a prior that there exists at least one prime toric divisor with volume in a range around $40$ in string units, such that it is possible in principle to host the QCD gauge group on a stack of D7-branes with a realistic gauge coupling in the UV, which further restricts $\mathcal{V}\lesssim 10^3$. Finally, we show the result of imposing cosmological likelihoods from the CMB and the Lyman alpha forest in a full forward model. The preference in this data combination for an ultralight axion dark matter (DM) component~\cite{Rogers:2023upm} strongly restricts the region in moduli space, leading to a sharply peaked posterior, visible here as a preferred value of the marginal distribution on the CY volume, and even more strikingly in the joint posterior on the CY volume and a selected divisor volume.

This paper is organised as follows. In \cref{sec:preliminaries} we discuss the \Kahler cone and WP metric of CY threefolds, as well as other relevant aspects of Calabi--Yau topology and geometry. In \cref{sec:sampling} we develop efficient sampling methods for the WP measure, including MCMC and normalising flows. In \cref{sec:geometry_results} we present geometric results under the WP prior, including the behaviour and distribution of volumes of the various divisor classes across different geometries. In \cref{sec:axion_results} we apply our sampling methods to axion physics, construct theory-informed priors for $(m_a,f_a)$, incorporate QCD-like and cosmological likelihoods, and analyse the resulting constraints on CY topology and moduli space. We summarise our findings in \cref{sec:conclusions}. The appendices contain technical material on the axion EFT, the normalising-flow architecture, aspects of the WP measure, and Calabi--Yau topology and geometry. All CY computations in this work employ \textsc{CYTools}~\cite{Demirtas:2022hqf}. Our results are publicly available on GitHub \github{https://github.com/AndreasSchachner/kahler_cone_sampler}.

\section{\Kahler moduli spaces of Calabi--Yau threefolds}\label{sec:preliminaries}

\subsection{\Kahler moduli and the \Kahler metric}

We are here concerned with the question of how physics varies across the moduli space of CY compactifications of string theory~\cite{Candelas:1990pi}. We consider the \Kahler moduli space of a CY orientifold $X$ given by the \Kahler parameters $t^i$ of dimension $h^{1,1}=h^{1,1}_+$. These parameters take values in the \Kahler cone $K_X\subset H^{2}(X,\mathbb{R})$ corresponding to the region where volumes of subvarieties are positive.
The complexified \Kahler moduli space of type IIB string theory can be parametrised by coordinates
\begin{equation}
    \label{eq:kahler_moduli}
    T_i=\tau_i+\mathrm{i}\, \phi_i = \frac{1}{2}\int_{X} J\wedge J\wedge\omega_i + \mathrm{i}\int_{X} C_4\wedge\omega_i\, , 
\end{equation}
where $C_4$ is the four-form with the associated axions $\phi_i$, $J= t^i\,\omega_i$ is the \Kahler form on $X$ in a basis $\omega_i \in H^2(X,\mathbb{Z})$ and
\begin{equation}\label{eq:divvols}
    \tau_{i} = \dfrac{1}{2}\kappa_{ijk}t^{j}t^{k}\; , \quad \kappa_{ijk} = \int_X \omega_{i}\wedge\omega_{j}\wedge\omega_{k}\, .
\end{equation}
Here, $\tau_i$ correspond to the volumes of the 4-cycles Poincar\'e dual to the $\omega_i$ and $\kappa_{ijk}$ are the triple intersection numbers of $X$.

The classical \Kahler potential for the \Kahler moduli is\footnote{The \Kahler potential in Eq.~\eqref{eq:KPcl} and the real parts of the Kähler coordinates in Eq.~\eqref{eq:kahler_moduli} receive both perturbative and non-perturbative corrections, see e.g. \cite{McAllister:2024lnt} for a discussion. These effects modify the classical WP metric, yielding a quantum-corrected geometry which becomes essential once the moduli are taken outside the strict geometric regime, i.e., outside the SKC. We will at points consider values of tree-level (i.e. purely geometric) quantities outside of the SKC (such as when diagnosing the existence of fibrations), but this is merely to infer topological properties of the geometry rather than physical quantities.} 
\begin{equation}\label{eq:KPcl}
    \mathcal{K}=-2\ln(\mathcal{V}) \; , \quad \mathcal{V}=\dfrac{1}{6}\kappa_{ijk}t^i t^j t^k
\end{equation}
where $\mathcal{V}$ is the CY volume in string units~\cite{Grimm:2004uq}. From the \Kahler potential, the \Kahler metric is given by $K^{ij} = \partial^2\mathcal{K}/\partial T_i\partial\bar{T}_j$,\footnote{We note that this convention differs from other references, such as \cite{Sheridan:2024vtt, Gendler:2023kjt}, which denote the \Kahler metric with lowered indices and its inverse with upper indices.} and the four-dimensional kinetic term for the divisor volumes $\tau_{i}$ is
\begin{align}
\label{eq:L_kin_1}
    \mathcal{L}_{\rm kin} \supset K^{ij}\,{\rm d}\tau_{i} \wedge\star {\rm d}\tau_{j}\,.
\end{align}

The classical metric on moduli space is independent of the axions, see also \cref{app:axionEFT}.
Using \eqref{eq:divvols}, the kinetic terms for the \Kahler parameters $t^i$ become
\begin{align}
\label{eq:L_kin_2}
    \mathcal{L}_{\rm kin} &\supset M_{ij}\,{\rm d}t^{i}\wedge\star {\rm d}t^{j}\,,\quad{\rm where}\nonumber\\
    M_{ij} &= K^{ab}\kappa_{aci}\kappa_{bdj}t^{c}t^{d} = \frac{1}{4\mathcal{V}^2}(2\tau_i\tau_j - \mathcal{V}\kappa_{ijk}t^k)\,.
\end{align}
Therefore, a natural metric defined on the space of \Kahler parameters $t^i$ is simply $M_{ij}$ referred to as the \emph{Weil--Petersson} (WP) metric (see also \cref{app:kahler_metric}),\footnote{The WP metric can be shown to be equivalent to other more fundamental metrics like the Zamolodchikov metric \cite{Candelas:1989qn} which is a special case of the information metric, see e.g. \cite{Stout:2022phm}.} giving the following probability measure
\begin{align}
\label{eq:WP_basic}
    {\rm D}_{\rm WP} \coloneqq \sqrt{|\det M|}\prod_i{\rm d}t^i\,.
\end{align}
This natural measure on the \Kahler cone $K_X$ is known as the WP measure.

\subsection{Parametrising the \Kahler moduli space}

Let us begin by parametrising the \Kahler cone $K_X$. It can be constructed as
\begin{align}
    K_X = \{t^{i} \;|\; t^{i} = \alpha^{\mu}\mathcal{G}^{i}_{\;\mu} \;;\; {\alpha^{\mu}} > 0\}\,,
\end{align}
where $\mathcal{G}^i_{\;\mu}$ is the $(N_g\times h^{1,1})$-dimensional matrix whose rows are the extremal generators of $K_X$ with $N_g$ denoting their total number. The measure induced by WP on the space of $\alpha$ coefficients becomes
\begin{align}
\label{eq:WP_measure_alpha}
    {\rm D}^{\mathcal{G}}_{\rm WP} = |\det \mathcal{G}|\sqrt{|\det M|}\prod^{N_g}_{\mu=1}{\rm d}{\alpha}^{\mu}\,\Theta(\alpha^{\mu}>0)\,.
\end{align}

For the purposes of traversing the cone efficiently, we further re-parametrise the positive $\alpha$ coefficients space, $\mathbb{R}^{N_g}_{+}$, into reals, $\mathbb{R}^{N_g}$, by exponentiation $\alpha^{\mu} = \exp(\theta^{\mu})$. Such a parametrisation and scanning in the space of $\theta$'s results in robust and efficient sampling of the \Kahler cone, since far away points that are deep within the cone --- points with exponentially large CY volumes --- are easily accessible:
in particular, multiplicative separations in the $\alpha$'s become additive separations in the $\theta$'s, so that regions corresponding to exponentially large $\alpha^\mu$ lie only a linear distance $|\Delta\theta^\mu|$ away. This parametrisation also removes the hard boundary at $\alpha^\mu = 0$ and allows proposal steps to explore large-volume interior regions with uniform numerical efficiency. The WP measure in terms of the target variables $\theta^{\mu}$ is therefore
\begin{align}
\label{eq:WP_extremalgeneratorsampling}
    {\rm D}^{\mathcal{G}}_{\rm WP} = |\det \mathcal{G}|\sqrt{|\det M|}\prod^{N_g}_{\mu = 1}{\rm d}{\theta}^{\mu}\,\alpha^{\mu}\,.
\end{align}

In general, especially as $h^{1,1}$ increases, the majority of \Kahler cones are non-simplicial ($N_g > h^{1,1}$). Thus, due to the large number of additional parameters, it becomes computationally difficult to sample non-simplicial, and consequently narrow, cones. Furthermore, it becomes computationally challenging to even determine all of the cone generators. Finally, for fibered CY manfolds, the support of the WP measure has non-compact regions (as we discuss in \cref{sec:geom_preq}) which can be narrow depending on the choice of basis, which also poses a challenge for sampling.

To circumvent these issues, we make use of the dual parametrisation of \Kahler cones, i.e., using their extremal facets (extremal codimension-one faces, or extremal rays of the dual cone known as the Mori cone). Let $\mathcal{H}$ be the $(N_h\times h^{1,1})$-dimensional matrix whose rows are the extremal facet normal vectors that carve out the \Kahler cone; that is,
\begin{align}
\label{eq:Kahlercone_facetnormlas}
    K_X = \{t^{i} \;|\; {\mathcal{H}^{\mu}_{\;\;i}t^{i}} > 0\}\,.
\end{align}
By extremal we mean that for any $h^{1,1}\times h^{1,1}$ sub-block that is a full rank square matrix, no other rows are expressible as \textit{positive} linear superpositions of the rows of this block. We pick the first such full rank $h^{1,1}$ rows of $\mathcal{H}$, call it $\bar{\mathcal{H}}^{j}_{\;i}$, and perform a change of basis:
\begin{align}\label{eq:tbarbasis}
    \bar{t}^{j} = \bar{\mathcal{H}}^{j}_{\;i}t^{i}\,,
\end{align}
In the new basis, the hyperplane matrix takes the form
\begin{align}
\label{eq:hyperplanes}
    \bar{\mathcal{H}}^{\mu}_{\;i} = \mathcal{H}^{\mu}_{\;j}\,[\bar{\mathcal{H}}^{-1}]^{j}_{\;i} = 
    \begin{pmatrix}
        \mathbb{I}_{h^{1,1}}\\
        \mathcal{B}
    \end{pmatrix}
\end{align}
where $\mathcal{B}$ is a $(N_h-h^{1,1})\times h^{1,1}$ dimensional matrix with $N_h$ being the number of extremal facets, rendering the \Kahler cone as
\begin{align}
\label{eq:Kahlercone_facetnormlas_2}
    \bar{K}_X = \{\bar{t}^{i} \;|\; \bar{t}^{i} > 0 \;{\rm and}\;{\mathcal{B}^{a}_{\;i}\bar{t}^{i}} > 0\}\,.
\end{align} 
That is, the transformed \Kahler cone is guaranteed to be contained within the positive orthant $\mathbb{R}^{h^{1,1}}_{+}$. As before, we can then parametrise the \Kahler parameters using an exponential map, $\bar{t}^i \coloneqq \exp(\bar{\theta}^i)$, in order to efficiently traverse it. The WP measure in the new basis and after using ${\rm d}\bar{t}^i={\rm d}\bar{\theta}^i\,\bar{t}^i$, is simply
\begin{align}
\label{eq:WP_directsampling}
    {\rm D}^{\mathcal{H}}_{\rm WP} &\coloneqq  \tfrac{\sqrt{|\det M|}}{|\det \bar{\mathcal{H}}|}\prod^{N_h - h^{1,1}}_{a=1}\Theta({\mathcal{B}^{a}_{\;i}\bar{t}^{i}})\prod^{h^{1,1}}_{i=1}{\rm d}\bar{\theta}^i\,\bar{t}^i
\end{align}

With the above parametrisations of the \Kahler cone and the associated WP measures (priors), we can impose additional constraints for physical/phenomenological considerations. In the context of a string compactification, we restrict ourselves to the \emph{stretched \Kahler cone} (SKC)~\cite{Demirtas:2018akl} --- the region of moduli space where volumes of all holomorphic curves are larger than unity (in string units). In fact, we shall also require that we remain within the \emph{divisor stretched \Kahler cone} (DSKC) --- the region of moduli space where all prime toric divisor volumes are also larger than unity (in string units). It is within this divisor stretched cone that we can be most confident that the four dimensional EFTs (see \cref{app:axionEFT}) are under perturbative control. From the \Kahler cone $K_X$, the DSKC can be obtained as
\begin{align}
    K_{X,s} 
    &= K_X\cap \{\, t^i \mid H^{a}_{i}\, t^i \ge 1 \,\}\cap \{\tau_{I} > 1\}\, ,
\end{align}
where $H$ is the matrix of hyperplane normal vectors associated with \emph{all} holomorphic curve constraints (not only the extremal facets), and the index $I$ runs over all prime toric divisors ($1$ to $h^{1,1}+4$). The inequalities $H^{a}_{i}\, t^i \ge 1$ enforce that the volumes of all effective curves exceed unity, thereby converting the conical region into an affine polyhedron.

In the absence of explicit extremal generators, we take
\begin{align}
    \bar{K}_{X,s} = \{\bar{t}^{i} \;|\; \bar{t}^{i} > 0 \,,\,{B^{a}_{\;\;i}\bar{t}^{i}} > 1\}\cap \{\tau_{I} > 1\},
\end{align}
where $B$ is the equivalent of~\cref{eq:hyperplanes}, but comprising \emph{all} the non-basis hyperplane vectors. That is, the non-identity (and non-square) matrix contained in $H^{a}_{\;j}\,[\bar{\mathcal{H}}^{-1}]^{j}_{\;i}$.

The restriction of $\tau_I > 1$ in both parametrisations can be imposed as an extra condition on top of the WP measure (\cref{eq:WP_extremalgeneratorsampling} and~\cref{eq:WP_directsampling} respectively).\\

The above discussion extends directly to $K_\cup$, a cone contained in the \Kahler cone of a toric hypersurface CY threefold and often used as an approximation to the true \Kahler cone (which is generally difficult to compute). In particular, $K_\cup$ for a CY hypersurface $X$ in a toric variety $V$ is defined as the union of the toric \Kahler cones of all toric varieties $V'$ related to $V$ by birational operations which do not affect $X$~\cite{CoxKatz1999}. In what follows, all CY threefolds we study are toric hypersurfaces, and we work exclusively with $K_\cup$ (in \textsc{CYTools}, $K_\cup$ can be constructed from a \texttt{CalabiYau} object \texttt{cy} as \texttt{cy.mori\_cone\_cap(in\_basis=True).dual()}).\\

How does the WP measure scale within the stretched \Kahler cone of a CY compactification of type IIB string theory? Using $\mathcal{K} = -2\ln(\mathcal{V})$, the entries of the \Kahler metric $K^{ij}$ scale like $\sim \tau^{-2} \sim t^{-4}$. Using \cref{eq:L_kin_2}, the entries of the $M$ matrix scale like $\sim t^{-2}$, giving $\sqrt{|\det \mathcal{M}|} \sim \sqrt{|\det M|} \sim t^{-n}$ where $n = h^{1,1}$. With this, we see that the WP measure scales with the CY volume in the following fashion: 
\begin{align}
\label{eqn:WP_measure_vol_scaling}
    {\rm D}_{\rm WP} \sim \mathrm{d}\Omega_{n-1} \frac{\mathrm{d}\mathcal{V}}{\mathcal{V}}\sim  \mathrm{d}\Omega_{n-1} \mathrm{d}\mathcal{K}\,,
\end{align}
where $\mathrm{d}\Omega_{n-1}$ is some angular measure that we need not compute explicitly. An important point to note is that the WP measure enforces scale-invariance in the CY volume. While this means that the WP prior is not normalisable a priori, there are lower and upper cut-off CY volumes that can be imposed on practical grounds. The lower cut-off of $\mathcal{V}_{\rm min}$ can simply come from the tip of the stretched \Kahler cone (TSKC), at which point the CY volume may even be larger than unity~\cite{Demirtas:2018akl}. On the other hand, the upper cut-off of $\mathcal{V}_{\rm max}$ could be imposed by, e.g., an experimental lower bound on the SUSY scale, or upper bound on the size of extra dimensions. The scale and diffeomorphism invariance properties of the WP measure appear to be attractive from a Bayesian perspective due to the relation with the Jeffreys prior.

\subsection{Topological and Geometric Prerequisites}
\label{sec:geom_preq}

Upon computing the parameter space for our \Kahler moduli --- the \Kahler cone --- and the axion EFT (see \cref{app:axionEFT}), one might be tempted to set aside the geometric interpretation of the subvariety volumes which determine the kinetic matrix and potential and just treat them as free independent parameters. However, the topology and geometry of the CY furnishes important constraints and correlations on the free parameters in the EFT --- namely, volumes of cycles. In particular, these will also be exploitable, enabling us to both engineer specific physical scenarios using topology, as well as constrain topology with phenomenology.

Three particular geometric phenomena will be of interest to us in this work: boundaries of the \Kahler cone where the overall CY volume vanishes (which correspond to fibrations), boundaries where divisor volumes vanish (which include but are not restricted to boundaries of the former kind), and how the dichotomy between divisors with and without normal bundle deformations imprints itself on axion decay constants. We summarise these three ideas in this section, but include a longer discussion in \cref{app:FibrationsDecayConstants}.

Let us begin with the two types of \Kahler cone boundaries. The classical, geometric moduli space of a Calabi--Yau threefold decomposes as a union of \Kahler cones, forming the extended \Kahler cone. The faces of a \Kahler cone where the CY volume or individual divisor volumes vanish are boundaries not only of that \Kahler cone but of the entire extended \Kahler cone. Even when our stretching condition is imposed, if one approaches these faces of the \Kahler cone, there are imprints on the axion physics. In this way, the near-boundary regions of \Kahler moduli space feature interesting axion phenomenology that can be dissimilar from the ``bulk'' phenomenology near a typical TSKC ray.

Boundaries where the CY volume vanishes are important to us first because they control the support of the WP measure.
As highlighted earlier, the WP measure is normalised only upon imposing bounds on the overall CY volume $\mathcal{V}$. The support of the WP measure --- i.e., the region where the CY volume $\mathcal{V}$ satisfies $\mathcal{V}_{\text{min}} \leq \mathcal{V} \leq \mathcal{V}_{\text{max}}$ --- is compact if and only if $\mathcal{V}$ does not vanish on any face of the \Kahler cone. In particular, the support of the WP measure asymptotically and non-compactly approaches any face where $\mathcal{V}$ vanishes. Such faces are in correspondence with fibrations of $X$, and correspond to limits where both the fiber volume and $\mathcal{V}$ vanish. While imposing stretching constraints will cut off this non-compactness, fibered geometries will still feature long, narrow regions in the \Kahler cone near the faces where $\mathcal{V}$ vanishes, corresponding to the limit of small fiber volume and large base volume. Beyond giving rise to qualitatively distinct axion physics compared to the interior of moduli space, these regions are additionally difficult to sample and thus pose computational challenges.

Fibrations of CY manifolds must have generic fibers that are themselves CY, so the fiber can either be a genus-one curve (i.e., a torus $T^2$) if the fiber is one-dimensional, or either a four-torus $T^4$ or $K3$ surface if the fiber is two-dimensional \cite{Oguiso}. When a one-dimensional fiber admits a section the fibration is elliptic; otherwise it is genus-one with a multisection. For brevity we will still say ``elliptic’’ unless the distinction matters. Likewise, we will abuse notation and refer to the two-dimensional fiber case as a $K3$ fibration. 

Now we turn our attention to \Kahler cone faces where divisors $D$ attain zero volume, $\tau = 0$. Inverse divisor volumes set important scales in string compactifications: for example, $1/\sqrt{\tau}$ sets the gauge coupling on worldvolume gauge theories arising from stacks of D7-branes wrapped on $\mathbb{R}^{1,3} \times D$. Similarly, the limit in which divisors shrink to points is also considered in constructions of Standard Model–like physics on D-branes located at the resulting singularities \cite{Cicoli:2012vw,Cicoli:2013mpa,Cicoli:2013zha,Cicoli:2013cha,Cicoli:2017shd,Cicoli:2021dhg}. Analogous to our discussion of the overall volume for fibrations, if $D$ shrinks on a face of the \Kahler cone, then the region in moduli space where $D$ has volume less than a certain value --- i.e., the region where a phenomenologically realistic gauge theory can be realised --- is non-compact, which is interesting and relevant for phenomenological purposes. 

Additionally, while the scale of divisors volumes are often set by the overall volume, $\tau \sim \mathcal{V}^{2/3}$ \cite{Cheng:2025ggf, Fallon:2025lvn}, some divisors can shrink along \Kahler cone faces where the CY volume is finite, and thus their volume can be characteristically smaller than the overall volume. Such divisors are known in the literature as \textit{blowup} divisors. We distinguish between divisors which shrink to points and to curves, distinguished by whether their quadratic volume polynomials vanish quadratically or linearly on a face. The decoupling with the overall volume is more prominent for the former, as we will see in \cref{sec:geometry_results}. By contrast, we will refer to divisors that shrink near faces where $\mathcal{V}$ does vanish (due to a shrinking fiber) as \textit{fibration divisors}. Fibration divisors can also be small at large overall volume, but must be compensated by a large base volume, while blowup divisors do not require such a compensation. Divisors with volume $\tau$ decoupled from the overall volume are important, as then $\tau$ behaves more like a free parameter, yielding useful freedom when engineering the axion EFT. Shrinkability at large overall volume will prove important when implementing QCD axions within our constructions in \cref{sec:top_from_halo}.

In summary, then, we will find it useful to distinguish five categories of divisors, organised by their shrinking behaviour.
\begin{itemize}
    \item \textbf{Non-shrinkable divisors}, which have finite volume everywhere in the \Kahler cone
    \item \textbf{Fibration divisors}, which can shrink in the limit that the CY volume vanishes (i.e., the fiber shrinks)
    \item \textbf{$\mathbf{K3}$ fiber divisors}, which (up to overall scaling) are the fibers of $K3$ fibrations.
    \item \textbf{Curve blowup divisors}, which can shrink to curves while the CY remains at finite volume (i.e., their volume vanishes linearly)
    \item \textbf{Point blowup divisors}, which can shrink to points while the CY remains at finite volume (i.e., their volume vanishes quadratically)
\end{itemize}
Divisors can belong to more than one of these categories (e.g., $K3$ divisors are a special case of fibration divisors, and divisors shrinking linearly along some face can shrink quadratically along another, etc.), so we will assign membership by iterating over this list from bottom to top and applying the first label whose condition is satisfied.

It is not an accident that this list resembles the taxonomy of facets of CY threefold \Kahler cones (reviewed in \cref{app:FibrationsDecayConstants}) but it's worth stressing that the limits where shrinkable divisors (i.e., members of the first three categories) have vanishing volume need not be at codimension-one in the \Kahler cone: higher codimension can and does happen, and such cases will be important for us.

Finally, we can discuss the relevance of normal bundle deformations for axion decay constants. If $D$ is a divisor belonging to the basis of leading instanton contributions, then in the limit of hierarchical instanton scales (which arises generically in \Kahler moduli space, and was studied in \cite{Gendler:2023kjt})\footnote{We are indebted to Sebastian Vander Ploeg Fallon for sharing \cref{eq:decay_const_bound} with us.}
\begin{equation}
    \label{eq:decay_const_bound}
    f \gtrsim \mathcal{O}(1) \cdot \frac{1}{\sqrt{\tau^2 - \mathcal{V}C}}.
\end{equation}
for $\tau$ the volume of $D$ and $C$ the formal volume of the curve class $D \cap D$ (though such a class need not have holomorphic representatives). 

The Hodge number $h^{2,0}(D)$\footnote{Convenient combinatorial formulae for Hodge numbers of certain divisors in toric hypersurface CY threefolds are presented in \cite{Braun:2017nhi}.} counts normal bundle deformations of a divisor $D$ in a CY threefold, and we say $D$ is \textit{geometrically rigid} if $h^{2,0}(D) = 0$, and \textit{movable} otherwise.\footnote{We say ``geometrically rigid'' rather than just ``rigid'' to avoid confusion with the divisors demonstrated in \cite{Witten:1996bn} to contribute to the ED3 non-perturbative superpotential in type IIB: those rigid divisors $D$ are also geometrically rigid, but additionally satisfy $h^{0,0}(D) = 1$ and $h^{1,0}(D) = 0$.} We explain in \cref{app:FibrationsDecayConstants} how, for geometrically rigid divisors $D$, $\tau^2 - \mathcal{V}C$ is often the sum of positive terms (as usually $C < 0$) and is essentially set by $\mathcal{V}$, while for movable $D$, $C > 0$, so $\tau^2 - \mathcal{V}C$ is smaller and bounded from above and below for fixed $\tau$ by positivity of $C$ and of $K^{ij}$, respectively. Thus, decay constants for geometrically rigid divisors are characteristically smaller, and have larger variance compared to their movable counterparts. 

An edge case worth noting is when $D$ is a $K3$ fiber divisor, for which $C = 0$ and the decay constant is totally decoupled from the overall volume, being purely set by the divisor volume $\tau$. Crucially, this can be kept $\mathcal{O}(1)$ even at large volume by taking the anisotropic limit of large base and small fiber. In this way, $K3$ divisors provide a mechanism for engineering anomalously large axion decay constants, in a manner independent of overall volume. This was exploited in \cite{Sheridan:2024vtt} to engineer decay constant hierarchies, which in turn was inspired by \cite{Cicoli:2008va, Cicoli:2011it}.

\section{Sampling Moduli Space}\label{sec:sampling}

In this section, we lay out two techniques for sampling the SKC: an MCMC method implemented via the Python-compatible package \textsc{emcee}~\cite{Foreman-Mackey:2012any}, and an NF method implemented using the \textsc{normflows} package \cite{Stimper2023}. 

\subsection{Classical Markov Chain Monte Carlo}

For the MCMC approach, we employ two distinct parameterisations of the SKC as outlined in the previous section. The first uses the extremal generators of the original \Kahler cone, while the second uses the extremal hyperplanes of the modified \Kahler cone. 

In the two MCMC strategies, the former has the advantage of no rejections beyond the stretching conditions—by construction, all sampled points lie within the \Kahler cone—but may require sampling in a space of dimension larger than $h^{1,1}$ (in the case of non-simplicial cones). The latter, in contrast, restricts sampling to an $h^{1,1}$-dimensional space independent of the number of generators, but introduces additional rejections when the cone is non-simplicial (since then the modified \Kahler cone is a strict subset of $\mathbb{R}^{h^{1,1}}_{+}$).

For both strategies, we sample according to the natural Weil-Petersson (WP) prior defined by the metric $M_{ij}$ in \cref{eq:L_kin_2}. Using the measures derived in~\cref{eq:WP_extremalgeneratorsampling} and~\cref{eq:WP_directsampling}, the corresponding log-probabilities—hereafter referred to as the WP Prior—are given by
\begin{align}
\label{eq:logWP_gen}
    \log P^{\mathcal{G}}_{\mathrm{WP}}
    &= \tfrac{1}{2}\ln|\det M| + \sum_{\mu=1}^{N_g}\theta^{\mu} + \ln|\det\mathcal{G}|\,,\\[3pt]
\label{eq:logWP_hyp}
    \log P^{\mathcal{H}}_{\mathrm{WP}}
    &= \tfrac{1}{2}\ln|\det M| + \sum_{i=1}^{h^{1,1}}\bar{\theta}^{i} - \ln|\det\bar{\mathcal{H}}|\,.
\end{align}
The above are defined up to additive constants. 
Points violating any facet or stretching constraint ($H^a_i t^i \ge 1$ or $\tau^{\rm pt}_{p} > 1$) are assigned $-\infty$. These log-prior expressions form the probability landscape explored by the MCMC chains.

\subsubsection{Implementation and convergence tests}
\label{sec:implementation_and_conv}

For purposes of this section, each MCMC chain was run within the divisor-stretched \Kahler cone (DSKC) of a given CY geometry, bounded between $\mathcal{V}_{\min}=\mathcal{V}_{\rm tip}$ and $\mathcal{V}_{\max}=10^{10}\,\mathcal{V}_{\rm tip}$, where $\mathcal{V}_{\rm tip}$ is the CY volume at the tip of the stretched \Kahler cone\footnote{The tip of the stretched \Kahler cone is obtained by minimising the Euclidean norm $\|t\|_2$ subject to the constraint $H^a_i t^i \ge 1$.}. For the generator-based sampling, the ensemble comprised $N_{\rm wlk}=10\,N_g$ walkers, while the hyperplane-based sampling used $N_{\rm wlk}=10\,h^{1,1}$. Convergence of each chain was determined through autocorrelation analysis of individual parameters: once the number of statistically uncorrelated samples in the slowest (most correlated) parameter exceeded $75$, the run was terminated. This uniform criterion allows a geometry-independent comparison of convergence efficiency.

\begin{figure*}
    \centering
    \includegraphics[width=0.99\linewidth]{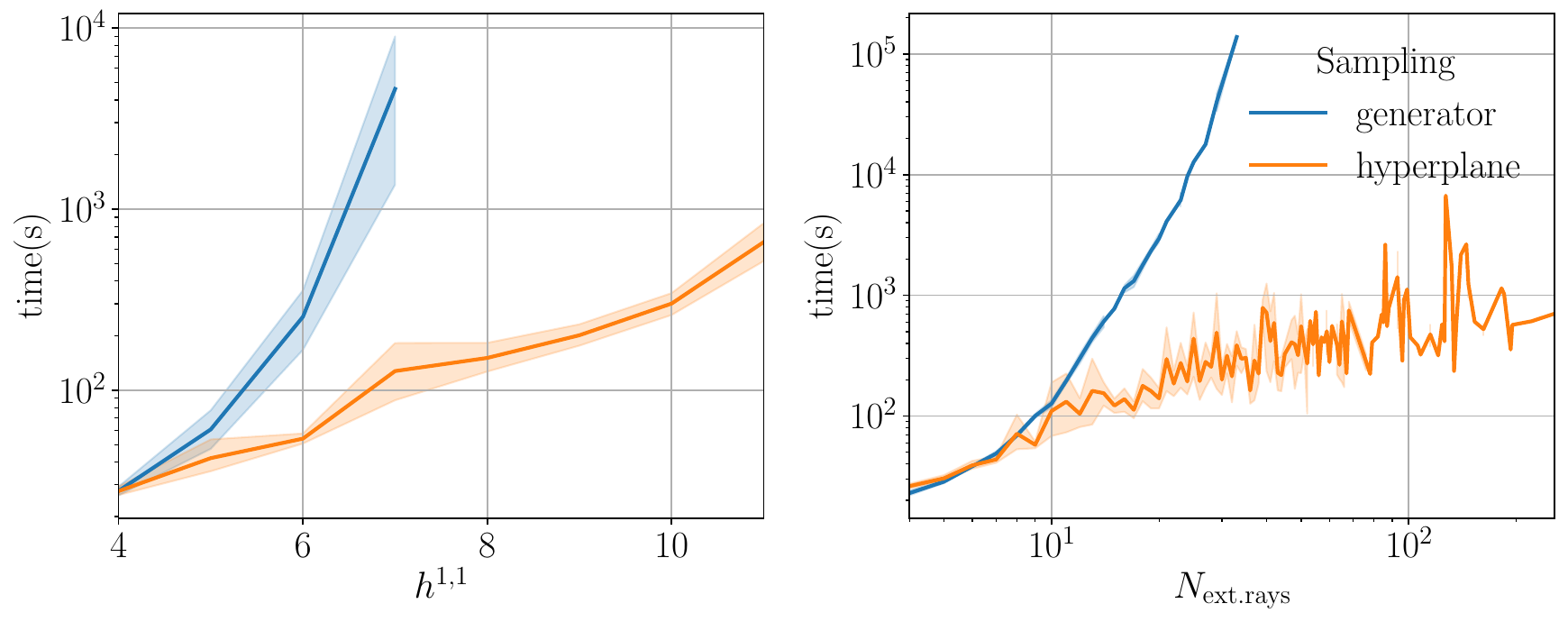}
    \caption{Convergence times for various values of $h^{1,1}$ and different numbers of generators for hyperplane sampling using \textsc{emcee}.  The full likelihood consists solely of the Weil--Petersson prior, together with an upper bound on the Calabi--Yau volume fixed at $10^{10}\times\mathcal{V}_{\mathrm{tip}}$. We declare convergence once the total number of MCMC steps performed for the slowest converging parameter, exceeds $75$ times its autocorrelation length.
    }
    \label{fig:convergencetime_hyperplanesampling}
\end{figure*}

The procedure was repeated across $100$ randomly selected geometries for each $h^{1,1}\leq 11$. For each $h^{1,1}$, we uniformly sampled $100$ polytopes from the KS list \cite{Kreuzer:2000qv} and then choose a random geometry by employing the methods of \cite{MacFadden:2023cyf,MacFadden:2024him}. For each geometry, we recorded the total time required to reach convergence, and compiled the results both as a function of $h^{1,1}$ and of the number of extremal generators $N_g$. These results are summarised in~\cref{fig:convergencetime_hyperplanesampling}.

\begin{figure*}
    \centering
    \includegraphics[width=0.99\linewidth]{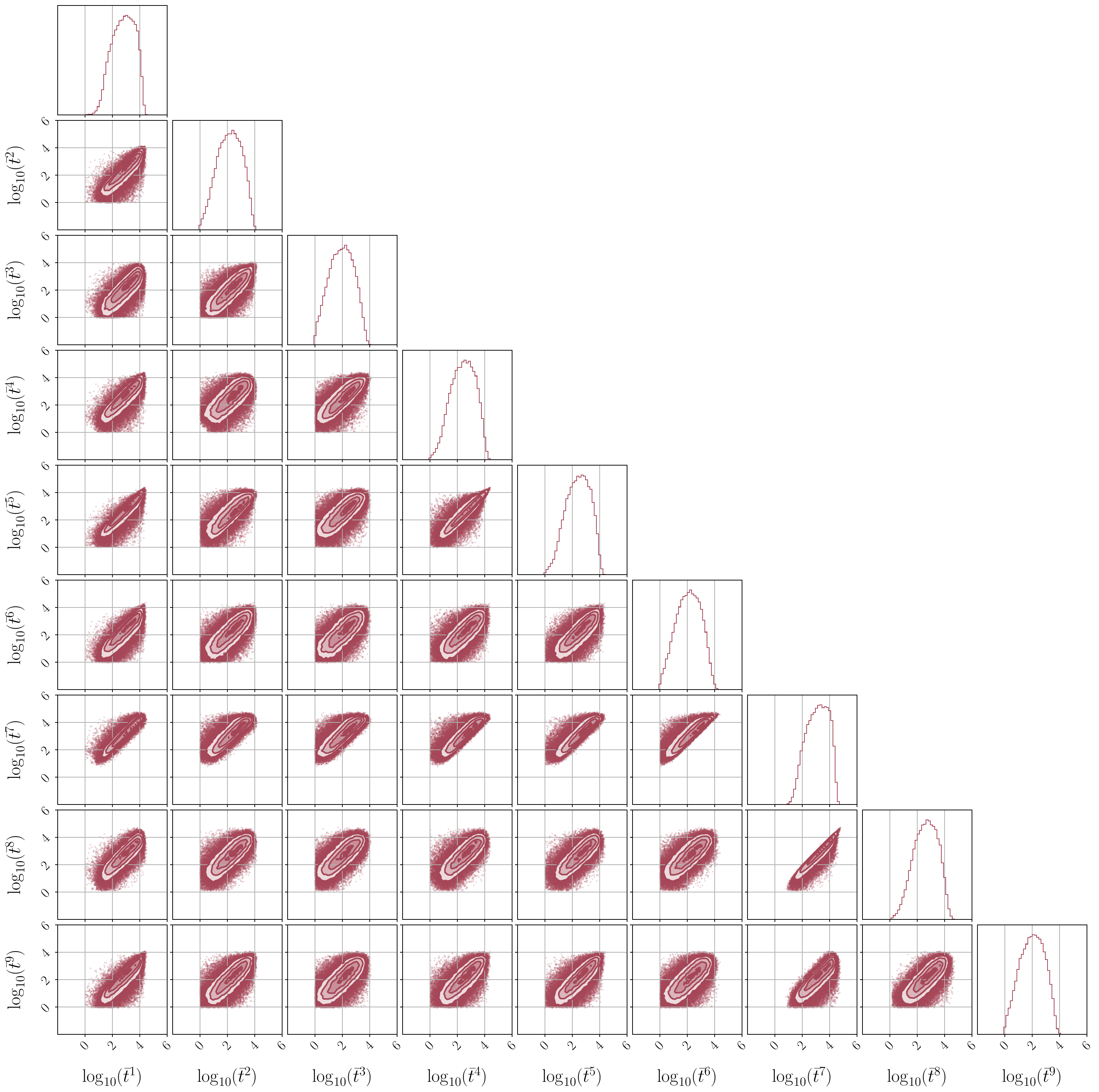}
    \caption{Marginal one-dimensional and two-dimensional distributions of the modified positive semi-definite \Kahler parameters, $\bar{t}^i \in \mathbb{R}_+$ basis as defined in \cref{eq:tbarbasis}, for a cone with $N_g = 28$ extremal generators, at $h^{1,1} = 9$ obtained using hyperplane sampling of the Weil--Petersson prior.}\label{fig:CornerPlot3}
\end{figure*}

\begin{figure}
    \centering
    \includegraphics[width=0.99\linewidth]{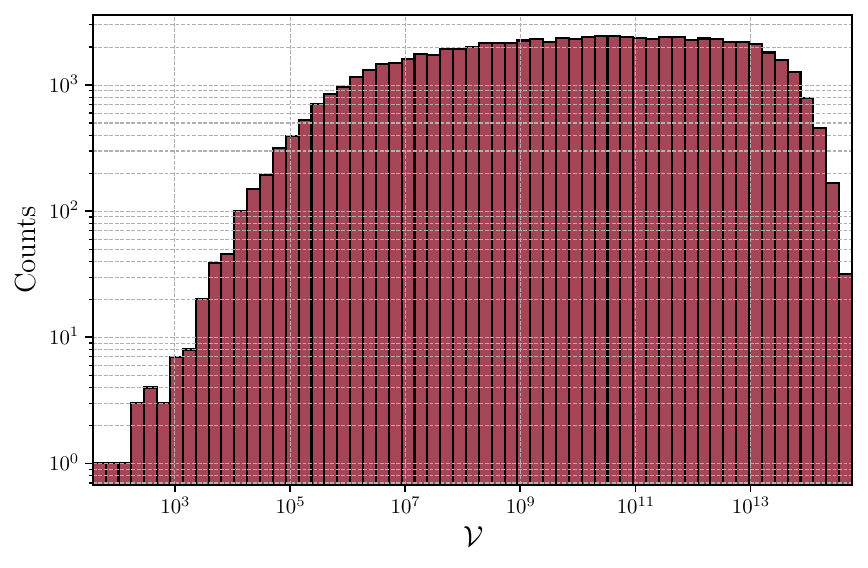}
    \caption{Converged histograms of CY volume for example at $h^{1,1} = 9$ with a
    \Kahler cone with $N_g = 28$ extremal generators. The likelihood is simply just the WP prior, with maximum CY volume set at $\mathcal{V}_{\rm max} = 10^{10} \times \mathcal{V}_o$ where $\mathcal{V}_o$ is the volume of the CY at the TSKC.}
    \label{fig:Histograms_4ExampleCases}
\end{figure}

In~\cref{fig:CornerPlot3}, we present the corner plot visualising the sampled region of the \Kahler cone for an example geometry, while in \cref{fig:Histograms_4ExampleCases} we show the associated distribution of CY volumes. The volume histogram exhibits logarithmic flatness, correctly reflecting the expected WP prior behaviour and providing an additional indication of convergence.

\subsubsection{Comparison of sampling performance}

The two strategies display complementary strengths. The generator-based method samples directly along the extremal directions of the cone, ensuring that all proposed points automatically belong to the \Kahler cone. Its primary cost, however, is dimensional: for non-simplicial cones with $N_g>h^{1,1}$, the effective sampling space grows quickly, leading to longer convergence times. In contrast, the hyperplane-based method operates in a fixed $h^{1,1}$-dimensional space and thus scales more mildly with $h^{1,1}$, but introduces additional rejections whenever the cone is a strict subset of the positive orthant. 

The convergence-time distributions reflect this trade-off. Across the ensemble, convergence time correlates more strongly with the number of extremal generators $N_g$ than with $h^{1,1}$ itself. At large $h^{1,1}$, the hyperplane-based approach exhibits superior numerical stability and faster convergence, whereas at low $h^{1,1}$ the generator-based sampling remains competitive and provides a more direct geometric interpretation of motion within the cone. For large-scale statistical scans of CY geometries, the hyperplane-based WP prior therefore constitutes the more efficient and robust choice. We will use the hyperplane-based sampling for cosmological purposes in later sections.

These observations motivate the use of more flexible, learnable priors for navigating highly non-simplicial cones, which we turn to next.

\medskip

\subsection{Normalising flows}

Normalising flows (NFs) \cite{rezende2015variational} are a method for constructing complex distributions by transforming a simple base distribution (e.g., a Gaussian) through a series of invertible mappings. 
Let $f:\mathbb{R}^{D}\rightarrow\mathbb{R}^{D}$ be an invertible and differentiable function whose inverse maps the input data $\mathbf{x}\in\mathbb{R}^{D}$ to the latent representation $\mathbf{z}\in\mathbb{R}^{D}$, i.e., $f^{-1}(\mathbf{x})=\mathbf{z}$ or equivalently $f(\mathbf{z})=\mathbf{x}$. The main idea is to express the distribution $p_{x}(\mathbf{x})$ in terms of the base distribution $p_{z}(\mathbf{z})$ and $f$. This is done using the change of variables formula:
\begin{equation}\label{eq:COV}
    p_{x}(\mathbf{x}) = p_{z}(\mathbf{z}) \left| \det{J_{f}(\mathbf{z})} \right|^{-1},
\end{equation}
where $J_{f}$ denotes the Jacobian of $f$. For more details on NF see~\cref{app:NFs}.

A base Gaussian distribution has full support, which means that $p_{z}(\mathbf{z})>0$ everywhere on $\mathbb{R}^{D}$. The invertible nature of $f$ means that using such a base distribution, the target distribution $p_{x}(\mathbf{x})$ must also have full support. For this reason, we choose the extremal generators parametrisation of the original \Kahler cone with target WP given in~\cref{eq:logWP_gen}, as opposed to the extremal hyperplanes parametrisation of the modified \Kahler cone with WP given in~\cref{eq:logWP_hyp}, since the latter does not have full support in the space of $\bar{\theta}$s. Unlike MCMC where the chains run inside the DSKC, for the NF we do not impose any stretching constraints during training. Instead, we train the NF to learn the WP measure on the full \Kahler cone. Once trained, we then use the NF to sample points and only keep those that lie inside the DSKC and whose volume is between $\mathcal{V}_{\min}=\mathcal{V}_{\rm tip}$ and $\mathcal{V}_{\max}=10^{10}\,\mathcal{V}_{\rm tip}$.

Specifically, we fit a real NVP model~\cite{Dinh:2016RealNVP} $p_{\theta}(\boldsymbol{\theta};\Theta)$, where $f$ is composed of a series of 16 masked affine flow layers $f_{i}$ with parameters $\Theta$, to the target distribution $p_{\theta}^{*}(\boldsymbol{\theta})$ defined by WP~\cref{eq:logWP_gen}.\footnote{For more details on the real NVP architecture, see \cref{app:NFs}.}
Since we can easily evaluate $p_{\theta}^{*}(\boldsymbol{\theta})$ for a given point $\boldsymbol{\theta}$ in \Kahler moduli space, but cannot easily sample from this distribution, we train our NF model using reverse Kullback--Leibler (KL) divergence $D_{KL}$~\cref{eq:KL_rev}. In evaluating $p_{\theta}^{*}(\boldsymbol{\theta})$ we must compute the matrix $M$ from~\cref{eq:L_kin_2}. This involves transforming $\boldsymbol{\theta}$ back into \Kahler parameters $\mathbf{t}$, by first computing $\alpha^{\mu}=\exp{(\theta^{\mu})}$ and then $t^{i}=\alpha^{\mu}\mathcal{G}^{i}_{\;\mu}$. In practice, if $\theta^{\mu}$ is too large then we encounter overflow errors during training from $\exp{(\theta^{\mu})}$, therefore we add an additional term to the loss function: $\text{ReLU}(||\boldsymbol{\theta}||-R)^{2}$, which is zero if $||\boldsymbol{\theta}||<R$ and $(||\boldsymbol{\theta}||-R)^{2}$ otherwise. This term penalises $\boldsymbol{\theta}$ larger than $R$ which we set to be $15$. The total loss function is thus
\begin{equation}
\label{eq:NF_loss}
    \begin{aligned}
        \mathcal{L}(\Theta) &= D_{KL}[p_{\theta}(\boldsymbol{\theta};\Theta) \, | \, p_{\theta}^{*}(\boldsymbol{\theta})] \\
        &\qquad + w\times\text{ReLU}(||\boldsymbol{\theta}||-R)^{2},
    \end{aligned}
\end{equation}
where $w$ is a tunable weight parameter that we set to $1$. 

In each of the 16 masked affine flow layers, we use an alternating binary mask and the translation and scale networks are MLPs consisting of 4 hidden layers each with 256 neurons. This gives a total of 6{,}373{,}472 learnable parameters. We train our NF model over 50{,}000 epochs using the Adam optimiser to find the optimal values for these parameters, which minimise the total loss~\cref{eq:NF_loss} on a batch of size 1{,}024. The initial learning rate is set to $10^{-5}$ and we use \textsc{PyTorch} learning rate scheduler \textsc{ReduceLROnPlateau} with factor 0.5 and patience 1000, which reduces the learning rate by one half if there is no improvement in the loss over 1{,}000 epochs. Training was performed on a 14-core CPU, 32-core GPU laptop equipped with 36 GB of unified memory, and required approximately 5 hours to complete 50{,}000 epochs. Due to the time taken to train the model, a limited amount of hyperparameter tuning was performed to determine these settings.

\begin{figure*}
    \centering
    \includegraphics[width=0.9\linewidth]{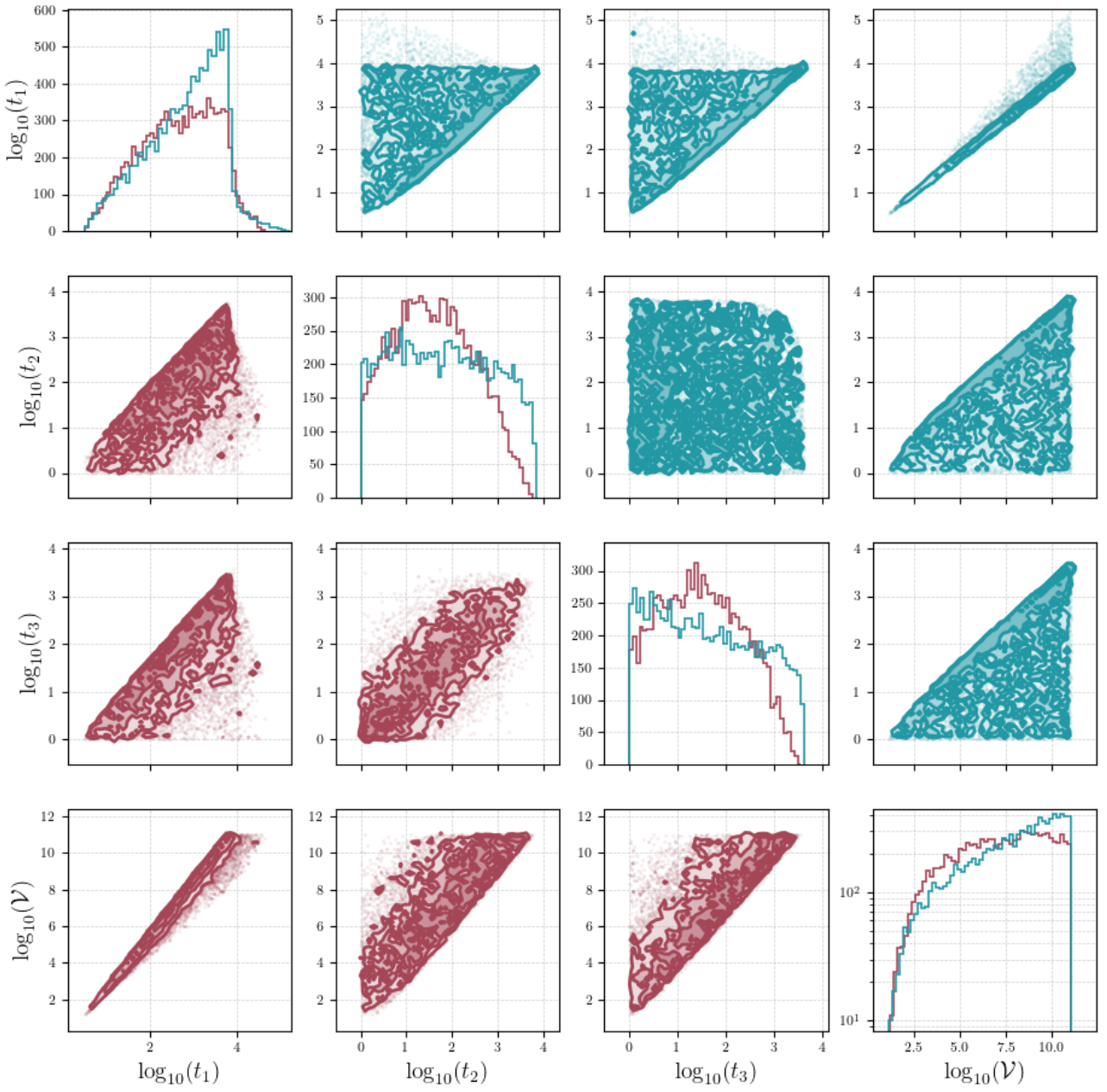}
    \caption{Distributions of the \Kahler parameters inside the DSKC and the associated distribution of Calabi--Yau volumes, for a cone with $N_{g}=4$, at $h^{1,1}=3$ obtained using MCMC (red) sampling with WP prior, and the trained NF model (cyan).}\label{fig:MCMC_NF_corner}
\end{figure*}

To compare the results of NF with MCMC, we generate samples (10{,}000) via both methods for the same geometry at $h^{1,1}=3$, which has $N_{g}=4$ generators, and compare their distributions. The corner plot of these samples with the associated distribution of CY volumes is shown in~\cref{fig:MCMC_NF_corner}, which demonstrates reasonable agreement between the methods.

\section{Geometric results}\label{sec:geometry_results}

Our primary application of the methods presented for sampling \Kahler moduli space will be the phenomenology of $C_4$ axions in type IIB string theory, but first we will briefly study something simpler: distributions of volumes of divisors under the Weil--Petersson measure. Beyond the importance of understanding the distribution of \Kahler moduli vevs --- recalling \cref{eq:kahler_moduli} --- under the natural moduli space metric, the statistics of divisor volumes $\tau$ is important as these geometric quantities set many scales in low energy string theory EFTs, including the gauge coupling in D7-brane worldvolume theories (as $1/\sqrt{\tau}$) and the $C_4$ axion masses (as $m \sim \mathrm{e}^{- 2\pi \tau}$). Indeed, efforts have been undertaken to model divisor volumes and apply such models to understand the universal behaviour of low-energy physics in string compactifications \cite{Cheng:2025ggf}. As discussed in \cref{sec:geom_preq}, it is especially worth understanding whether divisor volumes can be treated as free parameters, or if they are instead strongly correlated in a way that imprints on low energy phenomenology.\footnote{For divisor volumes in particular, we are essentially asking about the structure of the cone of dual coordinates of a CY threefold (in the sense of \cite{Gendler:2022ztv}), or the image of the extended \Kahler cone under the map $J \mapsto J \wedge J$, as when expanded in a basis of four-forms, the coefficients of $J \wedge J$ are interpreted as the volumes of the dual basis of divisors classes. Strong correlations among divisor volumes are encoded in the narrowness of this cone of dual coordinates.}

We will begin by presenting statistics of divisor volumes as a function of $h^{1,1}$ as well as the topological categories introduced in \cref{sec:geom_preq}, before exploring a specific example further illustrate the role of those topological categories. We will then interface with \cite{Cheng:2025ggf}, expanding on their result regarding the role of the minface dimension of a prime toric divisor (defined shortly). Finally, we will present divisor volume spectra for a geometry with $h^{1,1} = 30$ with the important property (optimised using a genetic algorithm \cite{MacFadden:2024him}) that its $K_\cup$ is simplicial, making sampling feasible in spite of the large Hodge number.

\begin{figure}
    \centering
    \includegraphics[width=\linewidth]{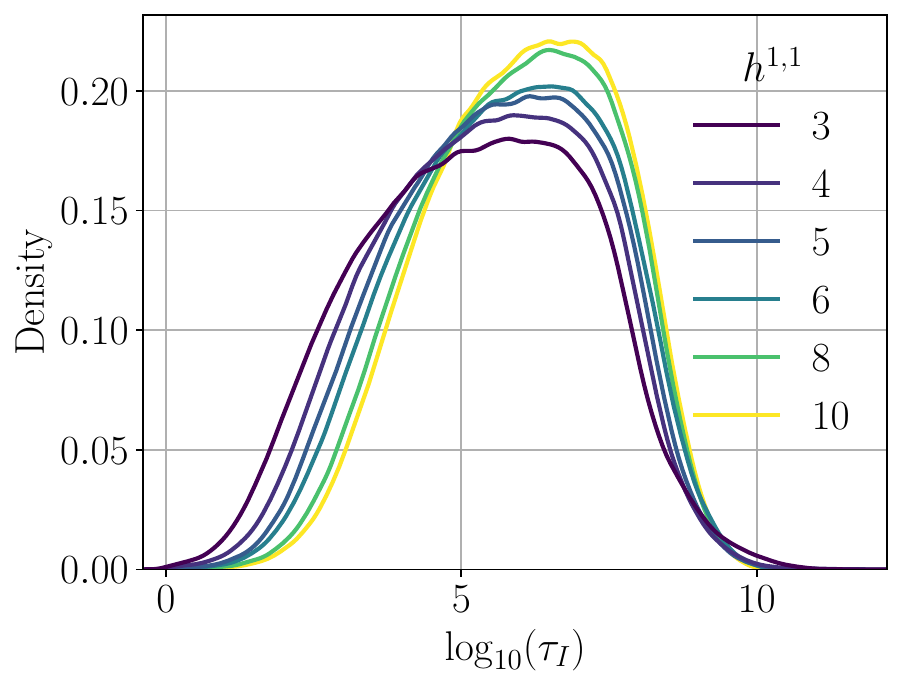}
    \caption{Distributions of prime toric divisor volumes, marginalised over the WP prior, all prime toric divisors, and over topologies, labelled by $h^{1,1}$. Marginalisation over geometries is exhaustive for $h^{1,1} = 3,4$, and otherwise is performed over the random geometries selected in \cref{sec:implementation_and_conv}.}
    \label{fig:tau_vs_h11}
\end{figure}

In \cite{Demirtas:2018akl}, distributions of divisor volumes on Kreuzer--Skarke CY threefolds were studied as a function of $h^{1,1}$, motivated by the desire to understand the string axiverse in this setting. This statistical study was undertaken by sampling points near the tip of the stretched \Kahler cone: in particular, points minimising divisior and CY volumes subjects to unit stretching. We are now equipped to explore these distributions under the WP measure. In particular, in \cref{fig:tau_vs_h11} we present the distribution of prime toric divisor volumes as a function of $h^{1,1}$, marginalised over geometries. More specifically, for $h^{1,1} = 3,4$ we marginalise over all FRST classes, while for larger $h^{1,1}$ we marginalise over the random geometries sampled in \cref{sec:implementation_and_conv}.

\begin{figure}
    \centering
    \includegraphics[width=\linewidth]{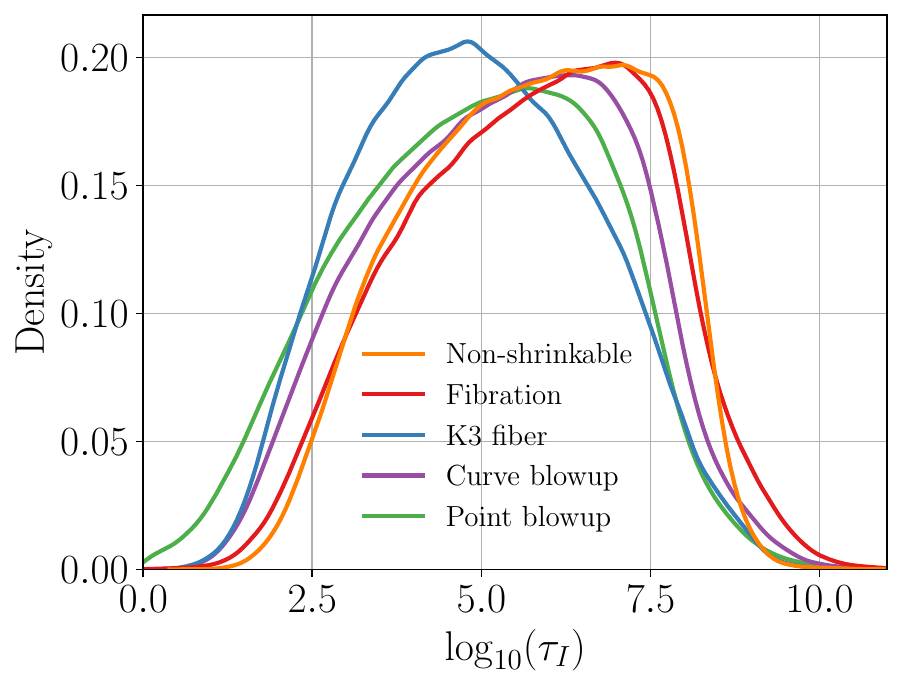}
    \caption{Distributions of prime toric divisor volumes, marginalised over the WP prior and over all topologies in Kreuzer--Skarke at $h^{1,1} = 4$, labelled by the divisor topology categories specified in \cref{sec:geom_preq}.}
    \label{fig:by_type}
\end{figure}

To illustrate the dependence of the divisor volume distribution on the topology of the divisor, in \cref{fig:by_type} we present a plot similar to the previous figure but now marginalised over all geometries at $h^{1,1} = 4$, coloured by the type of the divisor, as introduced in \cref{sec:geom_preq}. We find that $K3$ fibers and point blowup divisors, in particular, tend to be smaller, with the former featuring a smaller mean and the latter's distribution featuring a longer tail toward unit volume (recall our prior imposes that it cannot be smaller).

\begin{figure}
    \centering
    \includegraphics[width=\linewidth]{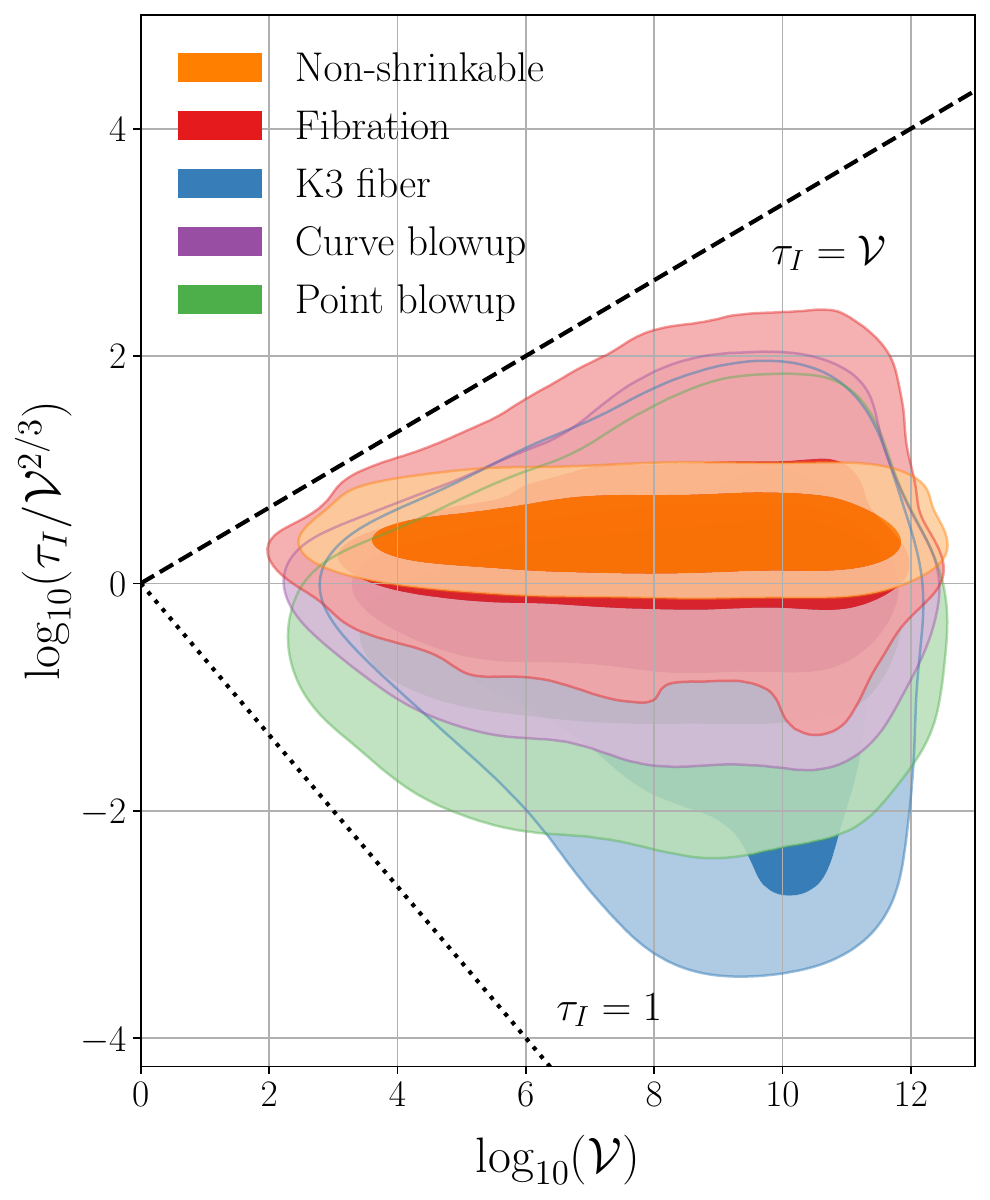}
    \caption{Joint distribution of normalised prime toric divisor volumes and overall volume, marginalised over the WP prior and over all topologies in Kreuzer--Skarke at $h^{1,1} = 4$, labelled by the divisor topology categories specified in \cref{sec:geom_preq}.}
    \label{fig:V_vs_tau}
\end{figure}

To probe the extent to which divisor volumes behave as free independent parameters or are instead set by scales such as the overall volume, in \cref{fig:V_vs_tau} we plot the distribution of the dimensionless parameter $\tau/\mathcal{V}^{2/3}$ as a function of $\mathcal{V}$ for all prime toric divisors for \Kahler parameters marginalised over all geometries in KS at $h^{1,1} = 4$, coloured by the type of the divisor in the same way as the previous plot. We see that while divisor volume distributions peak near the locus $\tau/\mathcal{V}^{2/3} = 1$ --- i.e., the divisor volume scale is set by the overall volume --- $K3$ and point blowup divisors in particular are able to be much smaller, even at large overall volume. Additionally, vertical and curve-blowup divisors are capable of smaller volumes than non-shrinkable divisors at each value of the overall volume. With the gray dashed line we have indicated our stretching constraint, requiring that all divisor volumes be at least one. With the black dashed line we emphasise an empirical finding that divisor volumes tend to obey $\tau \leq \mathcal{V}$. For $K3$ fibered geometries, for example, limits where the overall volume factorises as $\mathcal{V} \sim \tau \cdot C$ for $\tau$ and $C$ the volumes of the $K3$ fiber and base curve, respectively, requiring $C \geq 1$ imposes $\tau \leq \mathcal{V}$. However, we do not have a more general explanation for why this inequality holds for all divisor categories.

It is worth seeing in an example how divisor volume topology encodes correlations among volumes. Consider the 4D reflexive polytope from the Kreuzer--Skarke database whose vertices are the columns of the following matrix.
\begin{equation}
    \begin{pmatrix}
        -3 & \hphantom{-}0 & \hphantom{-}0 & \hphantom{-}1 & -1 & \hphantom{-}0 \\
        \hphantom{-}0 & \hphantom{-}0 & \hphantom{-}0 & \hphantom{-}0 & -1 & \hphantom{-}1 \\
        -1 & \hphantom{-}0 & \hphantom{-}1 & \hphantom{-}0 & \hphantom{-}0 & \hphantom{-}0 \\
        -1 & \hphantom{-}1 & \hphantom{-}0 & \hphantom{-}0 & \hphantom{-}0 & \hphantom{-}0
    \end{pmatrix}
\end{equation}
A triangulation of this polytope yields a toric variety with a CY hypersurface with $h^{1,1} = 3$. Of the seven prime toric divisors, there are only four unique divisor classes, which we will denote $D_1, D_4, D_5, D_7$ according to the default ordering of prime toric divisors set in \textsc{CYTools}. This geometry has a simplicial $\mathcal{K}_\cup$: two facets correspond to flops, while one corresponds to a boundary of \Kahler moduli space where $D_7$ shrinks to a point (i.e., $D_7$ is a point blowup divisor). The ray of $\mathcal{K}_\cup$ given by the intersection of the shrinking divisor facet and one of the flop facets corresponds to a K3 fibration over $\mathbb{P}^1$. The pullback of the hyperplane class on $\mathbb{P}^1$, which generates this ray and is the class of the $K3$ fiber, is $D_5$. The ray given by the intersection of the two flop facets is also a boundary of moduli space, and the divisor $D_1$ shrinks to a curve along this ray. Finally, $D_4$ is non-shrinkable.

\begin{figure*}
    \centering
    \includegraphics[width=0.8\linewidth]{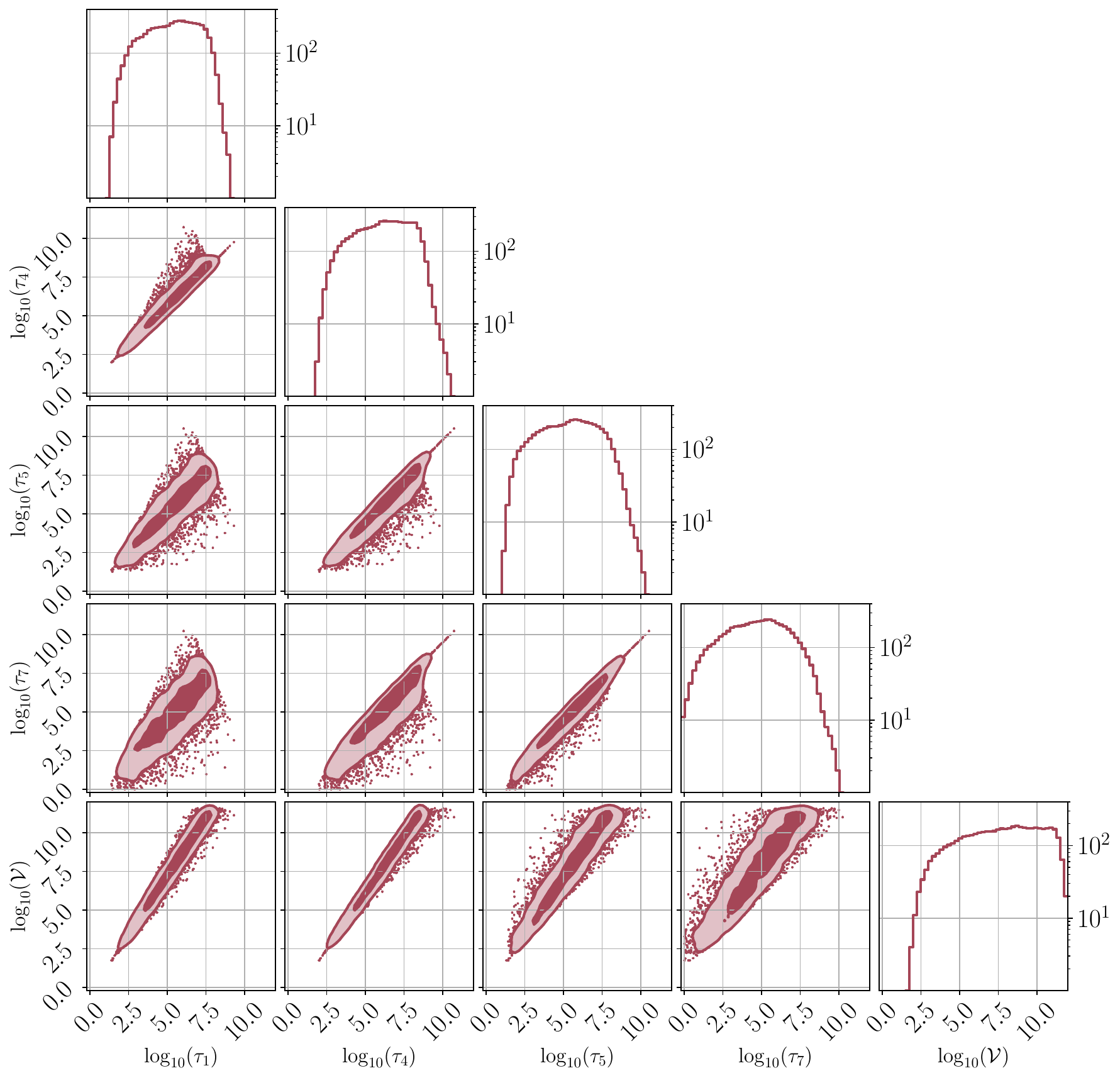}
    \caption{Distributions of prime toric divisor volumes and overall volume under Weil--Petersson measure for an example geometry at $h^{1,1} = 3$ (for which there were $4$ unique prime toric divisor classes).}
\label{fig:example_corner}
\end{figure*}

We can readily understand the corner plot of prime toric divisor and overall volumes featured in \cref{fig:example_corner} given this analysis of the \Kahler cone and divisor topologies. In particular, we see that the point blowup divisor $D_7$ and $K3$ fiber $D_5$ are less correlated with the overall volume and the other prime toric divisor volumes than the curve blowup divisor $D_1$ and non-shrinkable divisor $D_4$ (and $D_4$ is most correlated with the overall volume of them all). Additionally, we see that $\tau_5 > \tau_7$; one can verify this explicitly by comparing their volume polynomials, but intuitively this is because because the limit of small $K3$ fiber is along a ray of $K_\cup$ contained in the facet where $D_7$ shrinks, so $\tau_5$ cannot be small without $\tau_7$ being small as well. In this way, the face structure of $K_\cup$ --- i.e., its face lattice --- controls correlations between divisor volumes.

Finally, we employ our WP sampling capabilities to interface with the Cheng--Gendler model proposed in \cite{Cheng:2025ggf} for distributions of prime toric divisor volumes on CY threefolds arising from a polytope $\Delta^\circ$. There it was observed that when performing a random walk in \Kahler moduli space according to a particular scheme, volume distributions naturally organise into a trimodal distribution. In particular, the volume distribution of a prime toric divisor is characterised by its minface dimension \cite{Braun:2017nhi}, or the dimension of the smallest face of $\Delta^\circ$ containing the point associated to the prime toric divisor. Vertices, points strictly interior to one-faces (edges), and points strictly interior to two-faces have minface dimension $0, 1, 2$, respectively, and we call their associated divisors vertex, edge, and face divisors, respectively. For $h^{1,1} \lesssim 10$, the vast majority of prime toric divisors are of vertex type, but for larger $h^{1,1}$ more edge and face divisors begin to appear (and for $h^{1,1} \gg 50$, such divisors dominate). In \cite{Cheng:2025ggf}, the normalised logarithmic volumes $\hat{\tau}_{I}$ were defined in terms of the vector $\tau_{I}$ of prime toric divisors volumes, which was treated with an overall prefactor such that the smallest entry in $\tau_{I}$ was $1$. In particular, 
\begin{equation}
    \label{eq:log_normalized_vols}
    \hat{\tau}_{I} = \frac{\log(\tau_{I})}{\langle \log(\tau) \rangle} \, , \; \langle \log(\tau) \rangle = \dfrac{\sum_{I=1}^{h^{1,1} + 4}\, \log\bigl (\tau_{I} \bigl )}{h^{1,1} + 4}.
\end{equation}

\begin{figure}
    \centering
    \includegraphics[width=\linewidth]{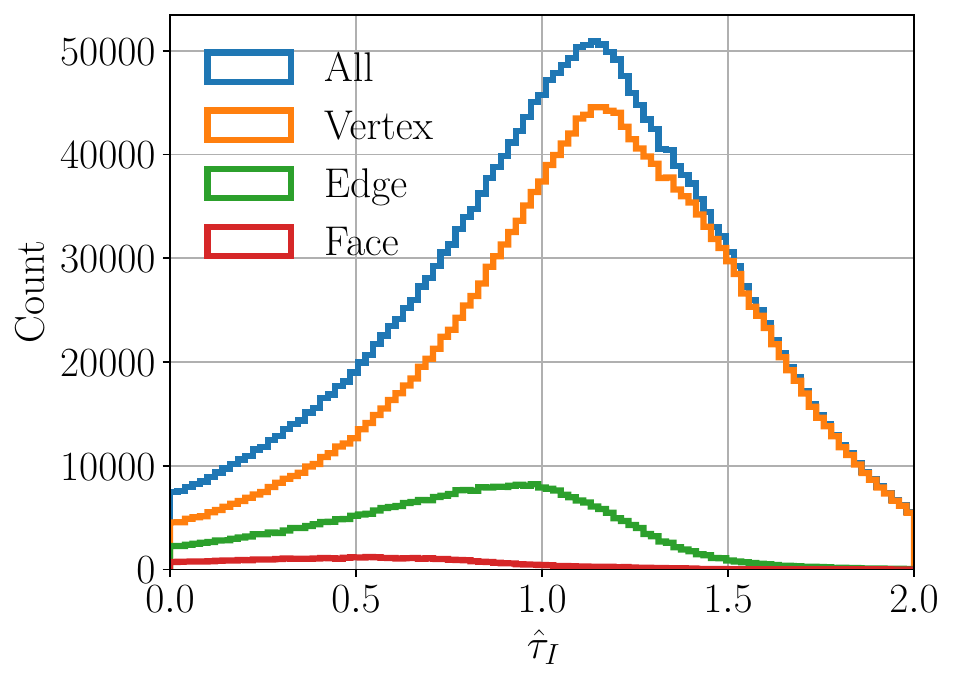}
    \caption{Distribution of logarithmically normalised volumes $\hat{\tau}$ (see \cref{eq:log_normalized_vols}) for vertex, edge, and face prime toric divisors marginalised the random geometries of \cref{sec:implementation_and_conv} at $h^{1,1} = 10$.}
    \label{fig:cheng_gendlder}
\end{figure}

In \cref{fig:cheng_gendlder}, for the random geometries selected in \cref{sec:implementation_and_conv} at $h^{1,1} = 10$ we aggregate each component of $\hat{\tau}$ across a WP chain (omitting the smallest entry in $\hat{\tau}$, following \cite{Cheng:2025ggf}), coloured by whether the associated divisor was of vertex, edge, or face type. We reproduce the trimodality, verifying that the observations of \cite{Cheng:2025ggf} continue to hold upon implementation of a Bayesian sampling methodology using the natural metric on \Kahler moduli space. We anticipate that the trimodality is related to the frequency with which vertex, edge, and face divisors belong to the various shrinkable divisor categories introduced in \cref{sec:geom_preq}, but we do not study this further here.

\begin{figure*}
    \centering
    \includegraphics[width=0.75\linewidth]{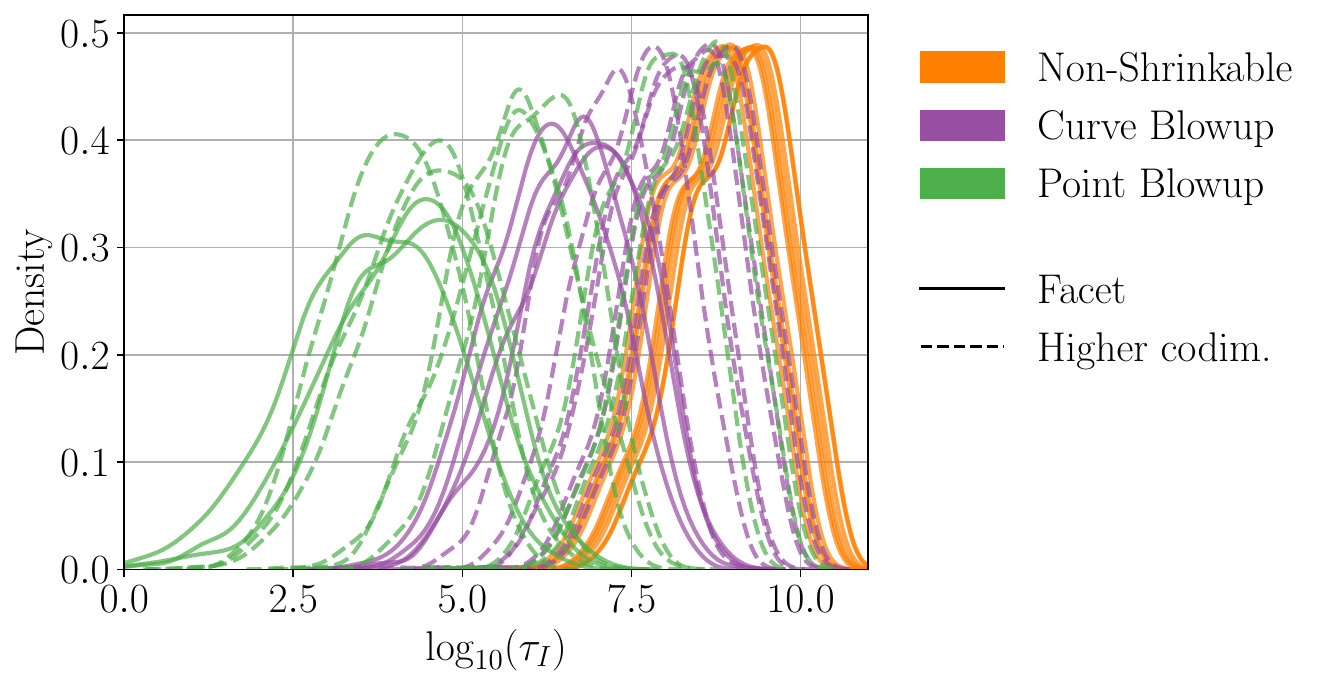}
    \caption{Distribution of prime toric divisor volumes marginalised over the WP measure within the SKC for a geometry at $h^{1,1} = 30$ with a simplicial $\mathcal{K}_\cup$, categorised according to the topological category of the divisor.}
    \label{fig:h11_30}
\end{figure*}

In \cref{fig:h11_30} we present the distributions of the volumes of the prime toric divisors under the WP measure for a geometry at $h^{1,1} = 30$. Because the computational complexity of sampling scales with both $h^{1,1}$ and the number of extremal rays of the \Kahler cone (which itself tends to grow exponentially with $h^{1,1}$), the sampling of generic geometries for $h^{1,1} \gtrsim 15$ is quite challenging without dedicated computational resources. However, if one can construct geometries at large $h^{1,1}$ with simplicial (or nearly simplicial) $K_\cup$, sampling can be performed on a laptop for even larger $h^{1,1}$. As a proof of concept, we employed the genetic algorithm of \cite{MacFadden:2024him} to minimise the number of extremal rays of $K_\cup$, and found geometries at $h^{1,1} = 30$ with simplicial $K_\cup$ (while typical geometries had $\gtrsim 50$ extremal rays). We do not expect such geometries to be representative: we merely endeavoured to maximise the $h^{1,1}$ for which we could sample \textit{some} geometry. 

We style the curves in \cref{fig:h11_30} according to the topological categories of the associated prime toric divisors introduced in \cref{sec:geom_preq}. In particular, this geometry has no fibrations (at least, none corresponding to faces of $K_\cup$) so that all shrinkable divisors are of blowup type, and we found it relevant to distinguish divisors based on whether they shrunk along facets (i.e., codimension one faces) or only along higher faces. It is unsurprising that divisors shrinking along facets have characteristically smaller volumes, as they are small in a greater fraction of the moduli space. More generally, the differences in distributions between point and curve blowup divisors vanishing on higher codimension faces can be accounted for by the exact codimension of those faces (the larger the codimension, the larger the volumes tend to be). 

\section{Application to axions}\label{sec:axion_results}

As above, throughout this section we scan SKCs with the WP prior, while now imposing additional likelihoods that we will define. We run MCMC scans using hyperplane sampling technique, with number of walkers equal to $10 \times h^{1,1}$. All our walkers start within the geometrically allowed divisor stretched \Kahler cone, randomly spread throughout the cone capped at the maximum volume $\mathcal{V}_{\rm max}$. To gauge convergence, we compute the autocorrelation times for each parameter, and run our MCMC until the total MCMC steps of the slowest converging parameter is equal to scale $\times$ its autocorrelation time, where we set the scale to be $75$. This makes sure that for this slowest converging parameter, we have $\sim 75$ statistically independent samples, thereby guaranteeing that all other parameters have \emph{at least} $\sim 75$ statistically independent samples. The minimum and maximum volume bounds that we impose (alongside the WP prior), are
\begin{align}
\label{eq:Vmin_Vmax_phys}
    \mathcal{V}_{\rm min} &= 1\nonumber\\
    \mathcal{V}_{\rm max} &= \left(\frac{g_s\,M_{\rm pl}}{\sqrt{4\pi} M_{\rm KK}}\right)^{3/2} \approx 6.2 \times 10^{12} \,,
\end{align}
where we have taken $g_s = 0.5$, and $M_{\rm KK} = 10^{9}\,{\rm GeV}$ to incorporate inflationary scales as low as this (since $H_{\rm inf} \geq M_{\rm KK}$). The choice of $\mathcal{V}_{\rm max}$ is a uniform and physically consistent upper bound across geometries (in contrast to the scaled value of $\mathcal{V}_{\rm tip}$ used in the previous section to more fairly measure computational efficiency across geometries).\footnote{We note that $\mathcal{V}_{\rm max}$ in~\cref{eq:Vmin_Vmax_phys} inherently assumes an isotropic CY hypersurface, an assumption that needs to be modified for anisotropic CY hypersurfaces (such as a K3 fibered CY). We briefly comment further on this in \cref{app:FibrationsDecayConstants}, but for our purposes in this paper, we use the simple estimate as the only purpose is to assign a well-defined and reproducible number to each geometry we study.}

\subsection{Theory informed priors on $(m_a,f_a)$}

Here, we present the collective distribution of axion masses and decay constants
for all inequivalent FRSTs with $h^{1,1}=3$.\footnote{We provide
distributions for $h^{1,1}=4$ and $5$ in the supplementary material \github{https://github.com/AndreasSchachner/kahler_cone_sampler}.}
On top of the WP prior on the \Kahler moduli space, we implement a likelihood
for a QCD-like axion. For each choice of \Kahler parameters, we scan over all
prime toric divisors and identify the one whose volume is closest to $40$ (in
string units). This is to achieve a realistic QCD gauge coupling in the UV.

The explicit divisor-volume likelihood that we impose on this
closest-volume prime toric divisor is
\begin{align}
\label{eq:div_likelihood}
    \ln \mathcal{L}_{\rm vol}
    = -\frac{1}{2 \times 5^2}\bigl(\tau_{\rm QCD} - \bar{\tau}_{\rm QCD} \bigr)^2 ,
\end{align}
with $\bar{\tau}_{\rm QCD}=40$. We then identify the instanton action associated to this divisor, and replace it with the QCD instanton action as discussed in \cref{app:axionEFT}.

\begin{figure}
    \centering
    \includegraphics[width=0.99\linewidth]{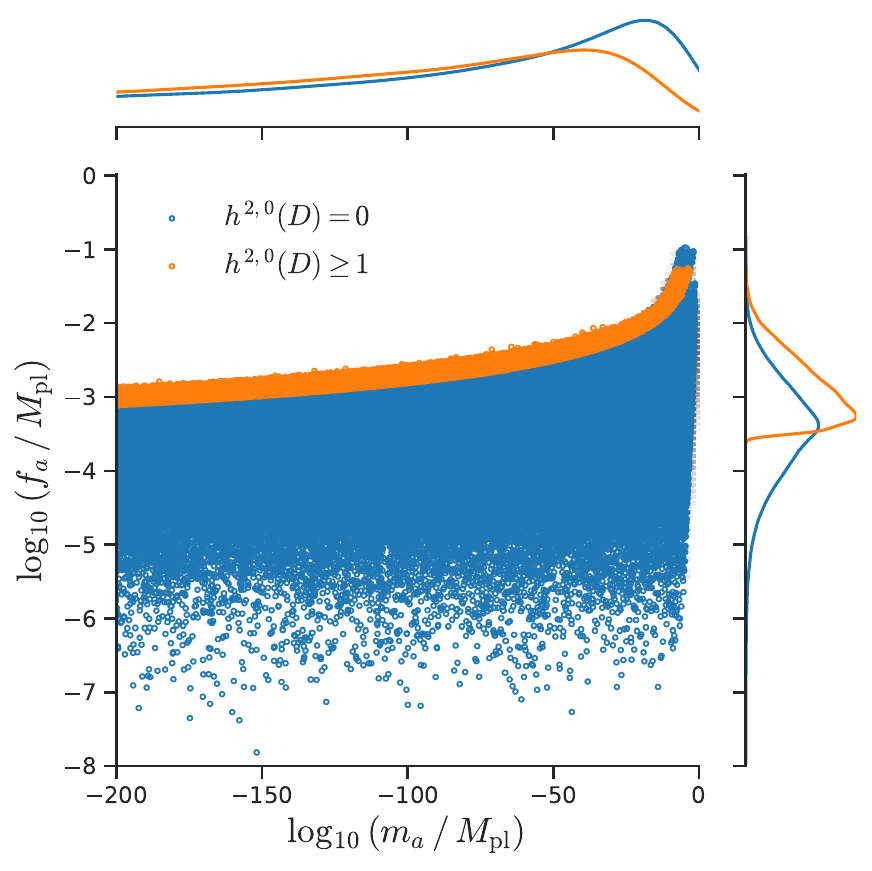}
    \caption{Spectra of all axions (for $h^{1,1}=3$) under the WP prior, without
    any imposition of a gauge-theory (QCD-like) axion. We also separate the
    distributions based on geometric rigidity of the three divisor classes that
    host the corresponding instantons.}
\label{fig:mandf_h113_noQCD}
\end{figure}

\begin{figure*}
    \centering
    \includegraphics[width=0.99\linewidth]{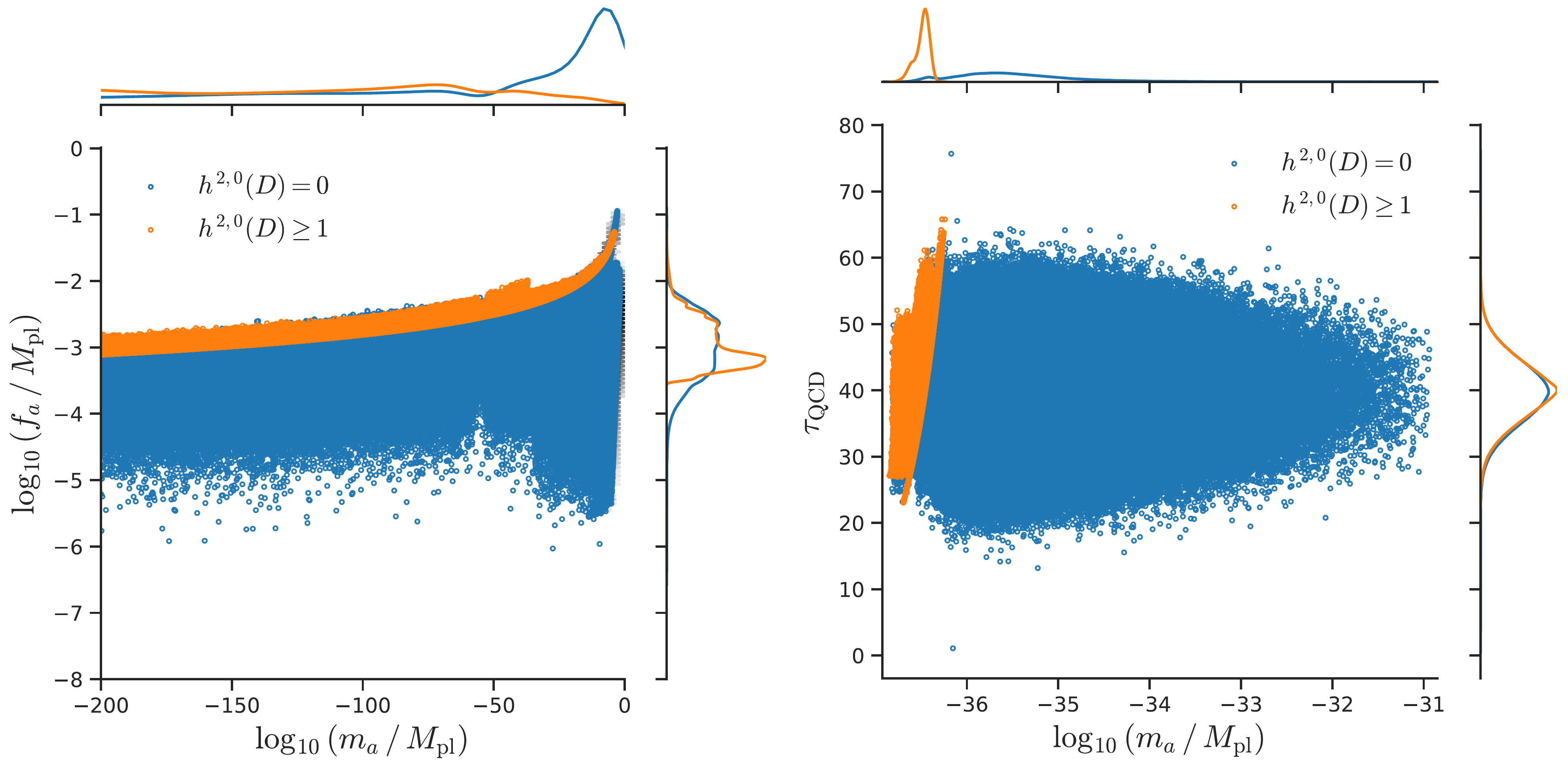}
    \caption{Spectra of all axions (for $h^{1,1}=3$) under the WP prior, with
    identification of a gauge-theory (QCD-like) axion according to the
    likelihood~\eqref{eq:div_likelihood}. Compare against
    \cref{fig:mandf_h113_noQCD} which shows the case with no such identification.
    The left panel shows the distributions of stringy axions (ALPs), while the
    right panel shows the spectrum of the QCD-like axion.}
\label{fig:mandf_h113_yesQCD}
\end{figure*}

In~\cref{fig:mandf_h113_noQCD,fig:mandf_h113_yesQCD} we display the collective
distribution of masses and decay constants for \emph{all} axions across all
inequivalent FRSTs with $h^{1,1}=3$, as obtained from our MCMC sampling.
We adopt the hierarchical approximation (see~\cref{app:axionEFT} and
also~\cite{Gendler:2023kjt}) to compute the axion spectrum, under which leading
instanton directions coincide with axion mass eigenstates. It is then useful to
further split the spectrum according to the geometric rigidity of the divisor
classes supporting the instantons, following the discussion of decay constants in \cref{sec:geom_preq}.

In~\cref{fig:mandf_h113_noQCD}, we do not impose any QCD-axion likelihood.
The spectrum therefore contains all three axions for every sampled point across
all \Kahler cones. For both geometrically rigid and movable divisors, the distribution on axion masses has a logarithmically flat behaviour along with a slight blue tilt. The corresponding decay constants for both type of divisor classes, on the other hand, are localised and do not spread across many orders of magnitude. Still, we observe that
geometrically rigid divisors ($h^{2,0}(D)=0$) can support lower decay constants compared to movable divisors ($h^{2,0}(D)\geq 1$). We recall that these two trends were explained geometrically in \cref{sec:geom_preq}.

In contrast, in~\cref{fig:mandf_h113_yesQCD} we impose the
divisor-volume Gaussian likelihood (on top of the WP prior), and we show the
spectrum of both the QCD-like axion and the remaining stringy axions (ALPs). While many qualitative features of the string–axion spectrum remain unchanged (left panel), two effects stand out. First, there is a clear `dip' in the ALP mass distribution around $\sim 10^{-55}\,M_{\rm pl}$. This occurs because the degree of freedom that would otherwise contribute an ALP at that mass has instead been modified to become the QCD-like axion, along with the mass distribution associated with movable divisors becoming noticeably flatter overall. Second, a QCD-like axion engineered on a movable divisor clusters in the neV mass range, whereas a significantly heavier QCD-like axion can arise when realised on a rigid divisor (right panel). This is because rigid divisors can support parametrically smaller decay constants. We return to this feature in more detail in \cref{sec:top_from_halo}.

We now employ a NF model to learn the ALP spectra without imposing any QCD likelihood~\cref{eq:div_likelihood}.
This allows us to extract marginalised distributions for the ALP masses and decay constants, yielding physically meaningful priors that can be directly used in cosmological and astrophysical analyses.

As earlier, we employ a real NVP model. In this case our model is comprised of 8 masked affine flow layers where the translation and scale networks are MLPs consisting of 2 hidden layers each with 128 neurons, which gives a total of 274{,}496 learnable parameters. We train our model over 1{,}000 epochs using the same optimisation procedure as before to minimise the forward KL divergence~\cref{eq:KL_forw} computed from a batch of 512.

\begin{figure*}
    \centering
    \includegraphics[width=0.99\linewidth]{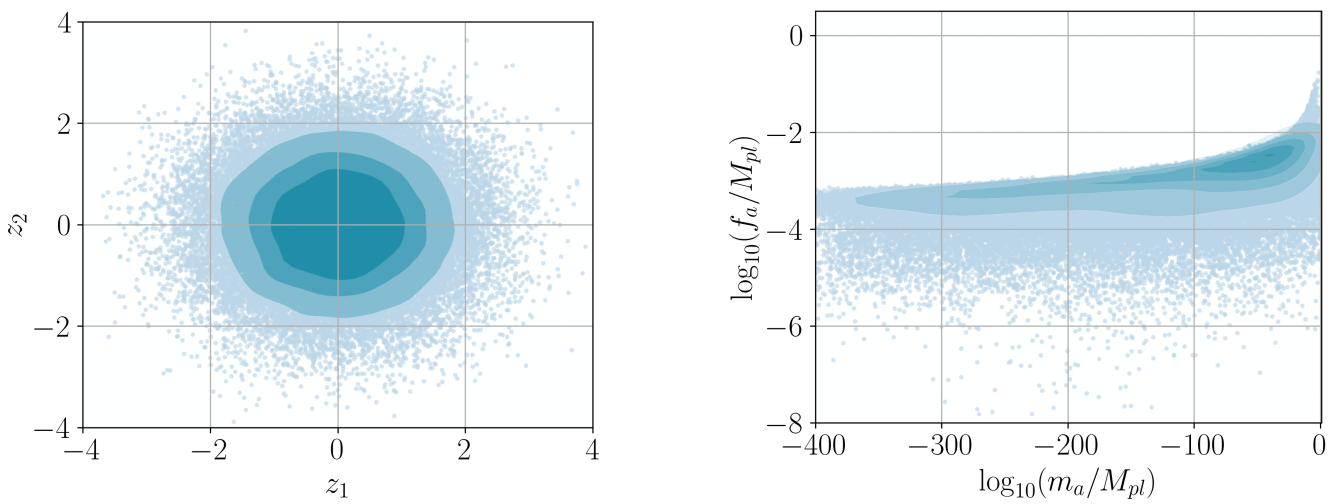}
    \caption{Shown are 10{,}000 samples drawn from the base Gaussian $\mathcal{N}(0,I)$ (left) and the corresponding distribution obtained after transforming these same samples through the normalising flow trained on the MCMC sample set in~\cref{fig:mandf_h113_noQCD} (right).}
\label{fig:mandf_h113_NF}
\end{figure*}

In~\cref{fig:mandf_h113_NF}, we show the initial Gaussian base distribution $\mathcal{N}(0,I)$ alongside the learned distribution for 50{,}000 generated samples. Comparing this with the full spectra shown in~\cref{fig:mandf_h113_noQCD} we can see that the NF model has successfully learned the $(m_{a},f_{a})$ distribution even down to the tails at lower values in the decay constant. 

While we do not attempt to learn the spectra shown in~\cref{fig:mandf_h113_yesQCD} here, we note that one could train a NF on the joint distribution of (1) the QCD-like axion mass and decay constant, (2) the volume of the QCD divisor, and (3) the ALP masses and decay constants, and then train a second NF model on the same variables but with the QCD-divisor volume likelihood imposed.
In principle, the latter could then be used to infer the distribution appearing in~\cref{fig:mandf_h113_yesQCD}.
We leave this experiment to future work.

\subsection{Inferring topology with haloscopes}
\label{sec:top_from_halo}

In this section we examine more specifically how the mass of the QCD axion is
related to the topology of the underlying CY manifold, with QCD
realised on a stack of D7-branes (see Appendix~\ref{app:axionEFT}). While
divisor volumes control the mass scales of generic axions receiving masses from
ED3 instantons, the mass of the QCD axion, $m_{\rm QCD}$, is determined only by
the decay constant $f_a$ once the D7-brane divisor volume has been fixed to
give the desired ultraviolet gauge coupling.

There is a strong statistical correlation between Hodge numbers and axion decay
constants, as seen in the existing literature: larger values of $h^{1,1}$ typically give smaller decay constants~\cite{Demirtas:2018akl,Halverson:2019cmy,Mehta:2021pwf,Gendler:2023kjt}.
The TSKC provides an indicator of how small the \Kahler parameters can
collectively become, and therefore of the largest possible decay constants, at
fixed $h^{1,1}$. This implies that at each value of $h^{1,1}$ there is a maximum decay constant and a corresponding minimum $m_{\rm QCD}$, both set roughly
by the CY volume at the tip, $\mathcal{V}_{\rm tip}$. If one moves
along a pure dilation of the TSKC, all decay constants decrease and all divisor
volumes increase. Demanding that there exists a divisor with volume close to
$40$ in string units, in order to give the desired gauge coupling, and
restricting attention to such dilations, leads to a minimum value of $f_a$ and
thus a maximum value of $m_{\rm QCD}$. These facts underlie the correlation
between QCD axion mass and topology discussed in
Refs.~\cite{Gendler:2024adn,Benabou:2025kgx}.

However, as we constructed explicitly in Ref.~\cite{Sheridan:2024vtt}, and
informed by the geometric picture we further developed in
Secs.~\ref{sec:preliminaries} and \ref{sec:geometry_results}, 
when exploring the full moduli space, one can find loci where the QCD divisor volume remains $40$ but the overall volume is parametrically larger than it was on the TSKC ray, resulting in much smaller decay constants.\footnote{In a similar view, there could also \textit{not} be a maximum far below the Planck scale. For instance K3 divisor classes may provide for large axion decay constants, in regions close to faces of the stretched \Kahler cone.}
In this
case the naive topology–mass correlation can be evaded, provided the geometry
contains divisor classes whose volumes are effectively decoupled from the
overall CY volume. With this in mind, we revisit some aspects of
Ref.~\cite{Gendler:2024adn} and ask to what extent axion haloscopes can make
inferences about CY topology.

For CY threefolds with small $h^{1,1}$, the decay constants are
typically large in regions that are not deep inside the \Kahler cone, with
representative values $f_a \gtrsim 10^{15}\,\mathrm{GeV}$. This leads, in
general, to a light QCD-like axion in the neV range of masses. As an explicit
example, for our ensemble with $h^{1,1}=3$ this generic behaviour is visible in
the right panel of Fig.~\ref{fig:mandf_h113_yesQCD}. The orange spectra,
which correspond to instantons on divisors with $h^{2,0}(D) \ge 1$, cluster in the
neV range. These non-rigid divisors have volumes that closely track the overall
\Kahler scale and therefore do not permit sufficiently small decay constants
for a heavier QCD-like axion at fixed gauge coupling.

In contrast, the rigid prime toric divisors with $h^{2,0}(D)=0$ can support a
high-mass QCD-like axion. This is the phenomenon anticipated at the end of the
previous subsection: the blue spectra in the right panel of
Fig.~\ref{fig:mandf_h113_yesQCD} arise from instantons on rigid divisors and
populate the high-mass tail of the QCD-axion distribution. For small $h^{1,1}$, such rigid divisors can remain at moderate volume even in
vacua that lie deep inside the \Kahler cone, so that some (or in special cases
all) axionic directions acquire parametrically small decay constants. This may
allow one axion to realise both a large mass and an appropriate gauge
coupling. The MCMC analysis of
Sec.~\ref{sec:geometry_results} shows that exactly these rigid divisors are
responsible for the high-mass QCD-like axions in our sample.

In summary, achieving a heavy QCD-like axion requires simultaneously controlling 
the divisor volumes and accessing regions of moduli space where certain decay 
constants become parametrically small. These two ingredients do not always 
coexist, and only arise when the underlying geometry satisfies a pair of sharp 
geometric conditions. For clarity, we summarise them here:
\begin{itemize}
    \item The vacuum must lie deep inside the (stretched) \Kahler cone; i.e., the overall CY volume is large.

    \item The geometry must contain at least one prime toric divisor whose constant $\tau (\sim \bar{\tau}_{\rm QCD})$ level sets extend towards this vacuum.

\end{itemize}
These two conditions together allow one axionic direction to realise both a
parametrically small decay constant, yielding a heavy QCD-like mass, and a
suitably small divisor volume that sets the gauge coupling.

Instantons on non-K3 shrinking divisors serve as possible candidates.\footnote{The decay constants of axions that are associated with instantons on K3 fibers, depend solely on their volumes (i.e. $f \sim 1/\tau$). Requiring volume to be close to 40 then, necessarily enforces a large decay constant. See~\cref{sec:Kahler_metric_decay_constants} for a discussion.} This
leaves us with two possible divisor types: (a) blow-up divisors, which can shrink while the CY has finite volume; (b) fibration divisors, which can only shrink if the whole CY also shrinks. 

In general for $h^{1,1} = 3$, we find that $\sim 90\%$ of inequivalent FRSTs from the KS database either have a blow up prime-toric divisor or an elliptic fibration. One then needs to exploit this structure. That is, be able to host the QCD instanton on a special shrinking divisor and also take an appropriate limit in the moduli space. We illustrate this by examining two explicit geometries at $h^{1,1} = 3$, each exhibiting these types of special divisors: a
non-fibered geometry with blow-up divisors that shrink on a facet and on a ray respectively, and an elliptically fibered geometry with non blow-up shrinking divisors.

For each of the two cases, we perform seven MCMC runs: one for every choice of the prime toric divisor for hosting QCD. We take a log-normal likelihood for the
QCD axion mass centred at $\bar{m}_{\rm QCD} = 6\,\mu\text{eV}$ with
sufficiently narrow width, and a Gaussian likelihood for the divisor volume
centred at $\bar{\tau}_{\rm QCD} = 40$ with a sufficiently narrow width.
Explicitly:
\begin{align}
\label{eq:likelihoods_qcd}
    \ln\mathcal{L} &= \ln\mathcal{L}_{m_{\rm QCD}} + \ln\mathcal{L}_{\rm vol}\,,\\
    \text{where}\quad
    \ln\mathcal{L}_{m_{\rm QCD}} &= - \frac{1}{2\times (0.5)^2}
    \left[\log_{10}\!\left(\frac{m_{\rm QCD}}{\bar{m}_{\rm QCD}}\right)\right]^2\,,\nonumber\\
   \ln\mathcal{L}_{\rm vol} &= -\frac{1}{2\times 5^2}(\tau_{\rm QCD} - \bar{\tau}_{\rm QCD})^2\,.\nonumber
\end{align}
This likelihood is meant to model a putative detection of the axion by the
haloscope ADMX~\cite{ADMX:2018gho}.\footnote{Computationally we were forced to impose a relatively wide likelihood on $\log_{10} m_{\rm QCD}$, while in reality the axion mass will be measured with a relative accuracy around $\sigma_m/m_{\rm QCD}\approx 1/Q \approx 10^{-5}$ by ADMX, where $Q$ is the cavity quality factor.} ADMX is sensitive in GHz frequencies, which count as ``high'' for $h^{1,1}=3$ CYs, i.e., not in the range of $h^{1,1}$ found to be accessible to ADMX by dilation of TSKC in Ref.~\cite{Gendler:2024adn}. 

\subsubsection{Blow-up divisors} 

We consider a 4D reflexive polytope $\Delta^{\circ}$ with points given by the columns of
\begin{align}
\label{eq:geometry1}
    \begin{pmatrix}
        0 & 0 & 1 & 0 & 1 & -1 & 1\\
        0 & 0 & -1 & 1 & 1 & 0 & 0\\
        0 & 1 & -1 & 0 & 0 & 0 & 0\\
        1 & 0 & -1 & 0 & 0 & 0 & 0
    \end{pmatrix}\,,
\end{align}
along with the following gauged linear sigma model (GLSM) charge matrix (which we take as the instanton charge matrix following e.g. \cite{Demirtas:2018akl}):
\begin{align}
    \begin{pmatrix}
        1 & 1 & 1 & 1 & 0 & 1 & 0\\
        0 & 0 & 0 & -1 & 1 & 1 & 0\\
        0 & 0 & 0 & 0 & 0 & 1 & 1
    \end{pmatrix}\,.
\end{align}
A choice of height vector induces a triangulation
that defines a toric variety in which the generic anticanonical hypersurface is a smooth CY
threefold with $h^{1,1} = 3$. The columns of the GLSM/instanton charge matrix correspond to prime toric divisor classes, and in particular our focus is on $D_5$ and $D_7$. There fare 3 extremal facets of the $K_\cup$ given by the rows of
\begin{align}
    \mathcal{H} = 
    \begin{pmatrix}
        0 & 1 & 0\\
        0 & -1 & 1\\
        1 & 1 & -2
    \end{pmatrix}\,,
\end{align}
and the two divisor volumes take the form
\begin{align}
    \tau_5 &= \frac{1}{2}\left(t_1^2 + t_2 + 4 t_2 t_3 - 4 t_3^2 + t_1(4t_3 - 6t_2)\right)\,,\nonumber\\
    \tau_7 &= (t_1 + t_2 - 2t_3)^2\,.
\end{align}

\begin{figure}
    \centering
    \includegraphics[width=0.99\linewidth]{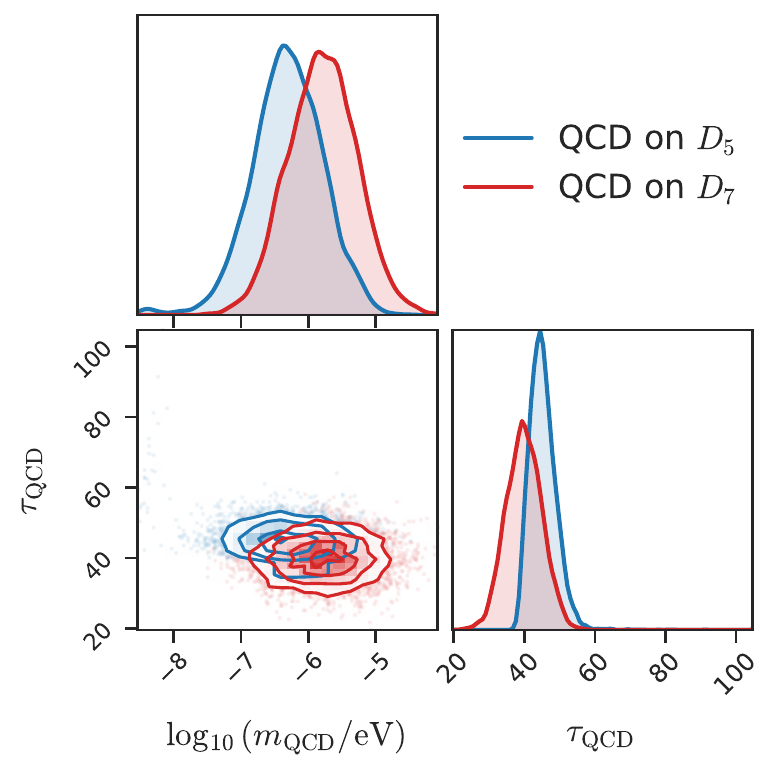}
    \caption{Posterior distributions of the QCD-axion mass and the corresponding QCD divisor volume for the two blow-up divisor choices of the geometry defined in~\eqref{eq:geometry1}. The results are obtained using the likelihood~\eqref{eq:likelihoods_qcd} with fiducial values $\bar{\tau}_{\rm QCD} = 40$ and $\bar{m}_{\rm QCD} = 6\,\mu$eV.}
\label{fig:corner_QCD_FRST44_indices_4_6}
\end{figure}

\begin{figure}
    \centering
    \includegraphics[width=0.99\linewidth]{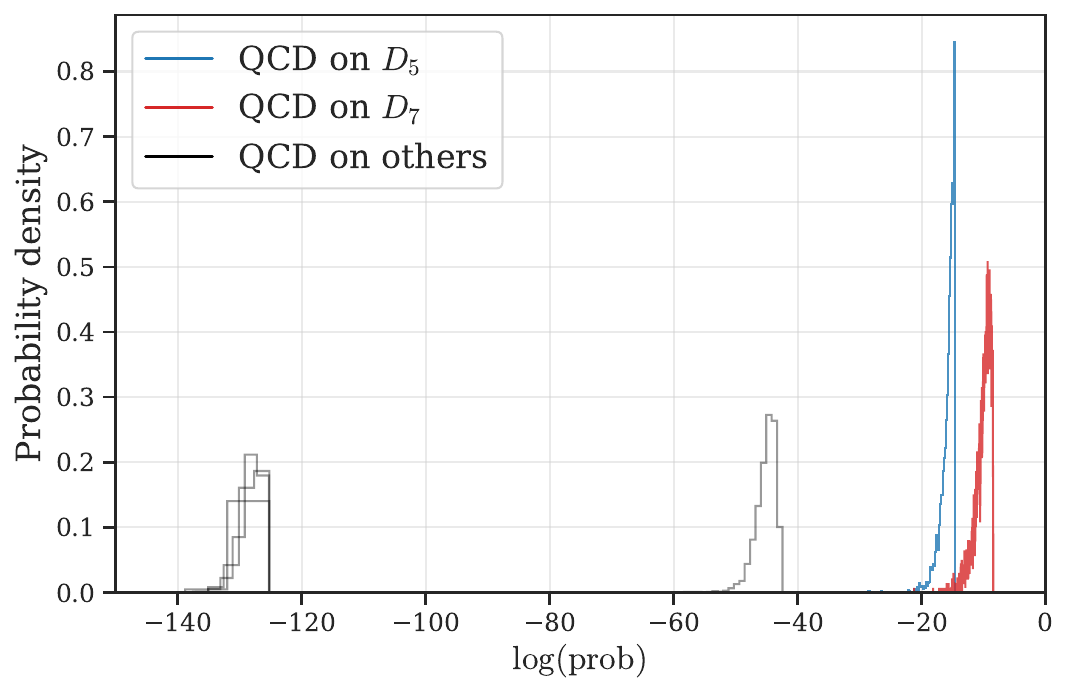}
    \caption{Posterior probability density for all seven choices of prime toric divisors, given a likelihood on the QCD-axion mass peaked at $6\,\mu$eV and its divisor volume peaked at $40$ (in string units). (Among the others, the log(prob) values for $D_6$ are too small and are not shown here).}
    \label{fig:QCD_posterior_FRST44}
\end{figure}

Note that $D_7$ shrinks to a point along the cone facet whose normal is $(1, 1, -2)$, while $D_5$ shrinks to a point along the codimension-two ray $(1,1,1)$. Both divisors correspond to blow-up loci, in the sense that they can shrink while the CY volume remains finite.\footnote{This geometry is in fact of `Swiss-Cheese' type, and the volume form has a simple decoupled structure $\mathcal{V} = \sqrt{8/5}\,\tau_6^{3/2} - \sqrt{1/2}\,(\tau_6 - \tau_4)^{3/2} - \tau_7^{3/2}$. Such a simplicity, however, is not essential for our purposes.} In~\cref{fig:corner_QCD_FRST44_indices_4_6} we show the derived parameters of QCD axion mass and divisor volumes, for both of the two choices of $D_5$ or $D_7$ hosting the QCD gauge group. To compare with the other five possibilities, in \cref{fig:QCD_posterior_FRST44} we show the posterior probability density for all seven prime toric divisors. The posterior probability is significantly higher for QCD realised on D5 or D7, indicating that an ADMX detection would prefer these discrete choices over the others available. 

Both figures also illustrate that, between $D_5$ and $D_7$, $D_7$ is the more suitable candidate. Its shrinking locus is a codimension-one surface (a facet), i.e.\ a higher-dimensional boundary of the \Kahler cone, whereas $D_5$ shrinks only along a codimension-two locus (an extremal ray), which is correspondingly smaller. The remaining prime toric divisors are not shrinkable.

\subsubsection{Non-blow-up shrinking divisors}

We consider a 4D reflexive polytope $\Delta^{\circ}$ whose points are given by the columns of
\begin{align}
\label{eq:geometry2}
    \begin{pmatrix}
        1 & -5 & 0 & 0 & 0 & -3 & -2 & -1\\
        0 & -1 & 0 & 0 & 1 & 0 & -1 & 0\\
        0 & -2 & 0 & 1 & 0 & -1 & 0 & 0\\
        0 & -1 & 1 & 0 & 0 & -1 & 0 & 0
    \end{pmatrix}\,,
\end{align}
along with the following instanton charge matrix
\begin{align}
    \begin{pmatrix}
        5 & 1 & 1 & 2 & 1 & 0 & 0\\
        3 & 0 & 1 & 1 & 0 & 1 & 0\\
        -3 & -1 & -1 & -2 & 0 & 0 & 1
    \end{pmatrix}\,.
\end{align}
A choice of height vector induces a triangulation
that defines a toric variety in which the generic anticanonical hypersurface is a smooth CY
threefold with $h^{1,1} = 3$, whose volume is
\begin{align}
    \mathcal{V} = \frac{1}{6}(t_1+t_3)(t_1(6t_2+t_3)-t_1^2 - 6t_2^2 - 6t_2t_3 - t_3^2)\,.
\end{align}
The columns of the instanton charge matrix correspond to prime toric divisor classes, and in particular our focus is on $D_2$ and $D_6$ whose volumes take the form
\begin{align}
    \tau_2 &= \frac{1}{2}(t_1+t_3)(-t_1 + 4t_2 + t_3)\,,\nonumber\\
    \tau_6 &= (t_1+t_3)(t_1 - 2t_2 - t_3)\,.
\end{align}
There are three extremal facets of $K_\cup$, given by the rows of
\begin{align}
    \mathcal{H} = 
    \begin{pmatrix}
        -1 & 2 & 0\\
        0 & -1 & -1\\
        1 & 0 & 1
    \end{pmatrix}\,.
\end{align}
From $\mathcal{V}$ and $\mathcal{H}$ it is clear that the facet of $K_\cup$ with normal vector $(1,0,1)$ corresponds to an elliptic fibration. Both $D_2$ and $D_6$ shrink to curves along this facet. Furthermore, they each shrink to points along codimension-two rays of $K_\cup$: $D_2$ shrinks along the intersection of this facet with the facet normal to $(-1,2,0)$, while $D_6$ shrinks along the intersection with the facet normal to $(0,-1,-1)$. These loci, where the two divisors shrink, are precisely the regions where a high-mass QCD instanton can be supported.

\begin{figure}
    \centering
    \includegraphics[width=0.99\linewidth]{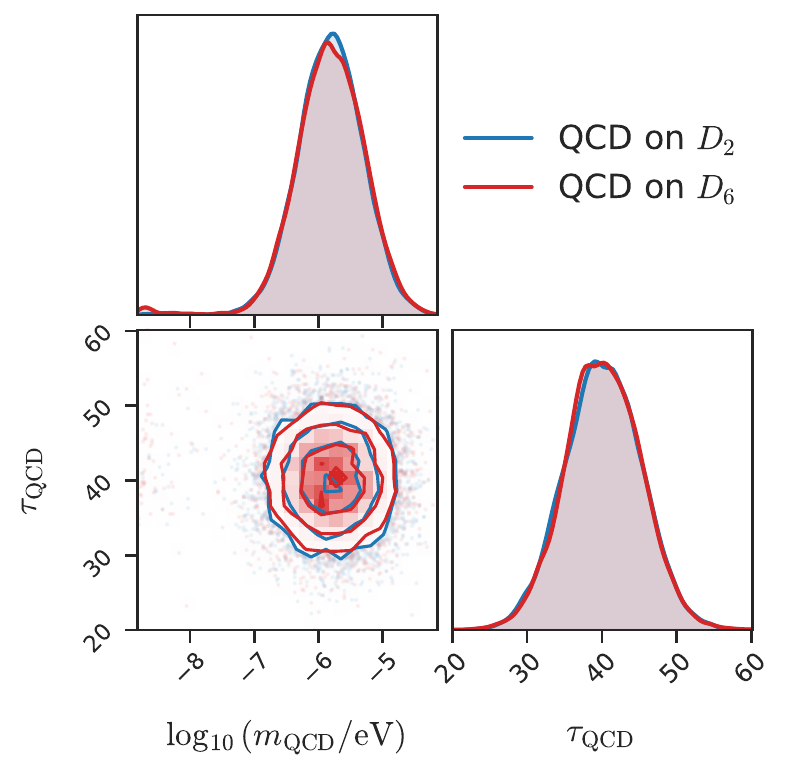}
    \caption{Posterior distributions of the QCD-axion mass and the corresponding QCD divisor volume for the two non-blow-up shrinking divisor choices of the geometry defined in~\eqref{eq:geometry2}. The results are obtained using the likelihood~\eqref{eq:likelihoods_qcd} with fiducial values $\bar{\tau}_{\rm QCD} = 40$ and $\bar{m}_{\rm QCD} = 6\,\mu{\rm eV}$.}
\label{fig:corner_QCD_FRST105_indices_1_5}
\end{figure}

\begin{figure}
    \centering
    \includegraphics[width=0.99\linewidth]{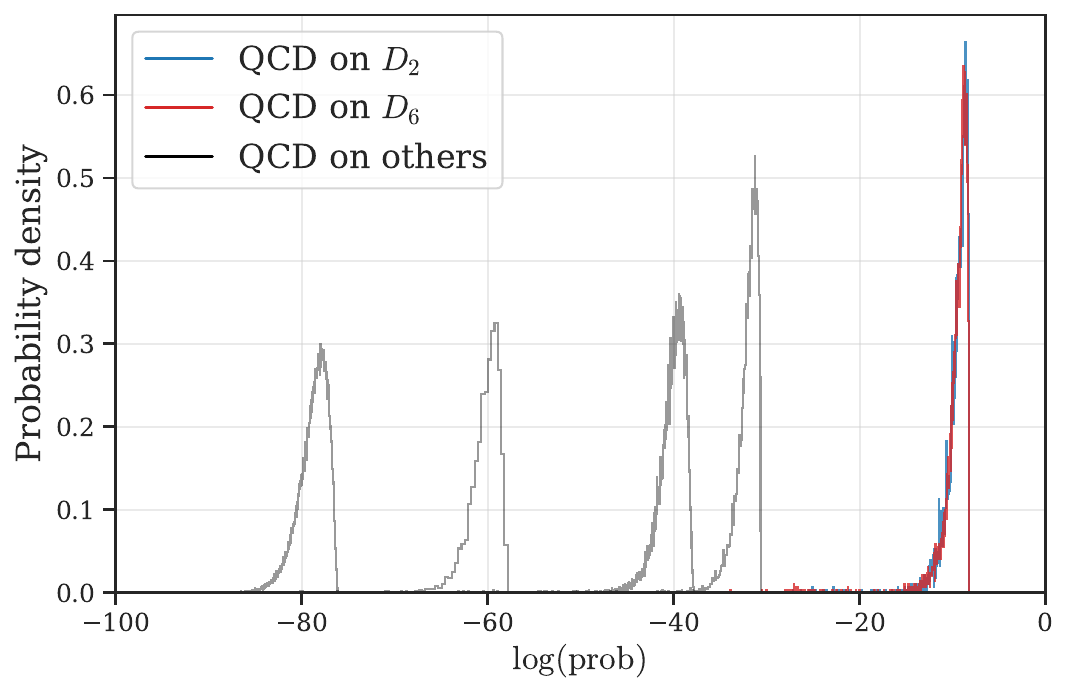}
    \caption{Posterior probability density for all seven choices of prime toric divisors, given a likelihood on the QCD-axion mass peaked at $6\,\mu$eV and its divisor volume peaked at $40$ (in string units). (Among the others, the log(prob) values for $D_1$ are too small and are not shown here).}    \label{fig:QCD_posterior_FRST105}
\end{figure}

In~\cref{fig:corner_QCD_FRST105_indices_1_5} we show the derived QCD-axion mass and divisor volume for the two choices $D_2$ and $D_6$ hosting the QCD gauge group. To compare with the five remaining choices, in~\cref{fig:QCD_posterior_FRST105} we show the posterior probability density for all seven possibilities.

To connect with the previous example, it is not surprising that in this case both $D_2$ and $D_6$ are equally good candidates (and exhibit similar statistics), as each shrinks along a codimension-two locus.

We fnally note that in this example there are two additional divisors ($D_3$ and $D_4$) that shrink to curves along the fibration facet. However, there is no locus where they can shrink to a point, and it turns out that they cannot support a high-mass QCD-like instanton. This aligns with \cref{fig:V_vs_tau}, where we see that point blowup divisors are capable of smaller volumes at larger overall volume than curve blowup divisors. \\

We also performed a subset of the analysis in this subsection with $\bar{m}_{\rm QCD}=20~\text{neV}$ to model a putative detection of the QCD axion by DMRadio~\cite{DMRadio:2022pkf}. Since masses in the neV range are ``typical'' for dilations of the TSKC at $h^{1,1}=3$, the DMRadio analysis yielded trivial results in which QCD could be realised on any type of divisor with high posterior probability. We have thus shown that, for CYs with low $h^{1,1}$, a detection by DMRadio would give little to no specific topological information; by contrast, a detection by ADMX has greater constraining power, statistically preferring certain CY topologies and divisor choices to host QCD.

\subsection{Fuzzy dark matter and the Lyman-alpha forest}

Ultralight axions with $m_a \lesssim 10^{-17}\,\text{eV}$ differ from standard cold DM (CDM) in their effects on cosmological observables. Ultralight axions suppress matter clustering below the de Broglie wavelength (which sets an effective Jeans scale), leading to a reduced amplitude of the linear matter power spectrum $P(k)$ on small scales~\cite{Frieman:1995pm,Hu:2000ke,Amendola:2005ad,Arvanitaki:2009fg,Marsh:2010wq,Lague:2021frh}, often dubbed ``fuzzy'' DM~ (see reviews in Refs.~\cite{Marsh:2015xka,OHare:2024nmr,Eberhardt:2025caq}). Cosmological observables are each individually consistent with the predictions of the CDM model. Thus, the absence of such suppression in $P(k)$ (and other effects) lead to upper limits on the allowed ultralight axion density $\rho_a=\Omega_a h^2\times 8\times 10^{-11}\text{ eV}^4$~\cite{Frieman:1995pm,Amendola:2005ad,Hlozek:2014lca,Kobayashi:2017jcf,Rogers:2020ltq,Rogers:2020cup,Lague:2021frh,Rogers:2023ezo,Winch:2024mrt}. When two observables, however, yield inconsistent inferences within the $\Lambda$CDM model, this implies either unmodeled astrophysical or instrumental systematics, or physics beyond CDM~\cite[e.g.,][]{Rogers:2023ezo,Rogers:2023upm}. Ref.~\cite{Rogers:2023upm} finds that a \(5 \sigma\) discrepancy in the inference of the small-scale (wavenumber \(k \sim 1\,h\,\mathrm{Mpc}^{-1}\)), high-redshift (\(z \sim 3\)) $P(k)$ given \textit{Planck} CMB and eBOSS Lyman-alpha forest (Ly\(\alpha\)F) data can be resolved by a contribution of ultralight axions (although see Refs.~\cite{Fernandez:2023grg,Walther:2024tcj}).

It is somewhat difficult to find CYs in type IIB that give rise to a large fuzzy DM relic density. Only a sub-percent fraction of the total number of CYs with $h^{1,1}\leq 7$ yield axion models within reach of current and upcoming cosmological observations~\cite{Sheridan:2024vtt,Preston:2025tyl}. In Ref.~\cite{Sheridan:2024vtt}, we constructed a specific example of fuzzy DM at $h^{1,1}=7$ that has the largest misalignment abundance of the lightest axion in the model, while also containing a divisor of volume near $40$ to host QCD.\footnote{The fuzzy axion with a large decay constant in this example corresponds to a $K3$ fibration. Interestingly, though, the associated divisor $D$ is not quite the $K3$ fiber, but rather a $\mathrm{dP}_9$ that is ``half'' of the $K3$ fiber (see, e.g., footnote 19 in \cite{Blumenhagen:2008zz}). The canonical class of $\mathrm{dP}_9$ is pure torsion, so $D \cap D$ still vanishes numerically and the discussion of $K3$ divisors and the \Kahler metric of \cref{sec:geom_preq} applies to $D$, explaining the large decay constant. Usefully, $\mathrm{dP}_9$ is also rigid, so it does not carry neutral fermionic zero modes that would obstruct a nonperturbative contribution to the superpotential.} Upon dilating the CY volume, this model furnishes a combination of $(m_a,\Omega_a)$ near the maximum posterior probability of these parameters inferred from the joint analysis~\cite{Rogers:2023upm} of \textit{Planck} CMB and eBOSS Lyman-$\alpha$ forest data~(alternative new physics resolutions to this parameter discrepancy include running of the primordial power spectrum spectral index, e.g. Refs.~\cite{Rogers:2023upm,Fairbairn:2025fko}). Thus, this particular CY model is a possible candidate to explain the discrepancy between these datasets.

We now employ the tools we have developed in the previous sections to sample the moduli space of this example CY, with the goal of inferring the region of moduli space preferred by a solution to the CMB-Ly$\alpha$F discrepancy. We sample the CY moduli space using the \Kahler parameters and the WP prior, and compute the axion masses \(m_a\) and decay constants \(f_a\) that arise at each point. We impose a prior that there exists a divisor with volume close to 40 [\cref{eq:div_likelihood}] and place QCD there in order to have the correct effect of the QCD axion mass on the overall spectrum of masses and decay constants.

We simultaneously sample cosmological parameters \((\omega_\mathrm{DM}, \mathrm{log}\frac{\Omega_a}{\Omega_\mathrm{DM}}, \omega_\mathrm{baryon}, h, A_s, n_s, \tau)\). These parameters are respectively: the total DM physical density, the logarithm of the fraction of the DM composed of ultralight axions, the physical baryon density, the amplitude and spectral index of the primordial power spectrum and the optical depth to reionization. Cosmological parameter priors are uniform other than a Gaussian prior on the optical depth to reionisation used as a compression of the \textit{Planck} \texttt{Sroll2} low-multipole polarization data \cite{Delouis:2019bub}. The only axion whose physics we treat precisely is the lightest axion in the spectrum, which in the parameter space of interest has \(m_a \in [10^{-27}, 10^{-23}]\,\mathrm{eV}\). For this axion, we use the Einstein-Boltzmann solver \textsc{AxiCLASS}~\cite{Blas:2011rf,Poulin:2018dzj,Smith:2019ihp,Lee:2020obi} to solve the Klein-Gordon equation for the relic density and the evolution of perturbations. We thus compute the CMB temperature \(T\) and polarisation \(E\) angular power spectra $(C_\ell^{TT}, C_\ell^{TE}, C_\ell^{EE})$ and the linear matter power spectrum $P(k)$.\footnote{With $m_a$ and $f_a$ as input, \textsc{AxiCLASS} uses a shooting method to find the initial misalignment angle $\langle \theta_i\rangle$ necessary to obtain the desired relic abundance $\Omega_a$. For a given $f_a$, this method only gives a solution for some minimum $m_a$ since arbitrary fine-tunings to $\langle \theta_i\rangle=\pi$ (``extreme axion'' models, e.g., Refs.~\cite{Zhang:2017dpp,Arvanitaki:2019rax,Winch:2023qzl}) are disallowed. This sets an effective prior $\langle \theta_i\rangle \sim \mathcal{O}(1)$.} The effect of all other axions is taken into account only through the CDM density $\Omega_c$. As discussed in detail in Ref.~\cite{Sheridan:2024vtt}, avoiding overproduction of CDM from heavier axions including the QCD axion implies a fine tuning of the intial misalignment for these axions and/or modified cosmology (low reheating scale), which we do not address here. We set the isocurvature amplitude (CDM and ultralight axion) to zero (for details on ultralight axion isocurvature, see Refs.~\cite{Marsh:2013taa,Hlozek:2017zzf}). Inflation is modeled only via the effective parameters $A_s$ and $n_s$. This simplified model with minimal tuning of the heavy axion misalignment angles and vanishing isocurvature is equivalent to an assumption of a low inflationary Hubble scale~\cite{Hertzberg:2008wr,Marsh:2015xka}. Going beyond these minimal assumptions is beyond the scope of the present work but can be addressed in future.

We jointly sample the posterior distribution of the \Kahler and cosmological parameters given \textit{Planck} CMB and eBOSS Ly\(\alpha\)F data using the Markov chain Monte Carlo sampler \textsc{emcee}~\cite{Foreman-Mackey:2012any}. For the CMB temperature and polarization data, we use the compressed version \textsc{Planck-lite-py} \cite{Prince:2019hse} of the \textit{Planck} 2018 low-multipole \(TT\) and high-multipole $TT$, $TE$, $EE$ data \cite{Planck:2019nip} (as discussed above, the low-multipole polarization data are compressed to the prior on \(\tau\)). For the eBOSS Ly\(\alpha\)F 1D flux power spectrum data \cite{eBOSS:2018qyj}, we use the compressed version \cite{Goldstein:2023gnw,Rogers:2023upm} of a Gaussian likelihood on the amplitude and spectral index of the linear matter power spectrum \(P(k = 0.009\,\mathrm{s}\,\mathrm{km}^{-1}, z = 3)\). Both these compressions are lossless for the axion cosmology we consider here. For comparison, we also perform a second analysis where we do not sample the \Kahler parameters and instead sample \((\mathrm{log}\,m_a, \mathrm{log}\,f_a)\) with an upper limit on \(f_a\) at the reduced Planck scale. This second analysis matches previous fuzzy DM studies and we dub the log-uniform prior on axion mass and decay constant an effective field theory (EFT) prior.

\begin{figure*}
    \centering
    \includegraphics[width=0.49\linewidth]{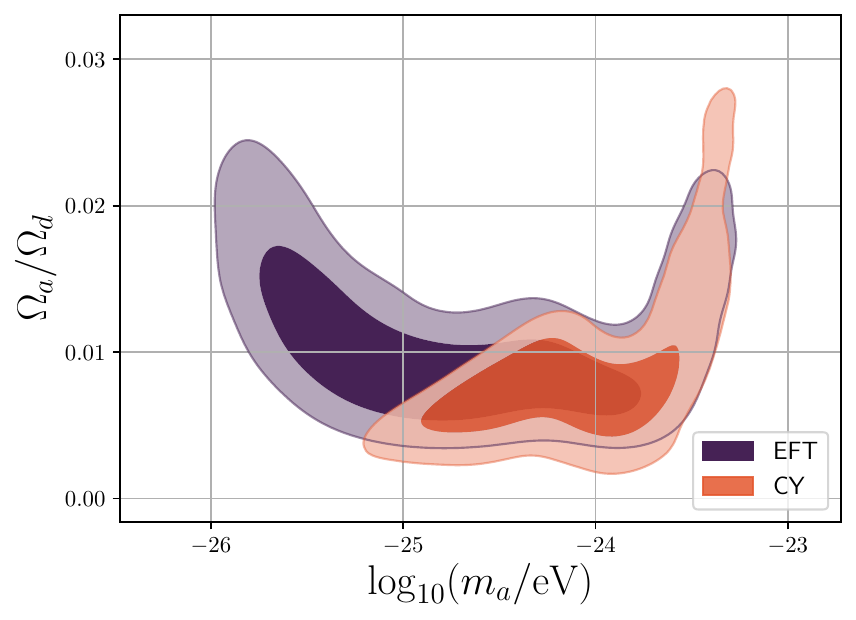}
    \includegraphics[width=0.49\linewidth]{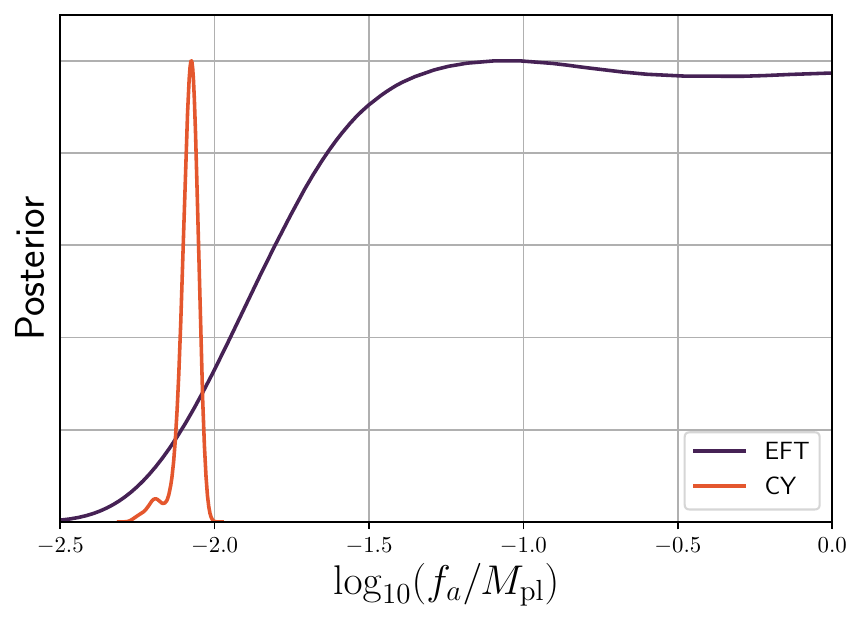}
    \caption{\textit{Left}: the marginalized posterior distribution of axion mass \(m_a\) and axion DM fraction \(\frac{\Omega_a}{\Omega_\mathrm{DM}}\). We indicate by the darker and lighter contours respectively the 68\% and 95\% credible regions. \textit{Right}: the marginalized posterior distribution of axion decay constant \(f_a\). The likelihood is a product of the \emph{Planck} 2018 cosmic microwave background temperature and polarization likelihood and the eBOSS Lyman-alpha forest 1D flux power spectrum compression. We show both an analysis that samples the CY moduli space for a model with $h^{1,1}=7$ with the WP prior and the QCD divisor volume prior, and an EFT analysis where the mass and decay constant of a single ultralight axion are independent free parameters with a log-uniform prior. In a pure CDM model, there is a parameter discrepancy between the CMB and Ly\(\alpha\)F data, which can be resolved by an ultralight axion causing a suppression in the linear matter power spectrum, leading to a posterior peak with non-zero $\Omega_a$. We find that sampling the CY moduli space for a chosen axion model prefers a lower-\(f_a\) part of the posterior.}
    \label{fig:cosmo_param_posterior}
\end{figure*}

\cref{fig:cosmo_param_posterior} (\textit{left}) shows the marginalized posterior distribution of the axion mass and DM fraction. The posteriors of the EFT and CY analyses are consistent. The EFT analysis (\textit{purple contours}) reproduces the results of Ref.~\cite{Rogers:2023upm}, which finds a preference for \(\sim (1-3)\%\) DM abundance in an axion with $10^{-26}\,\mathrm{eV}\lesssim m_a\lesssim 10^{-23}\,\text{eV}$. The CY analysis similarly finds a posterior peak in this region. However, the CY analysis (\textit{orange contours}) only allows for $10^{-25}\,\mathrm{eV}\lesssim m_a\lesssim 10^{-23}\,\text{eV}$. The reason for the CY result differing from the EFT result is shown in Fig.~\ref{fig:cosmo_param_posterior} (\textit{right}), which shows the marginalized posterior of the axion decay constant. The EFT analysis allows decay constants up to the Planck scale, which allows for larger axion DM abundance at lower mass. In contrast, the CY model does not allow for such large decay constants and the posterior is peaked near $f_a \sim 10^{-2} M_{\rm pl}$. This is the largest $f_a$ in the moduli space of the CY that we consider (given the prior and likelihood above). This result is consistent with the indications in Refs.~\cite{Sheridan:2024vtt,Cicoli:2021gss} concerning the maximum DM abundance of ultralight axions.

The power of this Bayesian analysis is demonstrated in Fig.~\ref{fig:money_plot}, which shows the marginalized posterior of the CY volume. Compared to the prior alone, the cosmological data lead to a sharply peaked posterior on the CY volume. In fact, the posterior of all seven \Kahler parameters is sharply peaked in comparison to the prior. This analysis demonstrates that it is possible to perform a rigorous statistical analysis using modern precision cosmological data to find the maximum posterior probability in CY moduli space, while jointly sampling cosmological model parameters.

\section{Conclusions}
\label{sec:conclusions}

In this work we have presented steps towards a long held ideal in string phenomenology: the ability to make robust statistical inference about string theory using observations. Steps towards this lofty goal required us to begin with many simplifications, the first of which was to take a ``modular approach'' to phenomenology, treating the vevs of the \Kahler moduli as continuous free parameters. With this assumption, we developed robust methodology to sample from a special diffeomorphism and scale invariant (and thus uninformative in the Bayesian sense) prior on moduli space: the Weil--Petersson (WP) prior. We used the classical \Kahler potential to construct the WP prior: we leave a study of quantum corrections (see, e.g., Ref.~\cite{McAllister:2023vgy}) to future work.

We employed two sampling methods: affine invariant MCMC, and normalising flows. With a laptop GPU, MCMC can achieve statistically reliable, converged samples from WP in geometries with up to $30$ \Kahler moduli in a modest amount of time. Sampling was also fast enough that we could exhaustively sample the WP measure in all CYs derivable from the Kreuzer--Skarke database with $h^{1,1}\leq 5$ with very modest (laptop time) computing resources. Using normalising flows, we can achieve sampling consistent with MCMC. Computation is more intensive, but once learned, sampling from the distribution can be applied again and again ``for free'', and such methods may therefore be of use in the future.

With a statistically robust method of moduli space sampling, we first explored distributions of geometric data of CY manifolds: namely, volumes of divisors and correlations between volumes of divisors and of the CY. In particular, we illustrate how the topology of a divisor --- especially how it shrinks along boundaries of the \Kahler cone --- controls the statistics of its volume distribution. This includes ensembles constructed by marginalising over geometries as well as over the prior within specific geometries, including a geometry at $h^{1,1} = 30$ optimised by a genetic algorithm~\cite{MacFadden:2024him} to have simplicial $K_\cup$. Additionally, we verify the expectation (stated, for example, in Ref.~\cite{Demirtas:2018akl}) that divisor volumes in the SKC tend to become larger as $h^{1,1}$ increases, and we show that the observations of Ref.~\cite{Cheng:2025ggf} regarding distributions of divisor volumes under a particular normalisation persist for samples from the WP prior. 

We applied our sampling methodology to the physics of axions. We first presented the theory informed prior on the joint distribution of axion masses and decay constants, marginalised over the WP measure for all FRST classes of CYs constructable from the Kreuzer--Skarke database at $h^{1,1}=3,4$, and demonstrated that this distribution can be learned using normalising flows. 

We next employed Bayesian inference with mock likelihoods designed to simulate the possible future detection of the QCD axion with a DM haloscope, extending the results of Ref.~\cite{Gendler:2024adn} across moduli space. We showed, in explicit examples at $h^{1,1}=3$, that detection of a relatively high-mass axion by ADMX is compatible only with certain CY topologies. Although these topologies comprise $\sim 90\%$ of all FRST classes at $h^{1,1}=3$, they nonetheless require the engineering of QCD-like instantons on special divisors. Within these topologies, such a detection would therefore constrain which divisor can host QCD on D7-branes and would localize the vacuum to a region of moduli space close to a wall of the \Kahler cone. The conclusion is powerful: detection of a high-mass QCD axion can provide detailed topological information about CYs with low Hodge number. It would be interesting to study the converse phenomenon at large $h^{1,1}$: for topologically complex CY manifolds, can one find decay constants parametrically larger than their typical TSKC values, thereby giving rise to light QCD axions at high $h^{1,1}$?

In the final part of this paper, we performed Bayesian inference on moduli space with observationally derived cosmological likelihoods. Specifically, we used the \emph{Planck} CMB anisotropies and eBOSS Ly-$\alpha$ forest flux power spectrum. We sampled the cosmological parameters in conjunction with the \Kahler parameters of a CY with $h^{1,1}=7$. The maximum posterior probability was localised in moduli space in a region favouring the existence of an ultralight axion contributing $\mathcal{O}(1\%)$ to the DM density. We interpret this result as being due to a tension between these data in a pure CDM model that is alleviated by the non-zero ultralight axion DM density: the axion Jeans scale causes a suppression of the matter power spectrum relative to CDM~\cite{Rogers:2023upm}. Whether the tension in these data continues to persist in future with a solution based on fundamental physics (rather than e.g. systematics) is not central to our conclusions: we have demonstrated the \emph{principle} of performing precise Bayesian analysis with a string axiverse model and cosmological data as a blueprint for future analyses. The next stage in such an analysis would be to compute the Bayesian evidence, in which case one could perform a model comparison between the CY and naive EFT models we have studied, and between these models and $\Lambda$CDM or other extensions of it. Such a computation could be carried out using nested sampling, and we leave this to a future work.

Taken as a whole, the methods and results we have presented open the door to future uses of observational data to perform Bayesian inference on CY moduli spaces, thus making tentative progress towards statistical model comparison both within string theory and relative to other theories. 

\acknowledgments
We are grateful to Naomi Gendler, Jim Halverson, Liam McAllister, Jakob Moritz, and Sebastian Vander Ploeg Fallon for helpful conversations. MJ and DJEM are supported by a Leverhulme Trust Research Project (Grant No. RPG-2022-145). DJEM is supported by an Ernest Rutherford Fellowship (Grant No. ST/T004037/1) and a consolidator grant (Grant No. ST/X000753/1) from the Science and Technologies Facilities Council. The research of ES and AS is supported by NSF grant PHY-2309456. KKR is supported by an Ernest Rutherford Fellowship from the UKRI Science and Technology Facilities Council (grant number ST/Z510191/1). This material is based upon work supported by the National Science Foundation Graduate Research Fellowship under Grant No. 2139899. This work was initiated at the Banff International Research Station during the workshop ``Progress in the String Axiverse''.  This work made use of the \textsc{python}-compatible packages \textsc{emcee}~\cite{Foreman-Mackey:2012any}, \textsc{getdist}~\cite{Lewis:2019xzd}, \textsc{normflows}~\cite{Stimper2023}, \textsc{matplotlib}~\cite{2007CSE.....9...90H}, \textsc{numpy}~\cite{2020Natur.585..357H}, and \textsc{scipy}~\cite{2020NatMe..17..261V}.

\appendix

\section{Axion effective field theory}
\label{app:axionEFT}

Here we briefly recapitulate the relevant structure of the four-dimensional axion EFT in type IIB Calabi--Yau orientifolds, following the conventions of~\cite{Gendler:2023kjt,Sheridan:2024vtt}.  Our aim is only to summarise definitions and scaling relations already implicit in Section~\ref{sec:preliminaries}, where the \Kahler moduli space and its Weil--Petersson measure were described.

\subsection{Moduli and \Kahler potential}

As recalled in Section~\ref{sec:preliminaries}, the complexified \Kahler moduli are
\begin{equation}
T_i = \tau_i + \mathrm{i}\,\phi_i 
   = \frac{1}{2}\int_{D_i} J\wedge J + \mathrm{i}\int_{D_i} C_4\, ,
\end{equation}
where $\tau_i$ are the divisor volumes and $\phi_i$ are the axions descending from the Ramond--Ramond four-form $C_4$.  
The Calabi--Yau volume and \Kahler potential are
\begin{align}
\mathcal{V} &= \frac{1}{6}\,\kappa_{ijk}\,t^i t^j t^k\, &
\mathcal{K} &= -2\log\mathcal{V}\, .
\end{align}
The metric on moduli space, $K^{ij} = \partial_{T_i}\partial_{\bar{T}_{j}}\mathcal{K}$, then leads to the kinetic term for the complex moduli,
\begin{equation}
\label{eq:L_kin_axions}
\mathcal{L}_{\rm kin}
= K^{ij}\,\partial_\mu T_i\,\partial^\mu \bar T_j
= K^{ij}\bigl(\partial_\mu \tau_i\,\partial^\mu \tau_j 
              + \partial_\mu \phi_i\,\partial^\mu \phi_j\bigr)\, .
\end{equation}
The axion sub-block of this metric therefore governs the kinetic mixing and canonical normalisation of the $\phi_i$.

The square root of the eigenvalues of $K$ are the periodicities of the axions (modulo $2\pi$), and therefore reflect the rough scaling of the axion decay constants $f_i$.
\begin{equation}
    f_i^2 \sim \frac{1}{\lambda_i}\,, \qquad    \lambda_i \propto (\tau_i)^2 - \mathcal{V}\,\kappa_{iik}t^k\, .
\end{equation}
In the large-volume regime, the typical scaling (for deformable divisors) is~\cite{Sheridan:2024vtt}
\begin{equation}
    f_i \sim \frac{M_{\rm Pl}}{\sqrt{\mathcal{V}}}\,,
\end{equation}
up to order-one factors determined by the \Kahler metric eigenvalues.

\subsection{Nonperturbative potential and axion masses}

Instanton and gaugino-condensation effects generate the nonperturbative axion potential
\begin{equation}
\label{eq:axion_potential}
V(\phi) = 
\sum_\alpha \Lambda_\alpha^4\,[1-\cos(q_\alpha^{~i}\phi_i)]\, ,
\quad
\Lambda_\alpha^4 \sim \mathrm{e}^{-2\pi\tau_\alpha}\, .
\end{equation}
Here the `$\alpha$' summation runs over deformable divisors (wrapping ED3 branes), and $q_\alpha^{~i}$ are the \emph{reduced} instanton charges. That is, $q = Q\cdot S$ where $Q$ agrees with the original integer-valued GLSM charge matrix, and $S$ is the matrix that canonicalises the kinetic term (i.e., it brings the \Kahler metric to the identity by a rotation and rescaling). The reduced charge matrix is lower triangular and therefore upon using a hierarchical approximation $\Lambda_\alpha \ll \Lambda_{\alpha+1}$, we have the following (for axions associated to deformable divisors)
\begin{equation}
m_i^2 \approx \frac{\Lambda_i^4}{f_i^2} 
          \approx \frac{M_{\rm Pl}^2}{\mathcal{V}}\,
                \mathrm{e}^{-2\pi\tau_i}\, .
\end{equation}

\subsection{The QCD axion}

A stack of D7-branes wrapping a prime toric divisor supports a gauge group, and the corresponding closed-string axion becomes the gauge-theory axion associated with that sector.\footnote{For concreteness, one may envision a stack of $n$ D7-branes wrapping a prime toric divisor, which in general supports a $\mathrm{U}(n)$ gauge group. If the stack coincides with an O7-plane---corresponding to local tadpole cancellation---the group is instead projected to $\mathrm{SO}(2n)$ or $\mathrm{Sp}(n)$ depending on the orientifold charge of the O7-plane. In this work we remain agnostic about the precise realisation, as it would require specifying the O7-plane configuration and global tadpole cancellation conditions, which lie beyond the present scope.} For a generic gauge theory, the mass scales as $m_a \sim \Lambda_c^2/f_a$, where $\Lambda_c$ is the confinement scale of the gauge group, and the gauge coupling satisfies $g^{2} \sim 1/\mathrm{vol}(D)$ at the compactification scale.

We model the QCD axion by replacing the corresponding instanton scale with the QCD scale inferred from chiral perturbation theory at zero temperature~\cite{GrillidiCortona:2015jxo} giving: 
\begin{equation}
    m_{\rm QCD} = 5.7 \,\mu\text{eV}\left(\frac{10^{12}\text{ GeV}}{f_a}\right)\,.
\end{equation}
We demand that $\mathrm{vol}(D)$ be close to 40, i.e., giving roughly the value of the QCD gauge coupling after running to high scales with the Standard Model particle content. The ``axion quality''~\cite{Demirtas:2021gsq} quantifies how closely the induced $\theta$ angle of QCD approaches zero---equivalently, how nearly the gauge-theory axion aligns with a mass eigenstate---and we impose a benchmark quality criterion of $10^{-10}$~\cite{Abel:2020pzs}.

\section{Normalising Flows}
\label{app:NFs} 

Normalising flows (NF) learn a complex distribution $p_{x}(\mathbf{x})$ by flowing a simple distribution (e.g., a Gaussian) $p_{z}(\mathbf{z})$ through a series of transformations.
Let $f:\mathbb{R}^{D}\rightarrow\mathbb{R}^{D}$ be an invertible and differentiable function such that $f(\mathbf{z})=\mathbf{x}$. Then one can express the distribution $p_{x}(\mathbf{x})$ in terms of the base distribution $p_{z}(\mathbf{z})$ and $f$ using the change of variables formula:
\begin{equation}\label{eq:COV}
    p_{x}(\mathbf{x}) = p_{z}(\mathbf{z}) \left| \det{J_{f}(\mathbf{z})} \right|^{-1},
\end{equation}
where $J_{f}$ denotes the Jacobian of $f$, from which we get the following log likelihood:
\begin{align}\label{eq:log_like}
    \log{p_{x}(\textbf{x})} &= \log{p_{z}(\textbf{z})} - \log{\left| \det{J_{f}(\mathbf{z})} \right|}. 
\end{align} 

Given two invertible and differentiable transformations, $f_{1}$ and $f_{2}$, their composition $f_{2}\circ f_{1}$ is also invertible and differentiable. Therefore, we can build complex transformations by composing multiple simpler transformations $f=f_{K}\circ\cdots\circ f_{1}$, where $f_{k}$ transforms $\mathbf{z}_{k-1}$ into $\mathbf{z}_{k}$, where $\mathbf{z}_{0}=\mathbf{z}$ and $\mathbf{z}_{K}=\mathbf{x}$. 

Fitting a flow-based model $p_{x}(\mathbf{x};\Theta)$ to a target distribution $p_{x}^{*}(\mathbf{x})$ can be done by minimising the Kullback-Leibler (KL) divergence between them, with respect to the model's parameters $\Theta$.
Here we have two options; we can take the KL divergence of $p_{x}(\mathbf{x};\Theta)$ with respect to $p_{x}^{*}(\mathbf{x})$ or vice versa. The former is called the reverse KL divergence, and the latter is called the forward KL divergence. 
By a change of variables we can write the reverse KL divergence loss as:
\begin{align}\label{eq:KL_rev}
\begin{split}
    \mathcal{L}(\Theta) &= D_{KL}[p_{x}(\mathbf{x};\Theta) || p_{x}^{*}(\mathbf{x})] \\
    &= \mathbb{E}_{p_{x}(\mathbf{x};\Theta)}[\log{p_{x}(\mathbf{x};\Theta)} - \log{p_{x}^{*}(\mathbf{x})}] \\
    &= \mathbb{E}_{p_{z}(\mathbf{z};\mathbf{\psi})}[\log{p_{z}(\mathbf{z})} - \log{|\det{J_{f}(\mathbf{z};\Theta)}|}\\
    &\qquad\qquad\qquad\qquad\qquad - \log{p_{x}^{*}(f(\mathbf{z};\Theta))}]
\end{split}
\end{align}
Similarly, we can write the forward KL divergence loss as
\begin{align}\label{eq:KL_forw}
\begin{split}
    \mathcal{L}(\Theta) &= D_{KL}[p_{x}^{*}(\mathbf{x}) || p_{x}(\mathbf{x};\Theta)] \\
    &= -\mathbb{E}_{p_{x}^{*}(\mathbf{x})}[\log{p_{x}(\mathbf{x};\Theta)}] + \text{const.} \\
    &= -\mathbb{E}_{p_{x}^{*}(\mathbf{x})}[\log{p_{z}(f^{-1}(\mathbf{x};\Theta))}\\
    &\qquad\qquad + \log{|\det{J_{f^{-1}}(\mathbf{x};\Theta)}|}] + \text{const.}
\end{split}
\end{align}
The forward KL divergence is often used when we can generate samples from the target distribution but we cannot evaluate the target density. On the other hand, the reverse KL divergence is used when we have the ability to evaluate the target density but we cannot easily sample from it. 

\subsection{Real NVP}

In our experiments, we use a Real NVP (real-valued non volume preserving) flow model which consists of a series of affine coupling layers. The idea here is to partition the latent-variable $\mathbf{z}$ into two parts $\mathbf{z}=(\mathbf{z}_{A},\mathbf{z}_{B})$, where $\mathbf{z}_{A}$ has dimension $d$ and $\mathbf{z}_{B}$ has dimension $D-d$. The output vector $f(\mathbf{z})=\mathbf{x}=(\mathbf{x}_{A},\mathbf{x}_{B})$ is then given as
\begin{equation}
     f(\mathbf{z}) =
    \begin{cases}
        \mathbf{x}_{A} = \mathbf{z}_{A} \\
        \mathbf{x}_{B} = \mathbf{z}_{B} \odot \exp{(S(\mathbf{z}_{A}))} +T(\mathbf{z}_{A})
    \end{cases},
\end{equation}
where $\odot$ is the element-wise product and $S$ and $T$ are called the scale and translation functions respectively, which are implemented using neural networks. 

The inverse function is given by
\begin{equation}
    f^{-1}(\mathbf{x}) =
    \begin{cases}
        \mathbf{z}_{A} = \mathbf{x}_{A} \\
        \mathbf{z}_{B} = (\mathbf{x}_{B}-T(\mathbf{x}_{A})) \oslash \exp{S(\mathbf{x}_{A})}
    \end{cases},
\end{equation}
where $\oslash$ denotes element-wise division, and the Jacobian $J_{f}$ is given by 
\begin{equation}
    J_{f} = \begin{pmatrix}
        \mathbf{I}_{d} & \mathbf{0}_{d\times(D-d)} \\ \frac{\partial\mathbf{x}_{B}}{\partial\mathbf{z}_{A}} & \text{diag}(\exp{(S(\mathbf{z}_{A}))})
    \end{pmatrix}
\end{equation}

Partitioning into sets $A$ and $B$ can be achieved using a binary mask function $B$. In our implementation, we employ an alternating binary mask $b$ where $b_{i}=1$ for even indices and $b_{i}=0$ for odd indices.

Using reverse KL divergence loss, the overall training procedure involves creating mini-batches of data points $\{\mathbf{z}_{i}\}$ by sampling the base distribution, flowing samples through the transformation layers to get $\{\mathbf{x}_{i}\}$  and computing~\cref{eq:KL_rev}. Similarly, using forward KL divergence loss, training involves sampling batches $\{\mathbf{x}_{i}\}$ from the target distribution and flowing samples through the inverse transformation layers to get $\{\mathbf{z}_{i}\}$ and computing~\cref{eq:KL_forw}. Gradients of the loss are evaluated using automatic differentiation and the network parameters are updated by stochastic gradient descent.

\section{Geometry of \Kahler Moduli of Calabi--Yau Threefolds}\label{app:kahler_metric}

In this appendix we summarise the classical description of the \Kahler moduli of a Calabi--Yau threefold.  
At tree level, the \Kahler class determines the volumes of holomorphic cycles and provides the natural expansion parameters of the low-energy effective theory. Beyond tree level, these geometric quantities receive perturbative and non-perturbative corrections which play an essential role in realistic moduli stabilisation scenarios; see \cite{McAllister:2024lnt} for a recent overview.

\subsection{Ricci-flat deformations and \Kahler moduli}

Ricci-flat metrics on Calabi--Yau threefolds occur in continuous families: for a fixed topology, the space of solutions to $R_{mn}=0$ is parametrised by a \emph{moduli space} of deformations.
Let $g_{\alpha\bar{\beta}}$ be a Hermitian Ricci-flat metric on $X$, expressed in complex coordinates $z^\alpha$ and $\bar z^{\bar\alpha}$ so that
\begin{equation}
    g_{\alpha \beta}=g_{\bar\alpha\bar\beta}=0, \qquad R_{\alpha \bar\beta}(g)=0\,.
\end{equation}
Consider an infinitesimal deformation $g\to g+\delta g$ preserving Hermiticity,
\begin{equation}
    (g+\delta g)_{\alpha \beta}=(g+\delta g)_{\bar\alpha\bar\beta}=0\,,
\end{equation}
and keeping the metric Ricci-flat to first order,
\begin{equation}
    R_{\alpha \bar\beta}(g+\delta g)=0\,.
\end{equation}
In the de Donder gauge,
\begin{equation}
    \nabla^m\delta g_{mn}-\frac12\nabla_n \delta g^m{}_{m}=0\,,
\end{equation}
this condition becomes the Lichnerowicz equation
\begin{equation}
     \nabla^m\nabla_m \delta g_{pq}+2R_{p}{}^{r}{}_{q}{}^{s}\,\delta g_{rs}=0\,.
\end{equation}
The fluctuation decomposes as
\begin{equation}
    \delta g = \delta g_{\alpha \bar\beta}\,{\rm d}z^\alpha {\rm d}\bar z^{\bar\beta} + \delta g_{\alpha \beta}\,{\rm d}z^\alpha{\rm d}z^{\beta} + \text{c.c.}
\end{equation}
Since on a Calabi--Yau the only non-vanishing Riemann components are $R_{\gamma\bar\alpha \delta\bar\beta}$, the $(\alpha\bar\beta)$ and $(\alpha\beta)$ sectors of the Lichnerowicz operator decouple. Thus metric deformations fall into two independent classes.

In this appendix we focus exclusively on the mixed variations $\delta g_{\alpha \bar\beta}$, which correspond to deformations of the \Kahler class.
Such a deformation defines a harmonic $(1,1)$-form,
\begin{equation}
    \delta J \coloneqq \delta g_{\alpha \bar\beta}\,{\rm d}z^\alpha\wedge{\rm d}\bar z^{\bar\beta}\,.
\end{equation}
By Hodge theory, harmonic $(1,1)$-forms represent cohomology classes in $H^{1,1}(X)$, whose dimension is the Hodge number $h^{1,1}$. Hence these deformations correspond precisely to the \emph{\Kahler moduli}.

\subsection{Weil--Petersson metric on \Kahler moduli space}

It is a general principle that if a theory features scalar fields $\varphi^i$, $1 \leq i \leq N$, valued in an $N$-dimensional target space, that the metric on that target space is the kinetic matrix $K_{ij}$ featuring in the Lagrangian as $K_{ij} \partial^\mu \varphi^i \partial_\mu \varphi^j$. In the special case these these scalar fields take values in a mathematical moduli space, including (but by no means limited to) either the \Kahler or complex structure moduli space of Ricci-flat metrics on a Calabi--Yau threefold, this is historically known as the Weil--Petersson metric. 

Thus, as our focus in this work is on \Kahler moduli space, the natural metric is the Weil--Petersson metric on that moduli space. In particular, by studying the variations of the Ricci-flat metric induced by deformations of the \Kahler class, we obtain the kinetic matrix of the \Kahler moduli and thus the Weil--Petersson metric on \Kahler moduli space. For the metric deformations $\delta g_{\alpha \bar{\beta}}$, this leads to
\begin{equation}\label{eq:WPMdefKM}
	2G_{ij}\, \delta t^{i}\delta t^{j} = \dfrac{1}{2\mathcal{V}}\int_{X}\, g^{\alpha\bar{\beta}}g^{\gamma\bar{\delta}}\delta g_{\alpha\bar{\delta}}\delta g_{\bar{\beta}\gamma}\, \sqrt{g}\, \mathrm{d}^{6}y\, ,
\end{equation}
We choose a basis $\omega_i\in H^{1,1}(X,\mathbb{R})$ so that
\begin{equation}
    \delta g_{\alpha \bar{\beta}} = (\omega_i)_{\alpha\bar{\beta}}\, \delta t^{i}
\end{equation}
which implies that \eqref{eq:WPMdefKM} can be written as
\begin{equation}\label{eq:WPKMS}
	G_{ij} = \dfrac{1}{4\mathcal{V}}\int_{X}\, \omega_i\wedge\star_6\omega_j\, ,
\end{equation}
which agrees with \cite{Strominger:1985ks}, see also \cite{Candelas:1990pi,Grimm:2004uq}.
One can then show that the action of the Hodge-star $\star_6$ on $\omega_j\in H^{1,1}(X,\mathbb{R})$ is given by \cite{Candelas:1990pi}
\begin{equation}
    \star_6\omega_j = -J\wedge \omega_j +\dfrac{\int_{X}\, J\wedge J\wedge \omega_j}{4\mathcal{V}}\, J\wedge J\, .
\end{equation}
Let us introduce the quantities
\begin{equation}\label{eq:Kahlercoordinatestree}
    \tau_j = \dfrac{1}{2}\int_{X}\, J\wedge J\wedge \omega_j = \dfrac{1}{2}\kappa_{jkl}t^kt^l\, ,
\end{equation}
corresponding to the (classical) string frame divisor volumes.
We can then write \eqref{eq:WPKMS} as \cite{Candelas:1990pi}
\begin{align}\label{eq:WPKMSfinal}
    G_{ij} &= \dfrac{1}{4\mathcal{V}}\biggl [-\int_{X}\, J\wedge\omega_i\wedge \omega_j+\dfrac{\tau_i}{2\mathcal{V}}\, \int_{X}\, \omega_i\wedge J\wedge J\biggl ]\nonumber\\
    &=\dfrac{1}{4\mathcal{V}}\biggl [-\kappa_{ijk}t^k+\dfrac{\tau_i\,\tau_j}{\mathcal{V}}\biggl ] \, .
\end{align}
The Weil--Petersson metric $G_{ij}$ on the \Kahler moduli space thus obtained is entirely determined by the intersection data of $X$, a choice of \Kahler parameters $t^i$ and the classical divisor volumes $\tau_{i}$. This classical form of $G_{ij}$ will provide the reference point for incorporating perturbative and non-perturbative corrections in the next sections.

Having expressed the moduli space metric entirely in terms of the classical geometric data $(t^{i}, \kappa_{ijk})$, we now identify the corresponding holomorphic coordinates appropriate for the four-dimensional $\mathcal{N}=1$ theory.  
This requires promoting the real \Kahler parameters to complex fields by incorporating the R-R axions.  
We define the complex coordinate
\begin{equation}\label{eq:Ttree}
    T_i \coloneqq \frac{1}{2}\int_{X}\, J \wedge J\wedge\omega_i + \mathrm{i}\int_{X}\, C_4\wedge\omega_i \equiv \tau_i + \mathrm{i} \theta_i
\end{equation}
where the classical string frame divisor volumes $\tau_i$ are given by \eqref{eq:Kahlercoordinatestree}. The $T_i$ defined by \eqref{eq:Ttree} are good \Kahler coordinates on the \Kahler moduli space, at leading order in $g_s$ and $\alpha'$. The corresponding \Kahler potential is  
\begin{equation}\label{eq:ktreeex}
    \mathcal{K} = - 2\,\mathrm{log}\bigl(\mathcal{V} \bigr)\,, \; \mathcal{V} \coloneqq \dfrac{1}{6}\int_X  \, J^3= \frac{1}{6}\kappa_{ijk}t^it^jt^k\, .
\end{equation}
Let us show that the tree-level \Kahler potential \eqref{eq:ktreeex} leads to the appropriate field space metric $\mathcal{K}_{i\bar{\jmath}}$ which is directly related to the Weil--Petersson metric $G_{ij}$ in \eqref{eq:WPKMSfinal} and corresponds a \Kahler metric defined as $\mathcal{K}_{i\bar{\jmath}}=\partial_{T_i}\partial_{\overline{T}_{\bar{\jmath}}}\mathcal{K}$. Noting that at the level of the tree level theory
\begin{equation}
    \dfrac{\partial T_i}{\partial t^j} = \kappa_{ijk}t^k =\kappa_{ij}\; , \quad \partial_{t^i}\partial_{t^j}\mathcal{K} = 8G_{ij}
\end{equation}
we obtain the expression
\begin{align}
    \partial_{T_i}\partial_{\overline{T}^{\bar{\jmath}}}\mathcal{K} &= \dfrac{1}{8\mathcal{V}^{2}}\biggl [-2 \mathcal{V} \kappa^{ij}+ t^it^j \biggl ] \, ,
\end{align}
where $\kappa^{ij}$ is the inverse of $\kappa_{ij}$.

\subsection{Detailed Geometric Prerequisites}
\label{app:FibrationsDecayConstants}

In this appendix, we take some time to more thoroughly introduce the geometric ideas summarised in \cref{sec:geom_preq}. In particular, we will discuss fibrations, shrinkable divisors, and geometric aspects of \Kahler metric eigenvalues and axion decay constants.

To organise the former two phenomena, it is useful to recall the classification of facets (codimension-one faces) of \Kahler cones of CY threefolds, based on the four types of contractions they give rise to~\cite{Wilson1992,hayakawa1995degenarationcalabiyaumanifoldwp,Morrison:1996pp,Chiang:1995hi,Cordova:2009fg} 
\begin{enumerate}
    \item The CY threefold shrinks to a variety of smaller dimension (in particular, the CY is fibered, and both the fiber and CY shrink to zero volume),
    \item A divisor shrinks to a curve,
    \item A divisor shrinks to a point,
    \item A curve shrinks to a point.
\end{enumerate}
Facets in the first three categories correspond to those which bound the geometric moduli space of the CY, while the final category corresponds to a flop transition, in which the adjoining \Kahler cone belongs to a birational Calabi--Yau threefold obtained by flopping a curve. Facets where the overall volume or individual divisor volumes vanish have been studied closely \cite{Gendler:2022ztv}, but the same behaviour can occur along higher-codimension faces of the \Kahler cone, so we will not restrict to the codimension-one case.

\subsubsection{Fibrations}

Fibered CY threefolds play a crucial role in string dualities \cite{Aspinwall:1995vk, Curio:1998bva}, furnish canonical examples of infinite distance limits in moduli spaces studied by the Swampland program \cite{Corvilain:2018lgw, Lee:2019wij}, and have historically played an important role in string phenomenology: e.g., $K3$ fibrations in the Large Volume Scenario (LVS) \cite{Cicoli:2008va, Cicoli:2008gp, Cicoli:2011it}. We mentioned in \cref{sec:geom_preq} that because WP measure is normalised by fixing minimum and maximum values of the overall Calabi--Yau volume, compactness of the support of the WP measure is related to faces where $\mathcal{V}$ vanishes. Because $\mathcal{V}$ is a monotonic under overall rescaling of the \Kahler parameters, along any ray where $\mathcal{V}$ doesn't vanish identically, the region satisfying $\mathcal{V}_\text{min} \leq \mathcal{V} \leq \mathcal{V}_\text{max}$ is bounded. As a homogeneous polynomial in the \Kahler parameters, for any given ray, $\mathcal{V}$ either vanishes only at the origin or vanishes identically, with the latter only possible on faces of the \Kahler cone. The region $\mathcal{V}_\text{min} \leq \mathcal{V} \leq \mathcal{V}_\text{max}$ supporting the WP measure near such faces extend to infinity, as the bounding level sets of $\mathcal{V}$ becomes asymptotically parallel to the faces. Compactness of the WP support, then, is determined by whether there are \Kahler cone faces where $\mathcal{V}$ vanishes. We now explain why faces where $\mathcal{V}$ vanishes correspond to with fibrations of $X$.

If $\mathcal{V} = 0$ for some choice of \Kahler form $J$, then $J$ lies on a boundary of the \Kahler cone, corresponding to a \Poincare-dual nef divisor $D$ whose triple self-intersection vanishes, $D^3 = 0$.\footnote{A nef divisor $D$ satisfies $D\cdot C \geq 0$ for all effective curves $C$.} A nef divisor $D$ is ``big'' if and only if $D^3 > 0$ \cite{lazarsfeld2017}. It is a standard fact in algebraic geometry that ample divisors $A$ define embeddings of $X$ into projective space by mapping $x \in X$ to $[s_0(x) : \dots : s_n(x)]$ for $s_i$ the global sections of $mA$ for $m$ a sufficiently large integer (indeed, this is often taken as the definition of an ample divisor) \cite{lazarsfeld2017}. Intuitively, relevance of the ``big'' condition is that the multiples of a big divisor $D$ carry enough global sections to map $X$ to a three-dimensional image in projective space --- without ampleness, this may not be an honest embedding of $X$, but the big property ensure that the powers of $D$ probe all directions of the threefold. When a nef $D$ is not big (i.e., $D^3 = 0$), its sections fail to vary along some directions, collapsing those dimensions to form a lower-dimensional base $B$. Consequently, $D^3 = 0$ signals a fibration rather than an embedding (sometimes known as a semi-ample or Iitaka fibration \cite{lazarsfeld2017}). The process of approaching the \Kahler form $J$ for which $\mathcal{V}$ vanishes implements the limit where the generic fiber of this fibration shrinks (while the base volume is free to take on any value). 

If $\mathcal{V}$ vanishes on a $k$-dimensional face of the \Kahler cone, then the integral \Kahler forms $J$ on this face are \Poincare dual to divisors $D$ which are pullbacks $f^* D'$ of nef divisors $D'$ on $B$, which satisfies $h^{1,1}(B) = k$.\footnote{For the case of elliptic fibrations, these divisors are known in the F-theory literature as ``vertical divisors'' \cite{Weigand:2018rez}.} That is, the $D'$ are curves if $B$ is a surface, or isolated points if $B$ is a curve. Because these divisors are pullbacks from the base, each $D = f^*D'$ contains the entire fiber above the corresponding divisor $D' \subset B$. Consequently, the intersection of $\ell$ such divisors in $X$ corresponds to the union of fibers lying above the intersection of the $\ell$ divisors $D'_i$ on the base. This explains why $D^3 = 0$: the base is at most two-dimensional, and an intersection of $\ell = 3$ nef divisors on a space of dimension at most two, is empty.

We reviewed the classification of fibrations of Calabi--Yau threefolds in \cref{sec:geom_preq}: the fiber can have complex dimension one, and be an elliptic fibration, or complex dimension two, and be a $K3$ fibration, where here we continue to follow the convention set in \cref{sec:geom_preq} where we ignore distinctions between genus-one and elliptic fibrations and also between $K3$ and $T4$ fibrations for notational simplicity. We can diagnose the fiber dimension (and hence the type of fibration) directly from intersections of pullbacks from the base. Recall our logic from the previous paragraph for why $D^3 = 0$: if $B$ is a surface ($\dim_{\mathbb{C}}B=2$, the elliptic case), the intersection of two base divisors is a finite set of points, so intersecting two pullback divisors in $X$ yields a union of fibers (a curve class). If $B$ is a curve ($\dim_{\mathbb{C}}B=1$, the $K3$ case), two distinct base divisors do not meet, so the square of a pullback class vanishes. We conclude that a fibration is $K3$ if $D^2 = 0$ and elliptic otherwise. Indeed, we have recovered the famous Koll\'ar condition for the existence of an elliptic fibration on a CY threefold \cite{kollar2015deformations}: namely, the existence of a so-called Koll\'ar divisor, or a nef divisor satisfying $D^3 = 0$ but $D^2 \neq 0$. This discussion has an additional corollary: for an elliptic fibration, the fiber class is proportional to $f^*D'\!\cdot f^*D''$ for any two base divisors $D',D''$ falling on the \Kahler cone face. For a $K3$ fibration, the fiber class is proportional to $f^*D'$ itself. Two fibrations are equivalent if and only if the classes of their fibers coincide. Computing the class of the base is much more difficult: this amounts to constructing a section of the fibration, which is a challenging problem. It is worth noting that the base of an elliptic fibration has unconstrained $h^{1,1}$, while a $K3$ fibration must have base $\mathbb{P}^1$. So, elliptic fibrations are free to correspond to \Kahler cone faces of any dimension while $K3$ fibrations can only correspond to one-faces. In this work, $h^{1,1}$ is sufficiently small that we can compute extremal faces of $K_\cup$, our approximation to the \Kahler cone, and test for fibrations in the above fashion, but we comment in passing that there additionally exist useful methods for computing the fibrations of toric hypersurface Calabi--Yau varieties inherited from fibrations of the ambient toric variety \cite{Kreuzer:2000qv}.

We conclude our discussion of fibrations with a more physical comment on Kaluza-Klein scales. In practice, we will bound our overall volume (in order to bound the WP measure) by the physical requirement that the overall KK scale $\sim g_s M_{\mathrm{Pl}} / \mathcal{V}^{2/3}$ is sufficiently large. In the limit of small fiber, however, the KK scale of the base $B$ can become dangerously small even when the KK scale of the Calabi--Yau is phenomenologically viable. In particular, imposing our stretching condition that the fiber must have volume $\geq 1$, the base KK scale for unit fiber volume is $g_s M_{\mathrm{Pl}} / \mathcal{V}^{1/2 + 1/\mathrm{dim} \, B}$, which is smaller by a factor of $\mathcal{V}^{-1/3}$ for elliptic fibrations and $\mathcal{V}^{-1/12}$ for K3 fibrations. Because it is generally difficult to compute the class of the base of a fibration, we defer the problem of incorporating relevant divisor and curve KK scales to future work, but acknowledge that constraints on these scales should in general be imposed.

\subsubsection{Shrinking Divisors}

Now we turn our attention to divisors that can shrink to zero volume at faces of the \Kahler cone. Divisor volumes are quadratic forms in the \Kahler parameters and have signature $(+, -, \dots, -)$ (this follows from the Hodge index theorem, at least for smooth divisors). Thus, a (possibly complex) basis of \Kahler parameters can always be chosen such that their volume is $\tau = (t^1)^2 - \sum_{i=2}^k (t^i)^2$ for $k$ an invariant of the divisor that counts the dimension of the subspace of $H^{1,1}(D)$ inherited from the ambient Calabi--Yau (equal to the number of independent ambient two-forms restricting non-trivially to $D$). 

Let us first consider divisors that shrink at facets of the \Kahler cone: i.e. codimension-one faces. If the divisor shrinks to a point along a facet, then its volume vanishes quadratically at that facet, so $k = 1$ and $\tau = t_1^2$. Likewise, if it shrinks to a curve along a facet, then its volume vanishes linearly, so $\tau = fg$ for $f,g$ two linear functions of the \Kahler parameters which we can formally write as $\tau = ([f + ig]/2\mathrm{e}^{i\pi/4})^2 - ([f - ig]/2\mathrm{e}^{i\pi/4})^2 = (t^1)^2 - (t^2)^2$, so $k = 2$. Divisors which can shrink at finite overall volume have been referred to as blowup divisors --- indeed, they are the exceptional divisors achieved by blowing up a point or curve --- while divisors whose volumes are a perfect square (i.e., $k = 1$) have been called diagonal divisors \cite{Cicoli:2011it}. We note that satisfying $k = 1$ ($k = 2$) is necessary but not sufficient for a divisor to shrink to a point (curve) along some facet: sometimes the locus where the shrinking would take place is strictly outside of the \Kahler cone.

Because all boundaries of the CY \Kahler moduli space must either correspond to fibrations or shrinking divisors, divisors that shrink at facets are actually relatively common. However, for the purposes of axion physics, it isn't strictly necessary that the shrinking happens at codimension one. The volume polynomials for divisors which shrink at higher codimension faces do not need to factor: it suffices for them to factor only in the limit where one approaches the relevant face while remaining in the \Kahler cone. We do not endeavour to provide a comprehensive classification of divisors shrinking at higher codimension faces: however, we present a numerical approach for computing which divisors shrink along each face, and we can also make the following remark. Recall from \cref{sec:geometry_results} that a divisor $D$ is geometrically rigid (i.e., has no normal bundle deformations) if $h^{2,0}(D)$ (which agrees with $h^0(X, \mathcal{O}_X(D_i)) - 1$ for Calabi--Yau threefolds $X$ \cite{Braun:2017nhi}) equals zero, and movable otherwise. If $D$ is movable, then the divisors in its class sweep out a dense subset of the Calabi--Yau threefold. Such divisors cannot shrink without the entire CY shrinking: that is, they cannot shrink independently from the overall volume. Geometrically rigid divisors, on the other hand, are only represented by a single holomorphic divisor. The geometric rigidity property is a necessary but insufficient condition for a divisor to be able to shrink independent of the CY. 

An important class of shrinking divisors actually arise from fibrations. If $F$ is a face of the \Kahler cone corresponding to a fibration, then any divisor $D$ which lies on the linear subspace generated by that face will have its volume vanish at least linearly along that face. Intuitively, these divisors are the pullback divisors from the base of the fibration, so they are themselves fibered, so their volume also shrinks to zero as the fiber shrinks. These correspond to the ``fibration divisor'' category defined in \cref{sec:geom_preq}.

We now take some time to describe how to identify and taxonomise divisors which shrink along a given face $F$ of the \Kahler cone generated by \Kahler forms $J_1, \dots, J_n$. A divisor $D$ with volume $\tau$ vanishes along $F$ if $\kappa_{ijk} J^i J^j D^k = 0$ for $J$ any linear combination of $J_1, \dots, J_n$. One can form a matrix $K$ whose columns are the $J_i$: then we require that $\kappa_{ijk} K^{ia} K^{jb} D^k = 0$ identically, for all $a, b$. If we flatten $a,b$ indices of the tensor $\kappa_{ijk} K^{ia} K^{jb}$ into a single index $c$, leaving a matrix $L_i^c$, then we merely have the condition that $D$ lies in the kernel of $L$. We can refine this approach to identify the divisors which contract to points, or have a volume that approaches zero quadratically near $F$. In particular, we now require that $D$ lie in the kernel of $M^c_k$ given as $\kappa_{ijk} K^{ka}$ with the $a,j$ indices flattened into $c$. 

With this in hand, we elaborate on how to assign one of the five categories named in \cref{sec:geom_preq} to a divisor $D$. If $D$ shrinks to a point along some face $F$ that doesn't correspond to a fibration (i.e., $\mathcal{V}$ doesn't vanish identically on $F$) then it is a point blowup divisor. Otherwise, if it shrinks at all along some non-fibration face $F$, it is a curve blowup divisor. Otherwise, if $D$ is proportional to the class of a $K3$ fiber, it will shrink in the limit of shrinking fiber and we say it is a $K3$ fiber divisor. Otherwise, if $D$ shrinks along a face $F$ that does correspond to a fibration, we say $D$ is a fibration divisor. Otherwise, $D$ does not shrink and is a non-shrinkable divisor.

The challenge here is knowing the faces of the \Kahler cone: we take an inner approximation to the \Kahler cone given by $K_\cup$, and computing all of the faces of this approximation is computationally hard for $h^{1,1} \gtrsim 10$. 
  
\subsubsection{\Kahler metric and decay constants}
\label{sec:Kahler_metric_decay_constants}

Consider the inverse \Kahler metric
\begin{equation}
\label{eq:inverse_Kmetric}
    [K^{-1}]_{ij} = 2(\tau_i \tau_j - \mathcal{V} \kappa_{ijk} t^k).
\end{equation}
This furnishes a metric and thus a norm $\| \cdot \|_{K}$ on divisors $D$.\footnote{This is a mildly subtle point: the \Kahler metric, as defined, acts on divisor \textit{volumes} $\tau_i$ (and more generally the \Kahler moduli $T_i$) as well as curve classes $C_i$, as these both belong to $H_2(X, \mathbb{R})$, while the inverse \Kahler metric acts on curve volumes (parametrised by \Kahler parameters $t^i$) and divisor \textit{classes} $D^i$, members of $H^2(X, \mathbb{R}) \cong H_4(X, \mathbb{R})$.} In particular,
\begin{equation}
    \| D \|^2_{K} \propto \tau^2 - \mathcal{V}C\,,
\label{eq:inv_kahler}
\end{equation}
for $\tau$ the volume of $D$, and $C$ the formal volume of the curve class of the self-intersection $D \cap D$. But, this class may have no holomorphic representatives: i.e., it may not be an effective curve class. In the limit of hierarchical instanton scales, one can consider the axion decay constant $f$ associated to a divisor $D$ belonging to the basis of leading instantons, and compute it perturbatively using \cite{Gendler:2023kjt}. This decay constant satisfies
\begin{equation}
    f \geq \frac{1}{2 \pi \| D \|_{K}}.
\end{equation}
Briefly, this is because $1/2 \pi f$ is a component of the divisor $D$ expressed in a basis where $K^{-1} = 1$ (i.e., the kinetic matrix is canonical), so that $1/(2\pi f)^2 + \text{positive} = \|D\|_K^2$. In this way, $\|D\|_K$ provides an important probe of axion decay constants.

There are two qualitatively different cases to consider. First, the case that $D$ is geometrically rigid. In this scenario there is no effective curve arising from self-intersection of representatives of $D$, because $D$ has a unique representative, so the intersection of two representatives is a divisor, not a curve. Because $D^2$ isn't an effective curve class, it is nearly always negative (as a numerical coincidence, it can sometimes be positive for certain choices of \Kahler forms, but it is empirically almost always negative). In this case the \Kahler metric is manifestly positive --- the sum of two positive numbers --- and either term may dominate. Second, consider the case that $D$ is movable: then the self-intersection curve associated to $C$ is effective, and in particular $C \geq 0$. Combining this with the constraint that $\|D\|_K$ be positive, we find $0 < \mathcal{V} C < (\tau_i)^2$, which is quite restrictive. Empirically, we find that $(\tau_i)^2 / \mathcal{V} C$ never comes too close to unity, so in the limit of large volume $(\tau_i)^2 - \mathcal{V} C$ tends to be large.

Let us consider the consequences of this discussion for axion physics. Firstly if some axions correspond to divisors $D$ with normal bundle deformations, then the associated decay constants will generally tend to be larger and have less spread than those of geometrically rigid divisors. This is because the $\|D\|_K$ are smaller --- a difference, rather than a sum --- and bounded from above and below. In the large-volume limit, the overall Calabi--Yau volume still sets the common scale of the decay constants, though the mechanism differs between the two classes of divisors. Geometrically rigid divisors can shrink without significantly affecting the total volume --- examples include divisors that contract to points or curves at finite $\mathcal{V}$. In this case, the two contributions in \cref{eq:inv_kahler} to $\|D\|_K$ are somewhat decoupled, and the $\mathcal{V}$ term dominates in the large volume regime, as was noted in \cite{Gendler:2023kjt} and exploited in \cite{Cheng:2025ggf}. By contrast, deformable divisors sweep out a dense subset of the Calabi--Yau and cannot shrink independently of the overall volume $\mathcal{V}$, implying a strong correlation between their volume and $\mathcal{V}$; in this sense, $\mathcal{V}$ acts as an overall factor in $\|D\|_K$.

We conclude by commenting briefly on the case where $D$ is the fiber of a $K3$ fibration. In this case, from the adjunction formula it follows that $D \cap D$ is the canonical class on $D$, which must vanish for $K3$, so $C = 0$. From this it follows that $\|D\|$ --- and thus, a lower bound on $f$ --- depends only on the volume of the $K3$ fiber.

\medskip

\bibliographystyle{JHEP}
\bibliography{refs}

@article{McAllister:2023vgy,
    author = "McAllister, Liam and Quevedo, Fernando",
    title = "{Moduli Stabilization in String Theory}",
    eprint = "2310.20559",
    archivePrefix = "arXiv",
    primaryClass = "hep-th",
    month = "10",
    year = "2023"
}

@article{Berg:2004ek,
    author = "Berg, Marcus and Haack, Michael and Kors, Boris",
    title = "{Loop corrections to volume moduli and inflation in string theory}",
    eprint = "hep-th/0404087",
    archivePrefix = "arXiv",
    reportNumber = "MIT-CTP-3486, NSF-KITP-04-41",
    doi = "10.1103/PhysRevD.71.026005",
    journal = "Phys. Rev. D",
    volume = "71",
    pages = "026005",
    year = "2005"
}

@article{Berg:2005ja,
    author = "Berg, Marcus and Haack, Michael and Kors, Boris",
    title = "{String loop corrections to Kahler potentials in orientifolds}",
    eprint = "hep-th/0508043",
    archivePrefix = "arXiv",
    reportNumber = "MIT-CTP-3671, NSF-KITP-2005-55",
    doi = "10.1088/1126-6708/2005/11/030",
    journal = "JHEP",
    volume = "11",
    pages = "030",
    year = "2005"
}

@article{Giddings:2001yu,
    author = "Giddings, Steven B. and Kachru, Shamit and Polchinski, Joseph",
    title = "{Hierarchies from fluxes in string compactifications}",
    eprint = "hep-th/0105097",
    archivePrefix = "arXiv",
    reportNumber = "SLAC-PUB-8807, NSF-ITP-01-37, SU-ITP-01-16",
    doi = "10.1103/PhysRevD.66.106006",
    journal = "Phys. Rev. D",
    volume = "66",
    pages = "106006",
    year = "2002"
}

@article{Cicoli:2021tzt,
    author = "Cicoli, Michele and Schachner, Andreas and Shukla, Pramod",
    title = "{Systematics of type IIB moduli stabilisation with odd axions}",
    eprint = "2109.14624",
    archivePrefix = "arXiv",
    primaryClass = "hep-th",
    doi = "10.1007/JHEP04(2022)003",
    journal = "JHEP",
    volume = "04",
    pages = "003",
    year = "2022"
}

@article{Grimm:2007xm,
    author = "Grimm, Thomas W.",
    title = "{Non-Perturbative Corrections and Modularity in N=1 Type IIB Compactifications}",
    eprint = "0705.3253",
    archivePrefix = "arXiv",
    primaryClass = "hep-th",
    reportNumber = "MAD-TH-07-07",
    doi = "10.1088/1126-6708/2007/10/004",
    journal = "JHEP",
    volume = "10",
    pages = "004",
    year = "2007"
}

@article{Grimm:2004uq,
    author = "Grimm, Thomas W. and Louis, Jan",
    title = "{The Effective action of N = 1 Calabi-Yau orientifolds}",
    eprint = "hep-th/0403067",
    archivePrefix = "arXiv",
    reportNumber = "LPTENS-04-14",
    doi = "10.1016/j.nuclphysb.2004.08.005",
    journal = "Nucl. Phys. B",
    volume = "699",
    pages = "387--426",
    year = "2004"
}

@article{Rogers:2023upm,
    author = "Rogers, Keir K. and Poulin, Vivian",
    title = "{5\ensuremath{\sigma} tension between Planck cosmic microwave background and eBOSS Lyman-alpha forest and constraints on physics beyond \ensuremath{\Lambda}CDM}",
    eprint = "2311.16377",
    archivePrefix = "arXiv",
    primaryClass = "astro-ph.CO",
    doi = "10.1103/PhysRevResearch.7.L012018",
    journal = "Phys. Rev. Res.",
    volume = "7",
    number = "1",
    pages = "L012018",
    year = "2025"
}

@article{Sheridan:2024vtt,
    author = "Sheridan, Elijah and Carta, Federico and Gendler, Naomi and Jain, Mudit and Marsh, David J. E. and McAllister, Liam and Righi, Nicole and Rogers, Keir K. and Schachner, Andreas",
    title = "{Fuzzy axions and associated relics}",
    eprint = "2412.12012",
    archivePrefix = "arXiv",
    primaryClass = "hep-th",
    reportNumber = "KCL-PH-TH/2024-75, KCL-PH-TH/2024-75",
    doi = "10.1007/JHEP09(2025)016",
    journal = "JHEP",
    volume = "09",
    pages = "016",
    year = "2025"
}

@article{Fallon:2025lvn,
    author = "Fallon, Sebastian Vander Ploeg and Halverson, James and McAllister, Liam and Zhu, Yunhao",
    title = "{F-theory Axiverse}",
    eprint = "2511.20458",
    archivePrefix = "arXiv",
    primaryClass = "hep-th",
    month = "11",
    year = "2025"
}

@article{Halverson:2019cmy,
    author = "Halverson, James and Long, Cody and Nelson, Brent and Salinas, Gustavo",
    title = "{Towards string theory expectations for photon couplings to axionlike particles}",
    eprint = "1909.05257",
    archivePrefix = "arXiv",
    primaryClass = "hep-th",
    doi = "10.1103/PhysRevD.100.106010",
    journal = "Phys. Rev. D",
    volume = "100",
    number = "10",
    pages = "106010",
    year = "2019"
}

@article{OHare:2024nmr,
    author = "O'Hare, Ciaran A. J.",
    title = "{Cosmology of axion dark matter}",
    eprint = "2403.17697",
    archivePrefix = "arXiv",
    primaryClass = "hep-ph",
    doi = "10.22323/1.454.0040",
    journal = "PoS",
    volume = "COSMICWISPers",
    pages = "040",
    year = "2024"
}

@article{Chadha-Day:2021szb,
    author = "Chadha-Day, Francesca and Ellis, John and Marsh, David J. E.",
    title = "{Axion dark matter: What is it and why now?}",
    eprint = "2105.01406",
    archivePrefix = "arXiv",
    primaryClass = "hep-ph",
    reportNumber = "KCL-PH-TH/2021-20, CERN-TH-2021-045, IPPP/20/91",
    doi = "10.1126/sciadv.abj3618",
    journal = "Sci. Adv.",
    volume = "8",
    number = "8",
    pages = "abj3618",
    year = "2022"
}

@inproceedings{Adams:2022pbo,
    author = "Adams, C. B. and others",
    title = "{Axion Dark Matter}",
    booktitle = "{Snowmass 2021}",
    eprint = "2203.14923",
    archivePrefix = "arXiv",
    primaryClass = "hep-ex",
    reportNumber = "FERMILAB-CONF-22-996-PPD-T",
    month = "3",
    year = "2022"
}

@article{Mehta:2021pwf,
    author = "Mehta, Viraf M. and Demirtas, Mehmet and Long, Cody and Marsh, David J. E. and McAllister, Liam and Stott, Matthew J.",
    title = "{Superradiance in string theory}",
    eprint = "2103.06812",
    archivePrefix = "arXiv",
    primaryClass = "hep-th",
    doi = "10.1088/1475-7516/2021/07/033",
    journal = "JCAP",
    volume = "07",
    pages = "033",
    year = "2021"
}

@article{Gendler:2024adn,
    author = "Gendler, Naomi and Marsh, David J. E.",
    title = "{Possible Implications of QCD Axion Dark Matter Constraints from Helioscopes and Haloscopes for the String Theory Landscape}",
    eprint = "2407.07143",
    archivePrefix = "arXiv",
    primaryClass = "hep-th",
    doi = "10.1103/PhysRevLett.134.081602",
    journal = "Phys. Rev. Lett.",
    volume = "134",
    number = "8",
    pages = "081602",
    year = "2025"
}

@article{Gendler:2023kjt,
    author = "Gendler, Naomi and Marsh, David J. E. and McAllister, Liam and Moritz, Jakob",
    title = "{Glimmers from the axiverse}",
    eprint = "2309.13145",
    archivePrefix = "arXiv",
    primaryClass = "hep-th",
    reportNumber = "KCL-PH-TH/2023-49",
    doi = "10.1088/1475-7516/2024/09/071",
    journal = "JCAP",
    volume = "09",
    pages = "071",
    year = "2024"
}

@article{MacFadden:2024him,
    author = "MacFadden, Nate and Schachner, Andreas and Sheridan, Elijah",
    title = "{The DNA of Calabi-Yau Hypersurfaces}",
    eprint = "2405.08871",
    archivePrefix = "arXiv",
    primaryClass = "hep-th",
    reportNumber = "{LMU-ASC 06/24}",
    month = "5",
    year = "2024"
}

@article{Cheng:2025ggf,
    author = "Cheng, Junyi and Gendler, Naomi",
    title = "{Universality in the Axiverse}",
    eprint = "2507.12516",
    archivePrefix = "arXiv",
    primaryClass = "hep-th",
    month = "7",
    year = "2025"
}

@article{Peccei:1977hh,
    author = "Peccei, R. D. and Quinn, Helen R.",
    title = "{CP Conservation in the Presence of Instantons}",
    reportNumber = "ITP-568-STANFORD",
    doi = "10.1103/PhysRevLett.38.1440",
    journal = "Phys. Rev. Lett.",
    volume = "38",
    pages = "1440--1443",
    year = "1977"
}

@article{Wilczek:1977pj,
    author = "Wilczek, Frank",
    title = "{Problem of Strong  $P$  and  $T$  Invariance in the Presence of Instantons}",
    reportNumber = "Print-77-0939 (COLUMBIA)",
    doi = "10.1103/PhysRevLett.40.279",
    journal = "Phys. Rev. Lett.",
    volume = "40",
    pages = "279--282",
    year = "1978"
}

@article{Weinberg:1977ma,
    author = "Weinberg, Steven",
    title = "{A New Light Boson?}",
    reportNumber = "HUTP-77/A074",
    doi = "10.1103/PhysRevLett.40.223",
    journal = "Phys. Rev. Lett.",
    volume = "40",
    pages = "223--226",
    year = "1978"
}

@article{Yin:2025amn,
    author = "Yin, Ziwen and Cheng, Hanyu and Di Valentino, Eleonora and Gendler, Naomi and Marsh, David J. E. and Visinelli, Luca",
    title = "{Constraining the axiverse with reionization}",
    eprint = "2507.03535",
    archivePrefix = "arXiv",
    primaryClass = "hep-ph",
    reportNumber = "CA21106; CA21136",
    month = "7",
    year = "2025"
}

@article{Leedom:2025mlr,
    author = "Leedom, Jacob M. and Putti, Margherita and Westphal, Alexander",
    title = "{Towards a Heterotic Axiverse}",
    eprint = "2509.03578",
    archivePrefix = "arXiv",
    primaryClass = "hep-th",
    reportNumber = "DESY 25-113",
    month = "9",
    year = "2025"
}

@article{Loladze:2025uvf,
    author = "Loladze, Vazha and Platschorre, Arthur and Reig, Mario",
    title = "{Higher axion strings}",
    eprint = "2503.18707",
    archivePrefix = "arXiv",
    primaryClass = "hep-ph",
    doi = "10.1007/JHEP08(2025)182",
    journal = "JHEP",
    volume = "08",
    pages = "182",
    year = "2025"
}

@article{Reig:2025dqb,
    author = "Reig, Mario and Weigand, Timo",
    title = "{Testing the Heterotic String with the Axion-Photon Coupling}",
    eprint = "2509.08042",
    archivePrefix = "arXiv",
    primaryClass = "hep-th",
    month = "9",
    year = "2025"
}

@article{Petrossian-Byrne:2025mto,
    author = "Petrossian-Byrne, Rudin and Villadoro, Giovanni",
    title = "{Open string axiverse}",
    eprint = "2503.16387",
    archivePrefix = "arXiv",
    primaryClass = "hep-ph",
    doi = "10.1007/JHEP07(2025)049",
    journal = "JHEP",
    volume = "07",
    pages = "049",
    year = "2025"
}

@article{Benabou:2025kgx,
    author = "Benabou, Joshua N. and Fraser, Katherine and Reig, Mario and Safdi, Benjamin R.",
    title = "{String theory and grand unification suggest a submicroelectronvolt QCD axion}",
    eprint = "2505.15884",
    archivePrefix = "arXiv",
    primaryClass = "hep-ph",
    doi = "10.1103/lthr-97lm",
    journal = "Phys. Rev. D",
    volume = "112",
    number = "6",
    pages = "066003",
    year = "2025"
}

@article{Demirtas:2021gsq,
    author = "Demirtas, Mehmet and Gendler, Naomi and Long, Cody and McAllister, Liam and Moritz, Jakob",
    title = "{PQ axiverse}",
    eprint = "2112.04503",
    archivePrefix = "arXiv",
    primaryClass = "hep-th",
    doi = "10.1007/JHEP06(2023)092",
    journal = "JHEP",
    volume = "06",
    pages = "092",
    year = "2023"
}

@article{Arvanitaki:2009fg,
    author = "Arvanitaki, Asimina and Dimopoulos, Savas and Dubovsky, Sergei and Kaloper, Nemanja and March-Russell, John",
    title = "{String Axiverse}",
    eprint = "0905.4720",
    archivePrefix = "arXiv",
    primaryClass = "hep-th",
    doi = "10.1103/PhysRevD.81.123530",
    journal = "Phys. Rev. D",
    volume = "81",
    pages = "123530",
    year = "2010"
}

@article{Witten:1984dg,
    author = "Witten, Edward",
    title = "{Some Properties of O(32) Superstrings}",
    reportNumber = "Print-84-0838 (PRINCETON)",
    doi = "10.1016/0370-2693(84)90422-2",
    journal = "Phys. Lett. B",
    volume = "149",
    pages = "351--356",
    year = "1984"
}

@article{Svrcek:2006yi,
    author = "Svrcek, Peter and Witten, Edward",
    title = "{Axions In String Theory}",
    eprint = "hep-th/0605206",
    archivePrefix = "arXiv",
    reportNumber = "SLAC-PUB-11894",
    doi = "10.1088/1126-6708/2006/06/051",
    journal = "JHEP",
    volume = "06",
    pages = "051",
    year = "2006"
}

@article{Conlon:2006tq,
    author = "Conlon, Joseph P.",
    title = "{The QCD axion and moduli stabilisation}",
    eprint = "hep-th/0602233",
    archivePrefix = "arXiv",
    reportNumber = "DAMTP-2006-17",
    doi = "10.1088/1126-6708/2006/05/078",
    journal = "JHEP",
    volume = "05",
    pages = "078",
    year = "2006"
}

@article{Demirtas:2022hqf,
    author = "Demirtas, Mehmet and Rios-Tascon, Andres and McAllister, Liam",
    title = "{CYTools: A Software Package for Analyzing Calabi-Yau Manifolds}",
    eprint = "2211.03823",
    archivePrefix = "arXiv",
    primaryClass = "hep-th",
    month = "11",
    year = "2022"
}

@article{Demirtas:2018akl,
    author = "Demirtas, Mehmet and Long, Cody and McAllister, Liam and Stillman, Mike",
    title = "{The Kreuzer-Skarke Axiverse}",
    eprint = "1808.01282",
    archivePrefix = "arXiv",
    primaryClass = "hep-th",
    doi = "10.1007/JHEP04(2020)138",
    journal = "JHEP",
    volume = "04",
    pages = "138",
    year = "2020"
}

@article{Marsh:2010wq,
    author = "Marsh, David J. E. and Ferreira, Pedro G.",
    title = "{Ultra-Light Scalar Fields and the Growth of Structure in the Universe}",
    eprint = "1009.3501",
    archivePrefix = "arXiv",
    primaryClass = "hep-ph",
    reportNumber = "OUTP-10-25P",
    doi = "10.1103/PhysRevD.82.103528",
    journal = "Phys. Rev. D",
    volume = "82",
    pages = "103528",
    year = "2010"
}

@article{Candelas:1989qn,
    author = "Candelas, Philip and Hubsch, Tristan and Schimmrigk, Rolf",
    title = "{Relation Between the Weil-petersson and Zamolodchikov Metrics}",
    reportNumber = "UTTG-17-89",
    doi = "10.1016/0550-3213(90)90072-L",
    journal = "Nucl. Phys. B",
    volume = "329",
    pages = "583--590",
    year = "1990"
}

@article{2-form,
    author = "Jain, Mudit and Carta, Federico and Marsh, David J. E.",
    title = "{Systematics of Axion Physics on Calabi-Yau Orientifolds (forthcoming)}"
}

@article{Candelas:1985en,
    author = "Candelas, P. and Horowitz, Gary T. and Strominger, Andrew and Witten, Edward",
    title = "{Vacuum configurations for superstrings}",
    reportNumber = "NSF-ITP-84-170",
    doi = "10.1016/0550-3213(85)90602-9",
    journal = "Nucl. Phys. B",
    volume = "258",
    pages = "46--74",
    year = "1985"
}

@article{Grimm:2011dj,
    author = "Grimm, Thomas W. and Kerstan, Max and Palti, Eran and Weigand, Timo",
    title = "{On Fluxed Instantons and Moduli Stabilisation in IIB Orientifolds and F-theory}",
    eprint = "1105.3193",
    archivePrefix = "arXiv",
    primaryClass = "hep-th",
    doi = "10.1103/PhysRevD.84.066001",
    journal = "Phys. Rev. D",
    volume = "84",
    pages = "066001",
    year = "2011"
}

@article{Denef:2004ze,
    author = "Denef, Frederik and Douglas, Michael R.",
    title = "{Distributions of flux vacua}",
    eprint = "hep-th/0404116",
    archivePrefix = "arXiv",
    doi = "10.1088/1126-6708/2004/05/072",
    journal = "JHEP",
    volume = "05",
    pages = "072",
    year = "2004"
}

@article{Stout:2022phm,
    author = "Stout, John",
    title = "{Infinite Distances and Factorization}",
    eprint = "2208.08444",
    archivePrefix = "arXiv",
    primaryClass = "hep-th",
    month = "8",
    year = "2022"
}

@article{Wilson1992,
  author    = {P. M. H. Wilson},
  title     = {The K{\"a}hler cone on Calabi--Yau threefolds},
  journal   = {Inventiones mathematicae},
  year      = {1992},
  volume    = {107},
  number    = {1},
  pages     = {561--583},
  doi       = {10.1007/BF01231902},
  url       = {https://doi.org/10.1007/BF01231902},
  issn      = {1432-1297}
}

@misc{hayakawa1995degenarationcalabiyaumanifoldwp,
      title={Degenaration of Calabi-Yau Manifold with W-P Metric}, 
      author={Yoshiko Hayakawa},
      year={1995},
      eprint={alg-geom/9507016},
      archivePrefix={arXiv},
      primaryClass={alg-geom},
      url={https://arxiv.org/abs/alg-geom/9507016}, 
}

@article{Morrison:1996pp,
    author = "Morrison, David R. and Vafa, Cumrun",
    title = "{Compactifications of F theory on Calabi-Yau threefolds. 2.}",
    eprint = "hep-th/9603161",
    archivePrefix = "arXiv",
    reportNumber = "DUKE-TH-96-107, HUTP-96-A012",
    doi = "10.1016/0550-3213(96)00369-0",
    journal = "Nucl. Phys. B",
    volume = "476",
    pages = "437--469",
    year = "1996"
}

@book{CoxKatz1999,
  author    = {David A. Cox and Sheldon Katz},
  title     = {Mirror Symmetry and Algebraic Geometry},
  series    = {Mathematical Surveys and Monographs},
  volume    = {68},
  publisher = {American Mathematical Society},
  address   = {Providence, RI},
  year      = {1999},
  isbn      = {978-0-8218-1059-5},
}

@article{2007CSE.....9...90H,
	adsurl = {https://ui.adsabs.harvard.edu/abs/2007CSE.....9...90H},
	author = {{Hunter}, John D.},
	doi = {10.1109/MCSE.2007.55},
	journal = {Computing in Science and Engineering},
	month = may,
	number = {3},
	pages = {90-95},
	title = {{Matplotlib: A 2D Graphics Environment}},
	volume = {9},
	year = 2007}

@article{2020Natur.585..357H,
	adsurl = {https://ui.adsabs.harvard.edu/abs/2020Natur.585..357H},
	archiveprefix = {arXiv},
	author = {{Harris}, Charles R. and others},
	doi = {10.1038/s41586-020-2649-2},
	eprint = {2006.10256},
	journal = {\nat},
	month = sep,
	number = {7825},
	pages = {357-362},
	primaryclass = {cs.MS},
	title = {{Array programming with NumPy}},
	volume = {585},
	year = 2020}

@article{Cicoli:2017shd,
    author = "Cicoli, Michele and Garc{\`\i}a-Etxebarria, I{\~n}aki and Mayrhofer, Christoph and Quevedo, Fernando and Shukla, Pramod and Valandro, Roberto",
    title = "{Global Orientifolded Quivers with Inflation}",
    eprint = "1706.06128",
    archivePrefix = "arXiv",
    primaryClass = "hep-th",
    doi = "10.1007/JHEP11(2017)134",
    journal = "JHEP",
    volume = "11",
    pages = "134",
    year = "2017"
}

@article{Cicoli:2013cha,
    author = "Cicoli, Michele and Klevers, Denis and Krippendorf, Sven and Mayrhofer, Christoph and Quevedo, Fernando and Valandro, Roberto",
    title = "{Explicit de Sitter Flux Vacua for Global String Models with Chiral Matter}",
    eprint = "1312.0014",
    archivePrefix = "arXiv",
    primaryClass = "hep-th",
    doi = "10.1007/JHEP05(2014)001",
    journal = "JHEP",
    volume = "05",
    pages = "001",
    year = "2014"
}

@article{Cicoli:2013zha,
    author = "Cicoli, Michele and Krippendorf, Sven and Mayrhofer, Christoph and Quevedo, Fernando and Valandro, Roberto",
    title = "{The Web of D-branes at Singularities in Compact Calabi-Yau Manifolds}",
    eprint = "1304.2771",
    archivePrefix = "arXiv",
    primaryClass = "hep-th",
    doi = "10.1007/JHEP05(2013)114",
    journal = "JHEP",
    volume = "05",
    pages = "114",
    year = "2013"
}

@article{Cicoli:2013mpa,
    author = "Cicoli, Michele and Krippendorf, Sven and Mayrhofer, Christoph and Quevedo, Fernando and Valandro, Roberto",
    title = "{D3/D7 Branes at Singularities: Constraints from Global Embedding and Moduli Stabilisation}",
    eprint = "1304.0022",
    archivePrefix = "arXiv",
    primaryClass = "hep-th",
    doi = "10.1007/JHEP07(2013)150",
    journal = "JHEP",
    volume = "07",
    pages = "150",
    year = "2013"
}

@article{Strominger:1985ks,
    author = "Strominger, Andrew",
    title = "{Yukawa Couplings in Superstring Compactification}",
    reportNumber = "NSF-ITP-85-105",
    doi = "10.1103/PhysRevLett.55.2547",
    journal = "Phys. Rev. Lett.",
    volume = "55",
    pages = "2547",
    year = "1985"
}

@article{MacFadden:2023cyf,
    author = "MacFadden, Nate",
    title = "{Efficient Algorithm for Generating Homotopy Inequivalent Calabi-Yaus}",
    eprint = "2309.10855",
    archivePrefix = "arXiv",
    primaryClass = "hep-th",
    month = "9",
    year = "2023"
}

@article{Cicoli:2012vw,
    author = "Cicoli, Michele and Krippendorf, Sven and Mayrhofer, Christoph and Quevedo, Fernando and Valandro, Roberto",
    title = "{D-Branes at del Pezzo Singularities: Global Embedding and Moduli Stabilisation}",
    eprint = "1206.5237",
    archivePrefix = "arXiv",
    primaryClass = "hep-th",
    reportNumber = "DAMTP-2012-47, ZMP-HH-12-10",
    doi = "10.1007/JHEP09(2012)019",
    journal = "JHEP",
    volume = "09",
    pages = "019",
    year = "2012"
}

@article{Cicoli:2021dhg,
    author = "Cicoli, Michele and Etxebarria, I{\~n}aki Garc{\'\i}a and Quevedo, Fernando and Schachner, Andreas and Shukla, Pramod and Valandro, Roberto",
    title = "{The Standard Model quiver in de Sitter string compactifications}",
    eprint = "2106.11964",
    archivePrefix = "arXiv",
    primaryClass = "hep-th",
    doi = "10.1007/JHEP08(2021)109",
    journal = "JHEP",
    volume = "08",
    pages = "109",
    year = "2021"
}

@article{2020NatMe..17..261V,
	adsurl = {https://ui.adsabs.harvard.edu/abs/2020NatMe..17..261V},
	archiveprefix = {arXiv},
	author = {{Virtanen}, Pauli and others},
	doi = {10.1038/s41592-019-0686-2},
	eprint = {1907.10121},
	journal = {Nature Methods},
	month = feb,
	pages = {261-272},
	primaryclass = {cs.MS},
	title = {{SciPy 1.0: fundamental algorithms for scientific computing in Python}},
	volume = {17},
	year = 2020}

@article{Foreman-Mackey:2012any,
    author = "Foreman-Mackey, Daniel and Hogg, David W. and Lang, Dustin and Goodman, Jonathan",
    title = "{emcee: The MCMC Hammer}",
    eprint = "1202.3665",
    archivePrefix = "arXiv",
    primaryClass = "astro-ph.IM",
    doi = "10.1086/670067",
    journal = "Publ. Astron. Soc. Pac.",
    volume = "125",
    pages = "306--312",
    year = "2013"
}

@article{Goldstein:2023gnw,
    author = "Goldstein, Samuel and Hill, J. Colin and Ir{\v{s}}i{\v{c}}, Vid and Sherwin, Blake D.",
    title = "{Canonical Hubble-Tension-Resolving Early Dark Energy Cosmologies Are Inconsistent with the Lyman-{\ensuremath{\alpha}} Forest}",
    eprint = "2303.00746",
    archivePrefix = "arXiv",
    primaryClass = "astro-ph.CO",
    doi = "10.1103/PhysRevLett.131.201001",
    journal = "Phys. Rev. Lett.",
    volume = "131",
    number = "20",
    pages = "201001",
    year = "2023"
}

@article{eBOSS:2018qyj,
    author = "Chabanier, Sol{\`e}ne and others",
    collaboration = "eBOSS",
    title = "{The one-dimensional power spectrum from the SDSS DR14 Ly$\alpha$ forests}",
    eprint = "1812.03554",
    archivePrefix = "arXiv",
    primaryClass = "astro-ph.CO",
    doi = "10.1088/1475-7516/2019/07/017",
    journal = "JCAP",
    volume = "07",
    pages = "017",
    year = "2019"
}

@article{Chiang:1995hi,
    author = "Chiang, Ti-ming and Greene, Brian R. and Gross, Mark and Kanter, Yakov",
    editor = "Gava, E. and Narain, K. S. and Vafa, C.",
    title = "{Black hole condensation and the web of Calabi-Yau manifolds}",
    eprint = "hep-th/9511204",
    archivePrefix = "arXiv",
    reportNumber = "CLNS-95-1376",
    doi = "10.1016/0920-5632(96)00010-2",
    journal = "Nucl. Phys. B Proc. Suppl.",
    volume = "46",
    pages = "82--95",
    year = "1996"
}

@article{Candelas:1990pi,
    author = "Candelas, Philip and de la Ossa, Xenia",
    title = "{Moduli Space of {Calabi-Yau} Manifolds}",
    reportNumber = "UTTG-07-90",
    doi = "10.1016/0550-3213(91)90122-E",
    journal = "Nucl. Phys. B",
    volume = "355",
    pages = "455--481",
    year = "1991"
}

@article{Demirtas:2021nlu,
    author = "Demirtas, Mehmet and Kim, Manki and McAllister, Liam and Moritz, Jakob and Rios-Tascon, Andres",
    title = "{Small cosmological constants in string theory}",
    eprint = "2107.09064",
    archivePrefix = "arXiv",
    primaryClass = "hep-th",
    doi = "10.1007/JHEP12(2021)136",
    journal = "JHEP",
    volume = "12",
    pages = "136",
    year = "2021"
}

@article{Abel:2020pzs,
    author = "Abel, C. and others",
    title = "{Measurement of the Permanent Electric Dipole Moment of the Neutron}",
    eprint = "2001.11966",
    archivePrefix = "arXiv",
    primaryClass = "hep-ex",
    doi = "10.1103/PhysRevLett.124.081803",
    journal = "Phys. Rev. Lett.",
    volume = "124",
    number = "8",
    pages = "081803",
    year = "2020"
}

@article{ADMX:2018gho,
    author = "Du, N. and others",
    collaboration = "ADMX",
    title = "{A Search for Invisible Axion Dark Matter with the Axion Dark Matter Experiment}",
    eprint = "1804.05750",
    archivePrefix = "arXiv",
    primaryClass = "hep-ex",
    reportNumber = "FERMILAB-PUB-18-101-AD-AE",
    doi = "10.1103/PhysRevLett.120.151301",
    journal = "Phys. Rev. Lett.",
    volume = "120",
    number = "15",
    pages = "151301",
    year = "2018"
}

@article{DMRadio:2022pkf,
    author = "Brouwer, L. and others",
    collaboration = "DMRadio",
    title = "{Projected sensitivity of DMRadio-m3: A search for the QCD axion below 1{\,}{\,}{\ensuremath{\mu}}eV}",
    eprint = "2204.13781",
    archivePrefix = "arXiv",
    primaryClass = "hep-ex",
    doi = "10.1103/PhysRevD.106.103008",
    journal = "Phys. Rev. D",
    volume = "106",
    number = "10",
    pages = "103008",
    year = "2022"
}

@article{ParticleDataGroup:2024cfk,
    author = "Navas, S. and others",
    collaboration = "Particle Data Group",
    title = "{Review of particle physics}",
    doi = "10.1103/PhysRevD.110.030001",
    journal = "Phys. Rev. D",
    volume = "110",
    number = "3",
    pages = "030001",
    year = "2024"
}

@article{Blas:2011rf,
    author = "Blas, Diego and Lesgourgues, Julien and Tram, Thomas",
    title = "{The Cosmic Linear Anisotropy Solving System (CLASS) II: Approximation schemes}",
    eprint = "1104.2933",
    archivePrefix = "arXiv",
    primaryClass = "astro-ph.CO",
    reportNumber = "CERN-PH-TH-2011-082, LAPTH-010-11",
    doi = "10.1088/1475-7516/2011/07/034",
    journal = "JCAP",
    volume = "07",
    pages = "034",
    year = "2011"
}

@article{Cicoli:2021gss,
    author = "Cicoli, Michele and Guidetti, Veronica and Righi, Nicole and Westphal, Alexander",
    title = "{Fuzzy Dark Matter candidates from string theory}",
    eprint = "2110.02964",
    archivePrefix = "arXiv",
    primaryClass = "hep-th",
    reportNumber = "DESY-21-153",
    doi = "10.1007/JHEP05(2022)107",
    journal = "JHEP",
    volume = "05",
    pages = "107",
    year = "2022"
}

@article{Poulin:2018dzj, author = "Poulin, Vivian and Smith, Tristan L. and Grin, Daniel and Karwal, Tanvi and Kamionkowski, Marc", title = "{Cosmological implications of ultralight axionlike fields}", eprint = "1806.10608", archivePrefix = "arXiv", primaryClass = "astro-ph.CO", doi = "10.1103/PhysRevD.98.083525", journal = "Phys. Rev. D", volume = "98", number = "8", pages = "083525", year = "2018" }

@article{Lewis:2019xzd,
   author = "Lewis, Antony",
   title = "{GetDist: a Python package for analysing Monte Carlo samples}",
   eprint = "1910.13970",
   archivePrefix = "arXiv",
   primaryClass = "astro-ph.IM",
   doi = "10.1088/1475-7516/2025/08/025",
   journal = "JCAP",
   volume = "08",
   pages = "025",
   year = "2025"
}

@article{Smith:2019ihp, author = "Smith, Tristan L. and Poulin, Vivian and Amin, Mustafa A.", title = "{Oscillating scalar fields and the Hubble tension: a resolution with novel signatures}", eprint = "1908.06995", archivePrefix = "arXiv", primaryClass = "astro-ph.CO", doi = "10.1103/PhysRevD.101.063523", journal = "Phys. Rev. D", volume = "101", number = "6", pages = "063523", year = "2020" }

@article{McAllister:2024lnt,
    author = "McAllister, Liam and Moritz, Jakob and Nally, Richard and Schachner, Andreas",
    title = "{Candidate de Sitter Vacua}",
    eprint = "2406.13751",
    archivePrefix = "arXiv",
    primaryClass = "hep-th",
    reportNumber = "CERN-TH-2024-090",
    month = "6",
    year = "2024"
}

@article{GrillidiCortona:2015jxo,
    author = "Grilli di Cortona, Giovanni and Hardy, Edward and Pardo Vega, Javier and Villadoro, Giovanni",
    title = "{The QCD axion, precisely}",
    eprint = "1511.02867",
    archivePrefix = "arXiv",
    primaryClass = "hep-ph",
    doi = "10.1007/JHEP01(2016)034",
    journal = "JHEP",
    volume = "01",
    pages = "034",
    year = "2016"
}

@article{Acharya:2010zx,
    author = "Acharya, Bobby Samir and Bobkov, Konstantin and Kumar, Piyush",
    title = "{An M Theory Solution to the Strong CP Problem and Constraints on the Axiverse}",
    eprint = "1004.5138",
    archivePrefix = "arXiv",
    primaryClass = "hep-th",
    doi = "10.1007/JHEP11(2010)105",
    journal = "JHEP",
    volume = "11",
    pages = "105",
    year = "2010"
}

@article{Cordova:2009fg,
    author = "Cordova, Clay",
    title = "{Decoupling Gravity in F-Theory}",
    eprint = "0910.2955",
    archivePrefix = "arXiv",
    primaryClass = "hep-th",
    doi = "10.4310/ATMP.2011.v15.n3.a2",
    journal = "Adv. Theor. Math. Phys.",
    volume = "15",
    number = "3",
    pages = "689--740",
    year = "2011"
}

@article{Hu:2000ke,
    author = "Hu, Wayne and Barkana, Rennan and Gruzinov, Andrei",
    title = "{Cold and fuzzy dark matter}",
    eprint = "astro-ph/0003365",
    archivePrefix = "arXiv",
    doi = "10.1103/PhysRevLett.85.1158",
    journal = "Phys. Rev. Lett.",
    volume = "85",
    pages = "1158--1161",
    year = "2000"
}

@article{Marsh:2015xka,
    author = "Marsh, David J. E.",
    title = "{Axion Cosmology}",
    eprint = "1510.07633",
    archivePrefix = "arXiv",
    primaryClass = "astro-ph.CO",
    reportNumber = "KCL-PH-TH-2015-50",
    doi = "10.1016/j.physrep.2016.06.005",
    journal = "Phys. Rept.",
    volume = "643",
    pages = "1--79",
    year = "2016"
}

@article{Marsh:2013taa,
    author = "Marsh, David J. E. and Grin, Daniel and Hlozek, Ren{\'e}e and Ferreira, Pedro G.",
    title = "{Axiverse cosmology and the energy scale of inflation}",
    eprint = "1303.3008",
    archivePrefix = "arXiv",
    primaryClass = "astro-ph.CO",
    doi = "10.1103/PhysRevD.87.121701",
    journal = "Phys. Rev. D",
    volume = "87",
    pages = "121701",
    year = "2013"
}

@article{Arvanitaki:2019rax,
    author = "Arvanitaki, Asimina and Dimopoulos, Savas and Galanis, Marios and Lehner, Luis and Thompson, Jedidiah O. and Van Tilburg, Ken",
    title = "{Large-misalignment mechanism for the formation of compact axion structures: Signatures from the QCD axion to fuzzy dark matter}",
    eprint = "1909.11665",
    archivePrefix = "arXiv",
    primaryClass = "astro-ph.CO",
    doi = "10.1103/PhysRevD.101.083014",
    journal = "Phys. Rev. D",
    volume = "101",
    number = "8",
    pages = "083014",
    year = "2020"
}

@article{Zhang:2017dpp,
    author = "Zhang, Ui-Han and Chiueh, Tzihong",
    title = "{Cosmological Perturbations of Extreme Axion in the Radiation Era}",
    eprint = "1705.01439",
    archivePrefix = "arXiv",
    primaryClass = "astro-ph.CO",
    doi = "10.1103/PhysRevD.96.063522",
    journal = "Phys. Rev. D",
    volume = "96",
    number = "6",
    pages = "063522",
    year = "2017"
}

@article{Eberhardt:2025caq,
    author = "Eberhardt, Andrew and Ferreira, Elisa G. M.",
    title = "{Ultralight fuzzy dark matter review}",
    eprint = "2507.00705",
    archivePrefix = "arXiv",
    primaryClass = "astro-ph.CO",
    month = "7",
    year = "2025"
}

@article{Winch:2023qzl,
    author = "Winch, Harrison and Hlozek, Renee and Marsh, David J. E. and Grin, Daniel and Rogers, Keir K.",
    title = "{Extreme axions unveiled: A novel fluid approach for cosmological modeling}",
    eprint = "2311.02052",
    archivePrefix = "arXiv",
    primaryClass = "astro-ph.CO",
    reportNumber = "KCL-PH-TH/2023-60",
    doi = "10.1103/PhysRevD.110.043517",
    journal = "Phys. Rev. D",
    volume = "110",
    number = "4",
    pages = "043517",
    year = "2024"
}

@article{Winch:2024mrt,
    author = "Winch, Harrison and Rogers, Keir K. and Hlo{\v{z}}ek, Ren{\'e}e and Marsh, David J. E.",
    title = "{High-redshift, Small-scale Tests of Ultralight Axion Dark Matter Using Hubble and Webb Galaxy UV Luminosities}",
    eprint = "2404.11071",
    archivePrefix = "arXiv",
    primaryClass = "astro-ph.CO",
    doi = "10.3847/1538-4357/ad7a73",
    journal = "Astrophys. J.",
    volume = "976",
    number = "1",
    pages = "40",
    year = "2024"
}

@article{Hertzberg:2008wr,
    author = "Hertzberg, Mark P and Tegmark, Max and Wilczek, Frank",
    title = "{Axion Cosmology and the Energy Scale of Inflation}",
    eprint = "0807.1726",
    archivePrefix = "arXiv",
    primaryClass = "astro-ph",
    reportNumber = "MIT-CTP-3950",
    doi = "10.1103/PhysRevD.78.083507",
    journal = "Phys. Rev. D",
    volume = "78",
    pages = "083507",
    year = "2008"
}

@article{Lague:2021frh,
    author = {Lagu\"e, Alex and Bond, J. Richard and Hlo\v{z}ek, Ren\'ee and Rogers, Keir K. and Marsh, David J. E. and Grin, Daniel},
    title = "{Constraining ultralight axions with galaxy surveys}",
    eprint = "2104.07802",
    archivePrefix = "arXiv",
    primaryClass = "astro-ph.CO",
    doi = "10.1088/1475-7516/2022/01/049",
    journal = "JCAP",
    volume = "01",
    number = "01",
    pages = "049",
    year = "2022"
}

@article{Crino:2020qwk,
    author = "Crin{\`o}, Chiara and Quevedo, Fernando and Valandro, Roberto",
    title = "{On de Sitter String Vacua from Anti-D3-Branes in the Large Volume Scenario}",
    eprint = "2010.15903",
    archivePrefix = "arXiv",
    primaryClass = "hep-th",
    doi = "10.1007/JHEP03(2021)258",
    journal = "JHEP",
    volume = "03",
    pages = "258",
    year = "2021"
}

@article{AbdusSalam:2020ywo,
    author = "AbdusSalam, S. and Abel, S. and Cicoli, M. and Quevedo, F. and Shukla, P.",
    title = {{A systematic approach to K{\"a}hler moduli stabilisation}},
    eprint = "2005.11329",
    archivePrefix = "arXiv",
    primaryClass = "hep-th",
    reportNumber = "IPPP/20/18",
    doi = "10.1007/JHEP08(2020)047",
    journal = "JHEP",
    volume = "08",
    number = "08",
    pages = "047",
    year = "2020"
}

@article{AbdusSalam:2025twp,
    author = "AbdusSalam, Shehu and Hughes, Christopher and Quevedo, Fernando and Schachner, Andreas",
    title = {{Coexisting Flux String Vacua from Numerical K{\"a}hler Moduli Stabilisation}},
    eprint = "2507.00615",
    archivePrefix = "arXiv",
    primaryClass = "hep-th",
    month = "7",
    year = "2025"
}

@article{Louis:2012nb,
    author = "Louis, Jan and Rummel, Markus and Valandro, Roberto and Westphal, Alexander",
    title = "{Building an explicit de Sitter}",
    eprint = "1208.3208",
    archivePrefix = "arXiv",
    primaryClass = "hep-th",
    reportNumber = "DESY-12-146, ZMP-HH-12-16",
    doi = "10.1007/JHEP10(2012)163",
    journal = "JHEP",
    volume = "10",
    pages = "163",
    year = "2012"
}

@article{Fairbairn:2025fko,
    author = "Fairbairn, Malcolm and Heurtier, Lucien and Olea-Romacho, Mar{\'\i}a Olalla",
    title = "{Is $\Lambda$CDM on the run? Reconciling the CMB with the Lyman-$\alpha$ Forest}",
    eprint = "2511.01612",
    archivePrefix = "arXiv",
    primaryClass = "astro-ph.CO",
    month = "11",
    year = "2025"
}

@article{Planck:2019nip,
    author = "Aghanim, N. and others",
    collaboration = "Planck",
    title = "{Planck 2018 results. V. CMB power spectra and likelihoods}",
    eprint = "1907.12875",
    archivePrefix = "arXiv",
    primaryClass = "astro-ph.CO",
    doi = "10.1051/0004-6361/201936386",
    journal = "Astron. Astrophys.",
    volume = "641",
    pages = "A5",
    year = "2020"
}

@article{Prince:2019hse,
    author = "Prince, Heather and Dunkley, Jo",
    title = "{Data compression in cosmology: A compressed likelihood for Planck data}",
    eprint = "1909.05869",
    archivePrefix = "arXiv",
    primaryClass = "astro-ph.CO",
    doi = "10.1103/PhysRevD.100.083502",
    journal = "Phys. Rev. D",
    volume = "100",
    number = "8",
    pages = "083502",
    year = "2019"
}

@article{Delouis:2019bub,
    author = "Delouis, J. -M. and Pagano, L. and Mottet, S. and Puget, J. -L. and Vibert, L.",
    title = "{SRoll2: an improved mapmaking approach to reduce large-scale systematic effects in the Planck High Frequency Instrument legacy maps}",
    eprint = "1901.11386",
    archivePrefix = "arXiv",
    primaryClass = "astro-ph.CO",
    doi = "10.1051/0004-6361/201834882",
    journal = "Astron. Astrophys.",
    volume = "629",
    pages = "A38",
    year = "2019"
}

@article{Hlozek:2017zzf,
    author = "Hlozek, Ren{\'e}e and Marsh, David J. E. and Grin, Daniel",
    title = "{Using the Full Power of the Cosmic Microwave Background to Probe Axion Dark Matter}",
    eprint = "1708.05681",
    archivePrefix = "arXiv",
    primaryClass = "astro-ph.CO",
    reportNumber = "KCL-PH-TH-2017-39",
    doi = "10.1093/mnras/sty271",
    journal = "Mon. Not. Roy. Astron. Soc.",
    volume = "476",
    number = "3",
    pages = "3063--3085",
    year = "2018"
}

@article{Lee:2020obi,
    author = {Lee, Nanoom and Ali-Ha{\"\i}moud, Yacine},
    title = "{HYREC-2: a highly accurate sub-millisecond recombination code}",
    eprint = "2007.14114",
    archivePrefix = "arXiv",
    primaryClass = "astro-ph.CO",
    doi = "10.1103/PhysRevD.102.083517",
    journal = "Phys. Rev. D",
    volume = "102",
    number = "8",
    pages = "083517",
    year = "2020"
}

@article{Rogers:2023ezo,
    author = {Rogers, Keir K. and Hlo\v{z}ek, Ren\'ee and Lagu\"e, Alex and Ivanov, Mikhail M. and Philcox, Oliver H. E. and Cabass, Giovanni and Akitsu, Kazuyuki and Marsh, David J. E.},
    title = "{Ultra-light axions and the S$_{8}$ tension: joint constraints from the cosmic microwave background and galaxy clustering}",
    eprint = "2301.08361",
    archivePrefix = "arXiv",
    primaryClass = "astro-ph.CO",
    doi = "10.1088/1475-7516/2023/06/023",
    journal = "JCAP",
    volume = "06",
    pages = "023",
    year = "2023"
}

@article{Hlozek:2014lca,
    author = "Hlozek, Ren\'ee and Grin, Daniel and Marsh, David J. E. and Ferreira, Pedro G.",
    title = "{A search for ultralight axions using precision cosmological data}",
    eprint = "1410.2896",
    archivePrefix = "arXiv",
    primaryClass = "astro-ph.CO",
    doi = "10.1103/PhysRevD.91.103512",
    journal = "Phys. Rev. D",
    volume = "91",
    number = "10",
    pages = "103512",
    year = "2015"
}

@article{Kobayashi:2017jcf,
    author = "Kobayashi, Takeshi and Murgia, Riccardo and De Simone, Andrea and Ir\v{s}i\v{c}, Vid and Viel, Matteo",
    title = "{Lyman-$\alpha$ constraints on ultralight scalar dark matter: Implications for the early and late universe}",
    eprint = "1708.00015",
    archivePrefix = "arXiv",
    primaryClass = "astro-ph.CO",
    reportNumber = "SISSA-33-2017-FISI",
    doi = "10.1103/PhysRevD.96.123514",
    journal = "Phys. Rev. D",
    volume = "96",
    number = "12",
    pages = "123514",
    year = "2017"
}

@article{Frieman:1995pm,
    author = "Frieman, Joshua A. and Hill, Christopher T. and Stebbins, Albert and Waga, Ioav",
    title = "{Cosmology with ultralight pseudo Nambu-Goldstone bosons}",
    eprint = "astro-ph/9505060",
    archivePrefix = "arXiv",
    reportNumber = "FERMILAB-PUB-95-066-A",
    doi = "10.1103/PhysRevLett.75.2077",
    journal = "Phys. Rev. Lett.",
    volume = "75",
    pages = "2077--2080",
    year = "1995"
}

@article{Amendola:2005ad,
    author = "Amendola, Luca and Barbieri, Riccardo",
    title = "{Dark matter from an ultra-light pseudo-Goldsone-boson}",
    eprint = "hep-ph/0509257",
    archivePrefix = "arXiv",
    reportNumber = "CERN-PH-TH-2005-163",
    doi = "10.1016/j.physletb.2006.08.069",
    journal = "Phys. Lett. B",
    volume = "642",
    pages = "192--196",
    year = "2006"
}

@article{Cicoli:2012sz,
    author = "Cicoli, Michele and Goodsell, Mark and Ringwald, Andreas",
    title = "{The type IIB string axiverse and its low-energy phenomenology}",
    eprint = "1206.0819",
    archivePrefix = "arXiv",
    primaryClass = "hep-th",
    reportNumber = "DESY-12-058, CERN-PH-TH-2012-153",
    doi = "10.1007/JHEP10(2012)146",
    journal = "JHEP",
    volume = "10",
    pages = "146",
    year = "2012"
}

@article{Dubey:2023dvu,
    author = "Dubey, Abhishek and Krippendorf, Sven and Schachner, Andreas",
    title = "{JAXVacua {\textemdash} a framework for sampling string vacua}",
    eprint = "2306.06160",
    archivePrefix = "arXiv",
    primaryClass = "hep-th",
    doi = "10.1007/JHEP12(2023)146",
    journal = "JHEP",
    volume = "12",
    pages = "146",
    year = "2023"
}

@article{Ebelt:2023clh,
    author = "Ebelt, Julian and Krippendorf, Sven and Schachner, Andreas",
    title = "{W0{\_}sample = np.random.normal(0,1)?}",
    eprint = "2307.15749",
    archivePrefix = "arXiv",
    primaryClass = "hep-th",
    doi = "10.1016/j.physletb.2024.138786",
    journal = "Phys. Lett. B",
    volume = "855",
    pages = "138786",
    year = "2024"
}

@article{Krippendorf:2023idy,
    author = "Krippendorf, Sven and Schachner, Andreas",
    title = "{New non-supersymmetric flux vacua in string theory}",
    eprint = "2308.15525",
    archivePrefix = "arXiv",
    primaryClass = "hep-th",
    reportNumber = "LMU-ASC 30/23",
    doi = "10.1007/JHEP12(2023)145",
    journal = "JHEP",
    volume = "12",
    pages = "145",
    year = "2023"
}

@article{Chauhan:2025rdj,
    author = "Chauhan, Aman and Cicoli, Michele and Krippendorf, Sven and Maharana, Anshuman and Piantadosi, Pellegrino and Schachner, Andreas",
    title = "{Deep observations of the Type IIB flux landscape}",
    eprint = "2501.03984",
    archivePrefix = "arXiv",
    primaryClass = "hep-th",
    doi = "10.1007/JHEP07(2025)271",
    journal = "JHEP",
    volume = "07",
    pages = "271",
    year = "2025"
}

@article{Bousso:2000xa,
    author = "Bousso, Raphael and Polchinski, Joseph",
    title = "{Quantization of four form fluxes and dynamical neutralization of the cosmological constant}",
    eprint = "hep-th/0004134",
    archivePrefix = "arXiv",
    reportNumber = "SU-ITP-00-12, NSF-ITP-00-40",
    doi = "10.1088/1126-6708/2000/06/006",
    journal = "JHEP",
    volume = "06",
    pages = "006",
    year = "2000"
}

@inproceedings{rezende2015variational,
    title={Variational Inference with Normalizing Flows},
    author={Rezende, Danilo and Mohamed, Shakir},
    booktitle={Proceedings of the 32nd International Conference on  Machine Learning},
    pages={1530--1538},
    year={2015},
    organization={PMLR},
    url={https://proceedings.mlr.press/v37/rezende15.html}
}

@article{Aspinwall:1995vk,
    author = "Aspinwall, Paul S. and Louis, Jan",
    title = "{On the ubiquity of K3 fibrations in string duality}",
    eprint = "hep-th/9510234",
    archivePrefix = "arXiv",
    reportNumber = "CLNS-95-1369, LMU-TPW-95-16, CLNS-95/1369, LMU-TPW 95-16",
    doi = "10.1016/0370-2693(95)01541-8",
    journal = "Phys. Lett. B",
    volume = "369",
    pages = "233--242",
    year = "1996"
}

@article{Curio:1998bva,
    author = "Curio, Gottfried and Donagi, Ron Y.",
    title = "{Moduli in N=1 heterotic / F theory duality}",
    eprint = "hep-th/9801057",
    archivePrefix = "arXiv",
    reportNumber = "IASSNS-HEP-97-139",
    doi = "10.1016/S0550-3213(98)00185-0",
    journal = "Nucl. Phys. B",
    volume = "518",
    pages = "603--631",
    year = "1998"
}

@article{Corvilain:2018lgw,
    author = "Corvilain, Pierre and Grimm, Thomas W. and Valenzuela, Irene",
    title = {{The Swampland Distance Conjecture for K{\"a}hler moduli}},
    eprint = "1812.07548",
    archivePrefix = "arXiv",
    primaryClass = "hep-th",
    doi = "10.1007/JHEP08(2019)075",
    journal = "JHEP",
    volume = "08",
    pages = "075",
    year = "2019"
}

@article{Lee:2019wij,
    author = "Lee, Seung-Joo and Lerche, Wolfgang and Weigand, Timo",
    title = "{Emergent strings from infinite distance limits}",
    eprint = "1910.01135",
    archivePrefix = "arXiv",
    primaryClass = "hep-th",
    reportNumber = "CERN-TH-2019-159",
    doi = "10.1007/JHEP02(2022)190",
    journal = "JHEP",
    volume = "02",
    pages = "190",
    year = "2022"
}

@article{Cicoli:2008gp,
    author = "Cicoli, M. and Burgess, C. P. and Quevedo, F.",
    title = "{Fibre Inflation: Observable Gravity Waves from IIB String Compactifications}",
    eprint = "0808.0691",
    archivePrefix = "arXiv",
    primaryClass = "hep-th",
    reportNumber = "DAMTP-2008-59",
    doi = "10.1088/1475-7516/2009/03/013",
    journal = "JCAP",
    volume = "03",
    pages = "013",
    year = "2009"
}

@article{Cicoli:2008va,
    author = "Cicoli, Michele and Conlon, Joseph P. and Quevedo, Fernando",
    title = "{General Analysis of LARGE Volume Scenarios with String Loop Moduli Stabilisation}",
    eprint = "0805.1029",
    archivePrefix = "arXiv",
    primaryClass = "hep-th",
    reportNumber = "DAMTP-2008-16",
    doi = "10.1088/1126-6708/2008/10/105",
    journal = "JHEP",
    volume = "10",
    pages = "105",
    year = "2008"
}

@article{Cicoli:2011it,
    author = "Cicoli, Michele and Kreuzer, Maximilian and Mayrhofer, Christoph",
    title = "{Toric K3-Fibred Calabi-Yau Manifolds with del Pezzo Divisors for String Compactifications}",
    eprint = "1107.0383",
    archivePrefix = "arXiv",
    primaryClass = "hep-th",
    reportNumber = "DESY-11-103",
    doi = "10.1007/JHEP02(2012)002",
    journal = "JHEP",
    volume = "02",
    pages = "002",
    year = "2012"
}

@book{lazarsfeld2017,
  title={Positivity in algebraic geometry I: Classical setting: line bundles and linear series},
  author={Lazarsfeld, Robert K},
  volume={48},
  year={2017},
  publisher={Springer}
}

@article{kollar2015deformations,
  title={Deformations of elliptic Calabi-Yau manifolds},
  author={Koll{\'a}r, J{\'a}nos},
  journal={Recent advances in algebraic geometry},
  volume={417},
  pages={254--290},
  year={2015},
  publisher={Cambridge Books Online, Cambridge University Press}
}

@article{Oguiso,
    author = {OGUISO, KEIJI},
    title = {ON ALGEBRAIC FIBER SPACE STRUCTURES ON A CALABI-YAU 3-FOLD},
    journal = {International Journal of Mathematics},
    volume = {04},
    number = {03},
    pages = {439-465},
    year = {1993},
    doi = {10.1142/S0129167X93000248},
    URL = {https://doi.org/10.1142/S0129167X93000248},
    eprint = {https://doi.org/10.1142/S0129167X93000248}
}

@article{Weigand:2018rez,
    author = "Weigand, Timo",
    title = "{F-theory}",
    eprint = "1806.01854",
    archivePrefix = "arXiv",
    primaryClass = "hep-th",
    reportNumber = "CERN-TH-2018-126",
    journal = "PoS",
    volume = "TASI2017",
    pages = "016",
    year = "2018"
}

@article{Witten:1996bn,
    author = "Witten, Edward",
    title = "{Nonperturbative superpotentials in string theory}",
    eprint = "hep-th/9604030",
    archivePrefix = "arXiv",
    reportNumber = "IASSNS-HEP-96-29",
    doi = "10.1016/0550-3213(96)00283-0",
    journal = "Nucl. Phys. B",
    volume = "474",
    pages = "343--360",
    year = "1996"
}

@article{Stimper2023, 
  author = {Vincent Stimper and David Liu and Andrew Campbell and Vincent Berenz and Lukas Ryll and Bernhard Schölkopf and José Miguel Hernández-Lobato}, 
  title = {normflows: A PyTorch Package for Normalizing Flows}, 
  journal = {Journal of Open Source Software}, 
  volume = {8},
  number = {86}, 
  pages = {5361}, 
  publisher = {The Open Journal}, 
  doi = {10.21105/joss.05361}, 
  url = {https://doi.org/10.21105/joss.05361}, 
  year = {2023}
}

@article{Dinh:2016RealNVP,
  author        = {Dinh, Laurent and Sohl-Dickstein, Jascha and Bengio, Samy},
  title         = {Density Estimation Using Real {NVP}},
  journal       = {arXiv preprint arXiv:1605.08803},
  year          = {2016},
  eprint        = {1605.08803},
  archivePrefix = {arXiv},
  primaryClass  = {cs.LG},
}

@article{Kreuzer:2000qv,
    author = "Kreuzer, Maximilian and Skarke, Harald",
    title = "{Reflexive polyhedra, weights and toric Calabi-Yau fibrations}",
    eprint = "math/0001106",
    archivePrefix = "arXiv",
    reportNumber = "HUB-EP-00-03, TUW-00-01",
    doi = "10.1142/S0129055X0200120X",
    journal = "Rev. Math. Phys.",
    volume = "14",
    pages = "343--374",
    year = "2002"
}

@article{Gendler:2022ztv,
    author = "Gendler, Naomi and Heidenreich, Ben and McAllister, Liam and Moritz, Jakob and Rudelius, Tom",
    title = "{Moduli space reconstruction and Weak Gravity}",
    eprint = "2212.10573",
    archivePrefix = "arXiv",
    primaryClass = "hep-th",
    reportNumber = "ACFI-T22-10",
    doi = "10.1007/JHEP12(2023)134",
    journal = "JHEP",
    volume = "12",
    pages = "134",
    year = "2023"
}

@article{Gendler:2023ujl,
    author = "Gendler, Naomi and MacFadden, Nate and McAllister, Liam and Moritz, Jakob and Nally, Richard and Schachner, Andreas and Stillman, Mike",
    title = "{Counting Calabi-Yau Threefolds}",
    eprint = "2310.06820",
    archivePrefix = "arXiv",
    primaryClass = "hep-th",
    reportNumber = "CERN-TH-2023-189",
    month = "10",
    year = "2023"
}

@article{Chandra:2023afu,
    author = "Chandra, Aditi and Constantin, Andrei and Fraser-Taliente, Cristofero S. and Harvey, Thomas R. and Lukas, Andre",
    title = "{Enumerating Calabi-Yau Manifolds: Placing Bounds on the Number of Diffeomorphism Classes in the Kreuzer-Skarke List}",
    eprint = "2310.05909",
    archivePrefix = "arXiv",
    primaryClass = "hep-th",
    doi = "10.1002/prop.202300264",
    journal = "Fortsch. Phys.",
    volume = "72",
    number = "5",
    pages = "2300264",
    year = "2024"
}

@article{Braun:2017nhi,
    author = "Braun, Andreas P. and Long, Cody and McAllister, Liam and Stillman, Michael and Sung, Benjamin",
    title = "{The Hodge Numbers of Divisors of Calabi-Yau Threefold Hypersurfaces}",
    eprint = "1712.04946",
    archivePrefix = "arXiv",
    primaryClass = "hep-th",
    month = "12",
    year = "2017"
}

@article{Blumenhagen:2008zz,
    author = "Blumenhagen, Ralph and Braun, Volker and Grimm, Thomas W. and Weigand, Timo",
    title = "{GUTs in Type IIB Orientifold Compactifications}",
    eprint = "0811.2936",
    archivePrefix = "arXiv",
    primaryClass = "hep-th",
    reportNumber = "MPP-2008-144, DIAS-STP-08-15, SLAC-PUB-13466",
    doi = "10.1016/j.nuclphysb.2009.02.011",
    journal = "Nucl. Phys. B",
    volume = "815",
    pages = "1--94",
    year = "2009"
}

@article{Rogers:2020ltq,
    author = "Rogers, Keir K. and Peiris, Hiranya V.",
    title = "{Strong Bound on Canonical Ultralight Axion Dark Matter from the Lyman-Alpha Forest}",
    eprint = "2007.12705",
    archivePrefix = "arXiv",
    primaryClass = "astro-ph.CO",
    doi = "10.1103/PhysRevLett.126.071302",
    journal = "Phys. Rev. Lett.",
    volume = "126",
    number = "7",
    pages = "071302",
    year = "2021"
}

@article{Rogers:2020cup,
    author = "Rogers, Keir K. and Peiris, Hiranya V.",
    title = "{General framework for cosmological dark matter bounds using $N$-body simulations}",
    eprint = "2007.13751",
    archivePrefix = "arXiv",
    primaryClass = "astro-ph.CO",
    doi = "10.1103/PhysRevD.103.043526",
    journal = "Phys. Rev. D",
    volume = "103",
    number = "4",
    pages = "043526",
    year = "2021"
}

@article{Fernandez:2023grg,
    author = "Fernandez, M. A. and Bird, Simeon and Ho, Ming-Feng",
    title = "{Cosmological constraints from the eBOSS Lyman-{\ensuremath{\alpha}} forest using the PRIYA simulations}",
    eprint = "2309.03943",
    archivePrefix = "arXiv",
    primaryClass = "astro-ph.CO",
    doi = "10.1088/1475-7516/2024/07/029",
    journal = "JCAP",
    volume = "07",
    pages = "029",
    year = "2024"
}

@article{Walther:2024tcj,
    author = "Walther, Michael and others",
    title = "{Emulating the Lyman-Alpha forest 1D power spectrum from cosmological simulations: new models and constraints from the eBOSS measurement}",
    eprint = "2412.05372",
    archivePrefix = "arXiv",
    primaryClass = "astro-ph.CO",
    doi = "10.1088/1475-7516/2025/05/099",
    journal = "JCAP",
    volume = "05",
    pages = "099",
    year = "2025"
}

@article{Preston:2025tyl,
    author = "Preston, Calvin and Rogers, Keir K. and Amon, Alexandra and Efstathiou, George",
    title = "{Prospects for disentangling dark matter with weak lensing}",
    eprint = "2505.02233",
    archivePrefix = "arXiv",
    primaryClass = "astro-ph.CO",
    doi = "10.1093/mnras/staf1321",
    journal = "Mon. Not. Roy. Astron. Soc.",
    volume = "2698",
    pages = "2713",
    year = "2025"
}
\end{document}